\documentclass[11pt,a4paper]{article}
\pdfoutput = 1
\usepackage{jheppub}

\usepackage{bm}
 
\newcommand{\mi}{\mathrm{i}}

\newcommand{\as}{\alpha_{\mathrm{s}}}

\newcommand{\LA}{\mathrm{A}}

\newcommand{\LE}{\mathrm{E}}
\newcommand{\LF}{\mathrm{F}}

\newcommand{\LR}{\mathrm{R}}

\newcommand{\La}{\mathrm{a}}
\newcommand{\Lb}{\mathrm{b}}
\newcommand{\Lc}{\mathrm{c}}

\newcommand{\Lf}{\mathrm{f}}

\def\ket#1{\big|{#1}\big\rangle}
\def\bra#1{\big\langle{#1}\big|}
\def\brax#1{\big\langle{#1}}   

\def\sket#1{\big|{#1}\big)}
\def\sbra#1{\big({#1}\big|}
\def\sbrax#1{\big({#1}}        

\title{Parton shower evolution with subleading color}

\author[a]{Zolt\'an Nagy}

\author[b]{and Davison E.\ Soper}

\affiliation[a]{
DESY\\
Notkestrasse 85\\
22607 Hamburg, Germany
}

\affiliation[b]{
Institute of Theoretical Science\\
University of Oregon\\
Eugene, OR  97403-5203, USA
}

\emailAdd{Zoltan.Nagy@desy.de}
\emailAdd{soper@uoregon.edu}

\abstract{
Parton shower Monte Carlo event generators in which the shower evolves from hard splittings to soft splittings generally use the leading color approximation, which is the leading term in an expansion in powers of $1/N_\Lc^2$, where $N_\Lc = 3$ is the number of colors. We introduce a more general approximation, the LC+ approximation, that includes some of the color suppressed contributions. There is a cost: each generated event comes with a weight. There is a benefit: at each splitting the leading soft$\times$collinear singularity and the leading collinear singularity are treated exactly with respect to color. In addition, an LC+ shower can start from a state of the color density matrix in which the bra state color and the ket state color do not match.
}

\preprint{DESY 12-010}

\begin{document}
\maketitle


\section{Introduction}
\label{sec:Intro}

Partons carry momentum, flavor, spin, and color. All of these quantum numbers are represented in parton shower Monte Carlo event generators like {\sc Pythia} \cite{Pythia}, {\sc Herwig} \cite{Herwig}, and {\sc Sherpa} \cite{Sherpa}. Spin and color do not fit easily into the event generator format because quantum interference between different spin and color states is important, even in the limit of very soft or collinear parton splittings. In this paper, we address the issue of color.

In previous work \cite{NSshower}, we have shown how the evolution equations for a parton shower can be formulated in a way that fully includes spin and color. The resulting integrals can, in principle, be evaluated by Monte Carlo integration. However, simple Monte Carlo methods will not be practical when there are many splittings to be represented. In ref.~\cite{NSspinless}, we found that these same evolution equations simplify if we average over spins and take the leading color limit, dropping all terms proportional to $1/N_\Lc^2$, where $N_\Lc = 3$ is the number of colors. Then one gets evolution equations that can be realized as a Markov process with positive probabilities. In ref.~\cite{NSspin}, we described a method for incorporating spin interference that we believe is practical. We are currently working on code to realize all of this. In order to keep the exposition in this paper as simple as possible, we average over spins and concentrate on color only.

The problem of treating color interference is, in our experience, more difficult. One method is to order gluon emissions according to emission angle, which takes into account quite a lot of the physics of color coherence at the cost of approximating the full dependence of the emissions on the emission angles \cite{AngleOrdered}. With the approximations, this does give positive splitting probabilities. The other main method is to simply drop terms that correspond to interference in the color space on the grounds that these terms are suppressed by factors $1/N_\Lc^2$. This is the standard leading color (LC) approximation.

In this paper, we describe a less restrictive approximation, which we call the LC+ approximation. There is a cost to using the LC+ approximation in place of the LC approximation: one then gets shower events with weights, which can be negative. However, we argue that the deleterious effect of the weights on the numerical performance of the algorithm can be controlled -- for instance by switching back to the LC approximation after some number of parton splittings. There are several advantages to the LC+ approximation. First, the color changes in collinear emissions and in emissions that are both soft and collinear are treated exactly. The only approximation is in wide angle soft emissions.   Second, LC+ evolution can be started from any partonic color state, including states in which partons in the quantum amplitude have one color configuration while partons in the conjugate quantum amplitude have a different color configuration. The LC approximation cannot work with this kind of interference state. Third, the LC+ approximation is quite flexible, so that it could be applied within any parton shower program based on color dipoles, such as {\sc Pythia} or {\sc Sherpa}. For this reason, we believe that the method is of quite general interest.

We find that to state the method precisely, we need a fairly elaborate formalism based on linear operators on a certain vector space. We borrow this formalism from Refs.~\cite{NSshower,NSspinless,NSspin}. However, the essence of the LC+ approximation can be understood from a consideration of standard Feynman-like diagrams that represent color flow. For this reason, we provide a heuristic introduction to the approximation in sections~\ref{sec:Statement}, \ref{sec:LCintro}, and \ref{sec:LCplusintro}. We then turn to the more detailed, operator based, analysis in sections~\ref{sec:operatorbased}, \ref{sec:LCplusfull}, and \ref{sec:weights}. In the LC+ approach, there are different color states for the amplitude $\ket{\{c\}_m}$ and the conjugate amplitude $\bra{\{c'\}_m}$. In section~\ref{sec:end}, we describe how to get back to diagonal configurations, $\{c'\}_m = \{c\}_m$ at the end of the shower. In section~\ref{sec:perturbativecolor} we describe how one can include in perturbation theory the effects that are neglected in the LC+ approximation. In section~\ref{sec:virtphase} we include the effects of the color phase induced by exchanging virtual soft gluons. Finally, in section~\ref{sec:conclusions}, we conclude the main part of the exposition. We treat some more technical topics in three appendices.

\section{Statement of the problem}
\label{sec:Statement}

In this section, we seek to describe how color evolves in a parton shower, illustrating why it is not simple to describe the color evolution exactly. First, let us note that, in a perturbative treatment, a parton carries momentum, flavor, spin and color. We can imagine that the momenta of partons at the end of the shower are quite precisely measured. Then the momenta of intermediate partons in a shower are also well determined. Thus we do not need to consider interference between states in which intermediate partons have different momenta $p$ and $p'$. Since the flavor of a mother parton in a splitting is determined by the flavors of its daughters, we also do not need to consider interference between states in which intermediate partons have different flavors $f$ and $f'$. However, this argument does not hold for spin and color. For instance, in a splitting $1 \to 2 + 3$ for gluons with colors $a_1$, $a_2$ and $a_3$, the matrix element is proportional to $f_{a_1,a_2,a_3}$ and for fixed $a_2$ and $a_3$, $f_{a_1,a_2,a_3}$ can be nonzero for more than one value of $a_1$. Thus we need to allow for quantum interference between different color states of the same parton. We also need to allow for quantum interference between different spin states. 

\subsection{Evolution without color or spin}

Let us consider what happens in a parton shower that evolves from hard splittings to soft splittings.\footnote{We have in mind something like {\sc Pythia} or {\sc Sherpa}. The program {\sc Herwig}, based on ordering in angles, is rather different.} To get started, we ignore both spin and color. We define a ``shower time'' variable $t$ such that an initial hard parton scattering happens at $t = 0$ and then at each interval $dt$ a parton has some probability of splitting to become two partons. Harder splittings happen at smaller $t$ values, successively softer splittings happen at larger $t$ values. For instance, $\exp(-t)$ can be proportional to the virtuality in a splitting or to the transverse momentum of the daughters relative to the mother parton direction. As implemented in a computer program, the partonic system always has a definite state. Ignoring spin and color, the state of the system is described by the momenta and flavors,
\begin{equation}
\label{eq:pfstate}
\{p,f\}_m
= \{p_\La,f_\La,p_\Lb,f_\Lb,p_1,f_1,\dots,p_m,f_m\}
\;\;.
\end{equation}
Here $p_\La$ is the momentum of the incoming parton from hadron A, equal to a fraction $\eta_\La$ of the hadron momentum, $p_\Lb$ is the momentum of the incoming parton from hadron B, and the $p_i$ are the momenta of $m$ outgoing partons. The flavors of the partons are denoted by discrete flavor variables $f$. In each shower time interval $dt$, there is a certain probability that the system will switch to a new state. What actually happens in a given simulated event is determined by generated pseudo-random numbers. This means that in an ensemble of simulated events, there is an probability $\rho(\{p,f\}_m,t)$ that the partonic system is in a certain state at shower time $t$. This probability distribution then evolves with $t$. Let us denote by $\sket{\rho(t)}$ the function $\rho$ at time $t$ considered as a vector in the space of functions of $m$ and $\{p,f\}_m$. Then the evolution equation for the probabilities has the form of a linear equation
\begin{equation}
\label{eq:evolution}
\frac{d}{dt}\sket{\rho(t)}
= [{\cal H}_I(t) - {\cal V}(t)]\sket{\rho(t)}
\;\;.
\end{equation}
Here ${\cal H}_I(t)$ is a linear operator on the space of probability distributions. The operator ${\cal H}_I(t)$ corresponds to the splitting probabilities chosen for shower evolution. (There are many possible choices.) Then ${\cal V}(t)$ is another linear operator that is constructed from ${\cal H}_I(t)$ so as to conserve probability in the shower evolution. The Sudakov factor that represents the probability that there was no splitting between times $t_1$ and $t_2$ is $\exp(-\int_{t_1}^{t_2} dt\ {\cal V}(t))$. We will return to this in later sections. For the moment, all that we need to know is that the evolution of the probability function $\rho$ is determined by parton splitting probabilities.

\subsection{Ignoring spin}

Now we need to consider spin and color. We have addressed the spin problem in ref.~\cite{NSspin}. Combining the spin treatment of ref.~\cite{NSspin} with the discussion of this paper is not difficult, but adds a layer of complexity. In order to keep this paper as simple as possible, we ignore spin by supposing that we sum over the spins of the daughter partons in a splitting and average over the spins of the mother parton. Thus in this paper we address the color problem but not the spin problem. 

\subsection{The color density operator}

The natural language for describing an evolving probability distribution is statistical mechanics. Indeed, eq.~(\ref{eq:evolution}) is a standard sort of equation for evolution in statistical mechanics. In this paper, we want to include quantum color, so we need {\em quantum} statistical mechanics. Thus we need a density operator that depends on $t$ and is an operator on the space of color states of a possibly large number of partons. The density operator then has the form
\begin{equation}
\label{eq:rhodef}
\rho(\{p,f\}_m,t) 
= \sum_{\{c\}_m ,\{c'\}_m}\rho(\{p,f,c',c\}_m,t)\,\ket{\{c\}_m}\bra{\{c'\}_m}
\;\;.
\end{equation}
Here $\ket{\{c\}_m}$ and $\ket{\{c'\}_m}$ are basis vectors for the quantum color space. The color configuration $\{c\}_m$ of the quantum amplitude is, in general, different from the color configuration $\{c'\}_m$ of the conjugate quantum amplitude. Thus the function $\rho(\{p,f,c',c\}_m,t)$ depends on two sets of color indices. The density operator $\rho(\{p,f\}_m,t)$ can be regarded as a vector in the space of functions of $\{p,f\}_m$ with values in the space of operators on the quantum color space. It obeys an evolution equation of the form (\ref{eq:evolution}). We simply have to reinterpret what $\sket{\rho(t)}$ means. Of course, with color involved, the detailed structure of eq.~(\ref{eq:evolution}) is now more complicated. To understand the color structure, we need to first say something about the color basis states.

\begin{figure}
\centerline{\includegraphics[width = 10 cm]{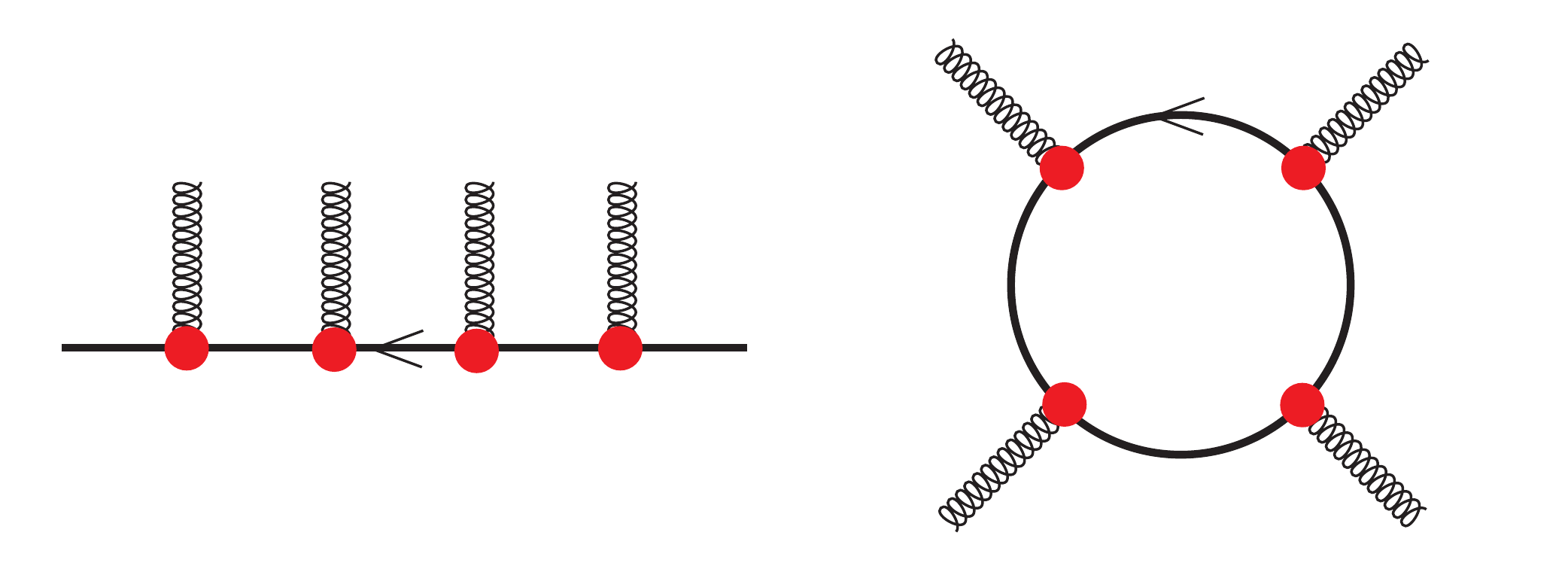}}

\caption{Illustration of an open string color basis state (left) and a closed string color basis state (right). These diagrams represent the color matrices without the normalization factors $N(S)^{-1/2}$.
}
\label{fig:strings}
\end{figure}

Our analysis in this paper makes use of the standard color basis defined by two sorts of vectors \cite{colorbasis,NSshower}. First, there are open string vectors
\begin{equation}
\label{eq:openstring}
\Psi(S)^{\{ a \}} 
= N(S)^{-1/2} [t^{a_2} t^{a_3}\cdots t^{a_{n-1}}]_{a_1,a_n}
\;\;.
\end{equation}
Here $a_1$ is a quark color index, $a_n$ is an antiquark color index, and the other $a_i$ are gluon color indices. The $t^a$ are standard SU(3) color matrices in the fundamental representation. Also, $N(S) = N_\Lc C_\LF^{n-2}$ is a normalization factor such that $\brax{\Psi}\ket{\Psi} = 1$. Second, there are closed string vectors
\begin{equation}
\label{eq:closedstring}
\Psi(S)^{\{ a\}} 
= N(S)^{-1/2}\, {\rm Tr}[t^{a_1} t^{a_2}\cdots t^{a_{n}}]
\;\;.
\end{equation}
Here all of the $a_i$ are gluon color indices. Also, $N(S) = C_\LF^{n}$ is a normalization factor such that $\brax{\Psi}\ket{\Psi} = 1 - [-1/(N_\Lc^2-1)]^{n-1}$. This normalization factor is close to 1, with a small correction. The most general color basis vector, which we denote by $\ket{\{c\}_m}$, is a product of these two kinds of units. The open string and closed string basis vectors are illustrated in figure~{\ref{fig:strings}}.

The color basis states are normalized to $\brax{\{c\}_m}\ket{\{c\}_m} = 1$ or to $\brax{\{c\}_m}\ket{\{c\}_m} \approx 1$ with very small corrections. They are not, however, generally orthogonal. However, when $\{c\}_m$ and $\{c'\}_m$ are different, one finds that $\brax{\{c'\}_m}\ket{\{c\}_m} = {\cal O}(1/N_\Lc^2)$. That is, the basis vectors are orthogonal in the $N_\Lc \to \infty$ limit.

\begin{figure}
\centerline{\includegraphics[width = 4.6 cm]{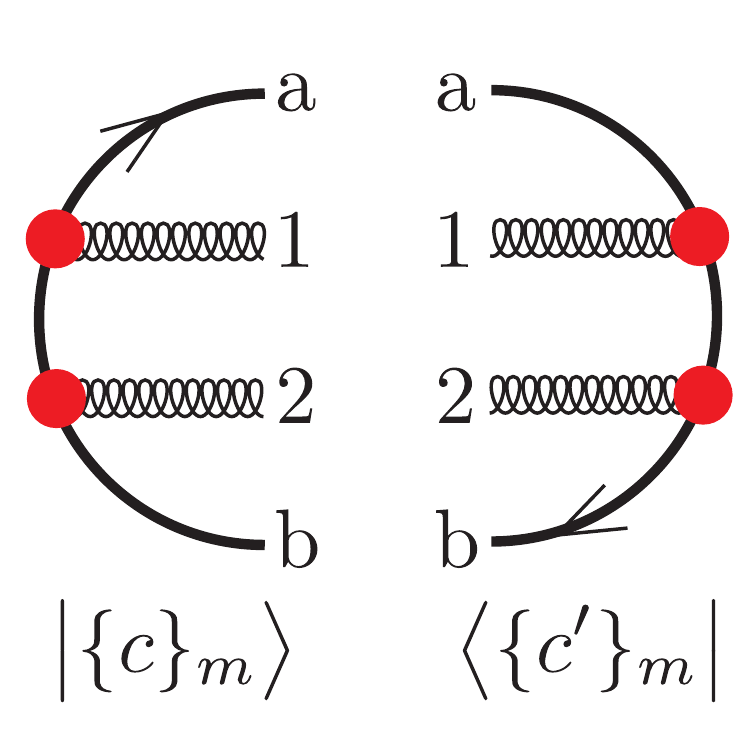}}

\caption{Illustration of the color density operator. We depict $\ket{\{c\}_m}\bra{\{c'\}_m}$ with a quark, an antiquark, and two gluons. Here $\{c\}_m = \{c'\}_m$.
}
\label{fig:colordensity1}
\end{figure}

Now let's look at an example. In figure~\ref{fig:colordensity1}, we depict $\ket{\{c\}_m}\bra{\{c'\}_m}$ for a color structure that arises in the hard scattering process $\bar q(\La) +  q(\Lb) \to g(1)+ g(2)$. Both $\ket{\{c\}_m}$ and $\bra{\{c'\}_m}$ show the color state that corresponds to $t$-channel $\bar q$ exchange. Thus we have a diagonal contribution to the color density operator, with $\{c\}_m = \{c'\}_m$. 

\begin{figure}
\centerline{\includegraphics[width = 4.6 cm]{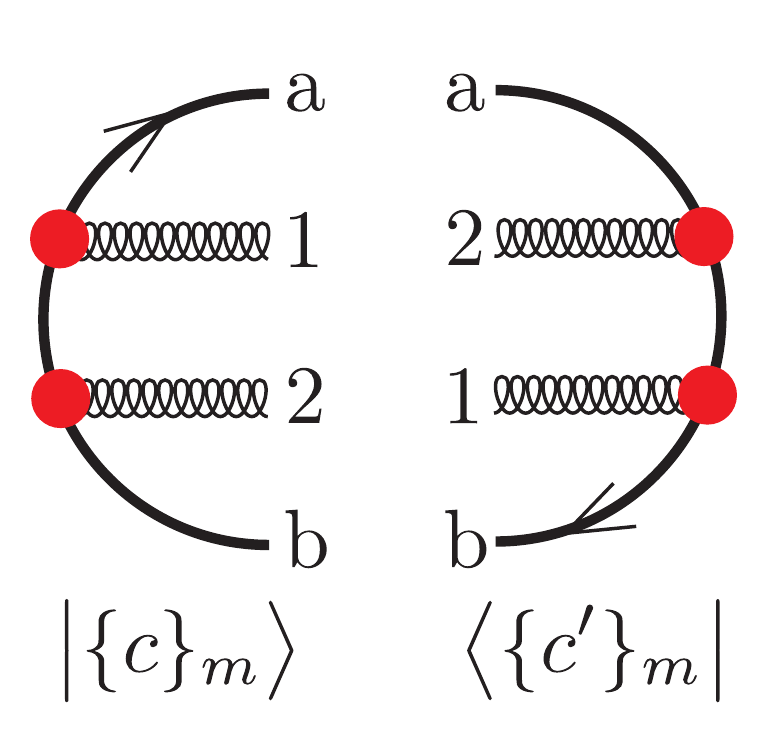}}

\caption{Illustration of the color density operator as in figure~\ref{fig:colordensity1}. In the case illustrated, we have $\{c\}_m \ne \{c'\}_m$ because we have switched the positions of gluons 1 and 2 in $\{c'\}_m$. 
}
\label{fig:colordensity2}
\end{figure}

One can also have $u$-channel $\bar q$ exchange, which amounts to exchanging gluons 1 and 2. In figure~\ref{fig:colordensity2}, we still have the $t$-channel diagram for $\{c\}_m$, but now we illustrate the contribution from the $u$-channel diagram for $\{c'\}_m$. This gives an off-diagonal contribution to the color density operator, with $\{c\}_m \ne \{c'\}_m$. Each of the two contributions to the color density operator shown in the two figures can serve as the starting point for a parton shower. Since their color structures are different, they should generate different showering.

\begin{figure}
\centerline{\includegraphics[width = 9.4 cm]{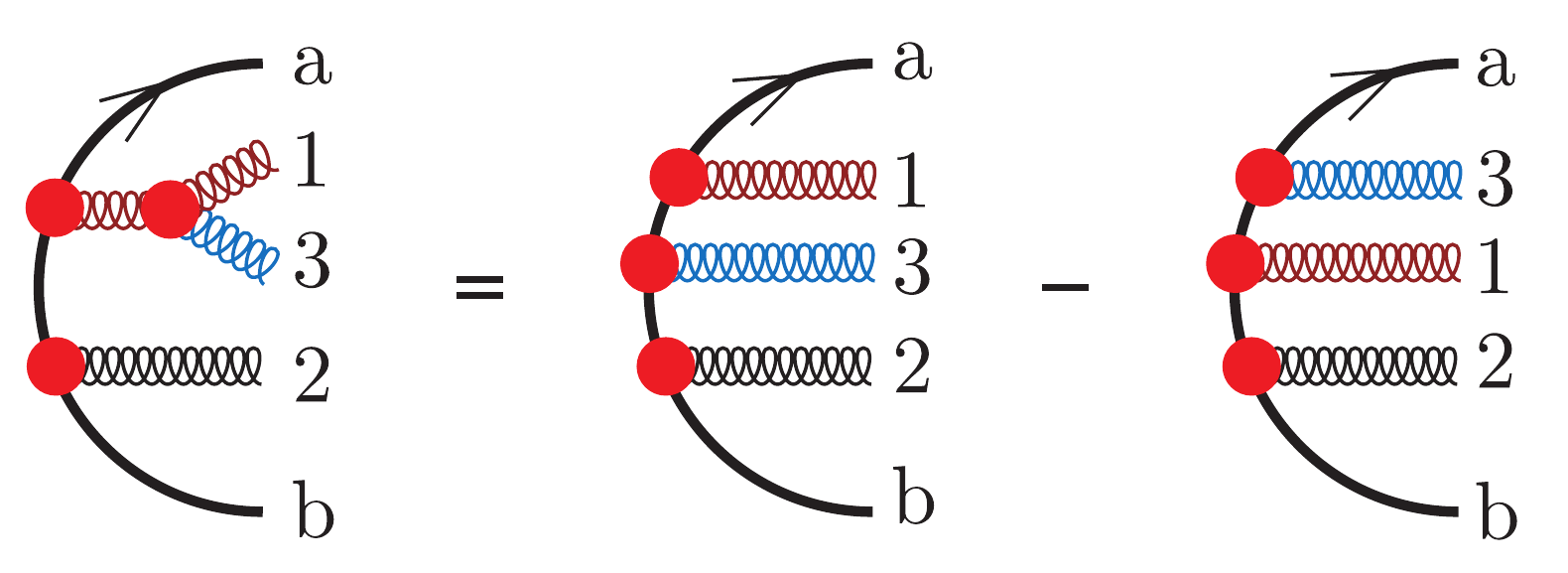}}

\caption{Identity for the color dependence of gluon splitting.
}
\label{fig:splittingidentity}
\end{figure}

\subsection{Color structure of the parton splitting operator}

Having seen the meaning of the color density operator, we can now consider what happens in shower evolution when a gluon splits. In figure~\ref{fig:splittingidentity}, we show the color structure when mother gluon 1 splits into daughter gluon 1 and daughter gluon 3. Using the identity $\mi f_{a_1 a_3 c}\, t_c = t_{a_1} t_{a_3} - t_{a_3} t_{a_1}$, we find that there is a term in which the new gluon 3 is inserted below gluon 1 on the quark line minus a term in which the new gluon is inserted above gluon 1 on the quark line.\footnote{In our diagrams, the three gluon vertex is $\mi f_{a_1 a_3 c}$ with $(a_1 a_3 c)$ ordered clockwise around the vertex.}

\begin{figure}
\centerline{\includegraphics[width = 13.4 cm]{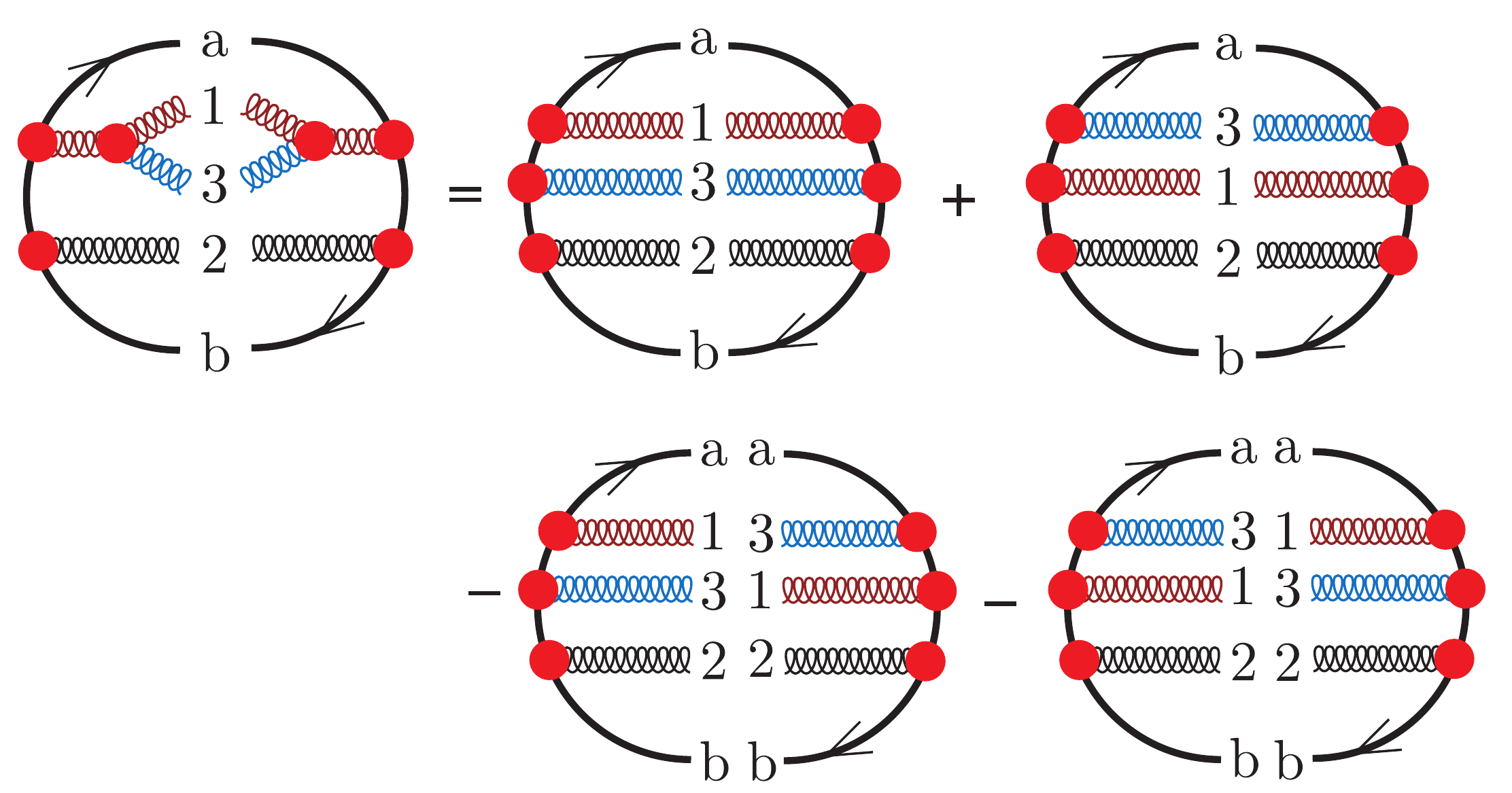}}

\caption{Identity for the color dependence of the splitting of gluon 1 in both the bra state and the ket state.
}
\label{fig:splittingidentityketbra1}
\end{figure}

\begin{figure}
\centerline{\includegraphics[width = 14.4 cm]{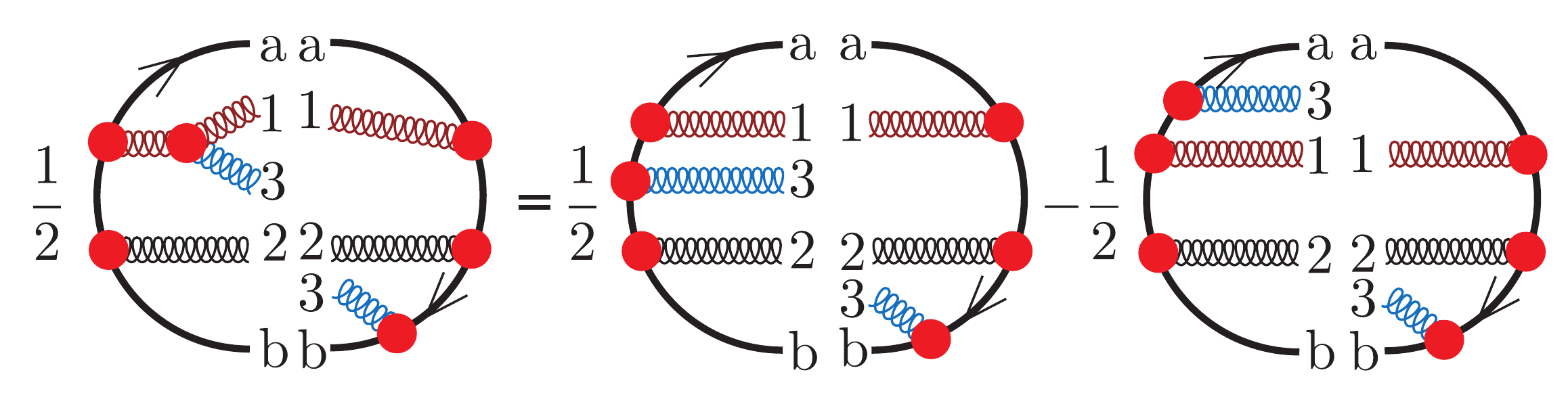}}

\caption{Identity for the color dependence of the splitting of gluon 1 in the ket state with the participation of helper parton b in the bra state.
}
\label{fig:splittingidentityketbra2}
\end{figure}
\begin{figure}
\centerline{\includegraphics[width = 14.8 cm]{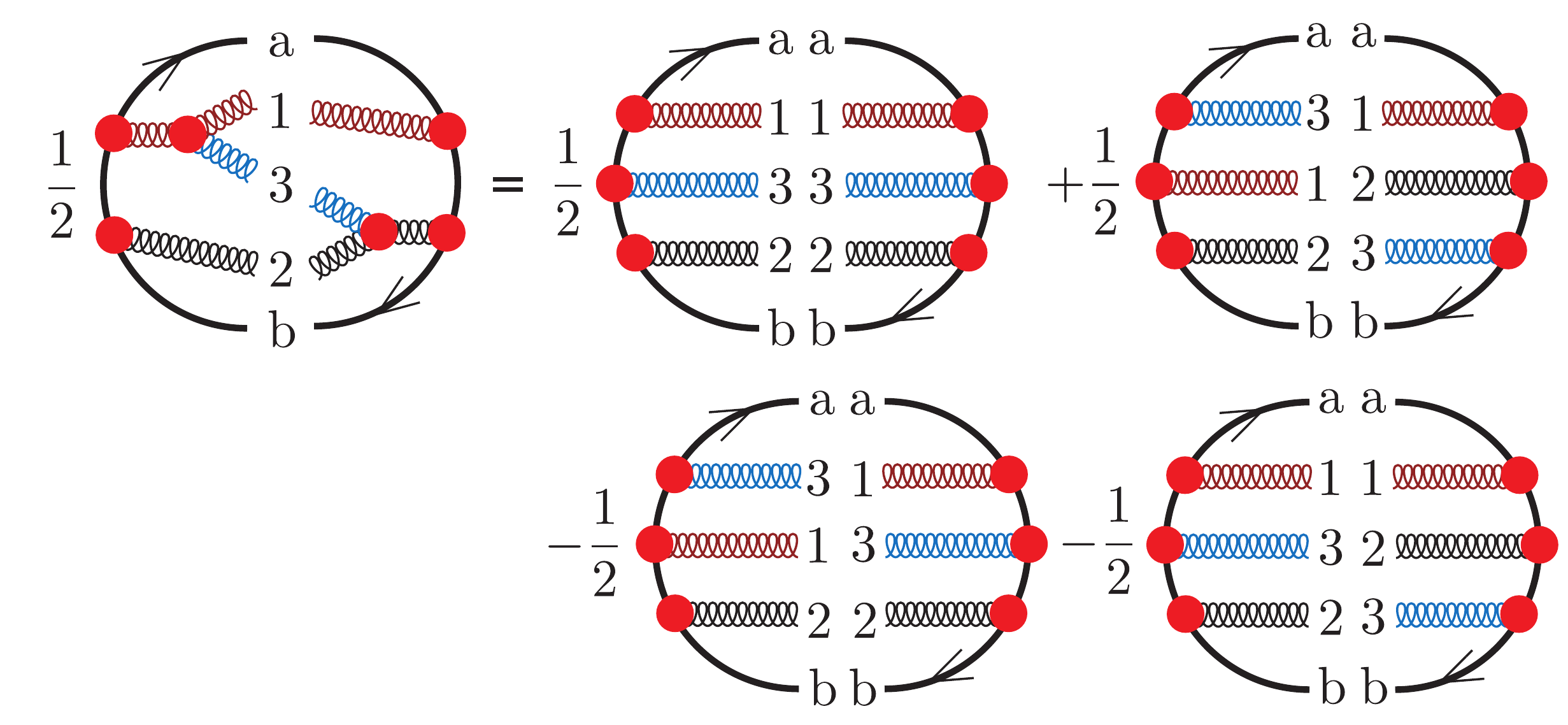}}

\caption{Identity for the color dependence of the splitting of gluon 1 in the ket state with the participation of helper parton 2 in the bra state.
}
\label{fig:splittingidentityketbra3}
\end{figure}

Applying this identity, we see in figure~\ref{fig:splittingidentityketbra1} that letting mother gluon 1 split in both the ket state and the bra state in figure~\ref{fig:colordensity1} gives contributions with four new $\ket{\{c\}_{m+1}}\bra{\{c'\}_{m+1}}$ terms. Even when we use only the simplest kind of splitting and even though we start with a color diagonal density operator, we generate off diagonal terms in the color density operator.

There can also be quantum interference between emissions of a soft gluon from two different partons. We illustrate this in figure~\ref{fig:splittingidentityketbra2}. Mother parton 1 emits gluon 3 in the ket state, while the antiquark with label b emits gluon 3 in the bra state. In a physical gauge, the matrix elements for this are large as long as gluon 3 is soft. One can consider that a color dipole consisting of partons 1 and b emits the gluon. In a dipole shower, one wants to account for emission from each dipole (insofar as complications from color allow). In a dipole antenna shower like {\sc Vincia} \cite{Vincia}, one keeps the dipoles together as a unit. In this paper, we follow the method of a partitioned dipole shower like {\sc Pythia}, which partitions the radiation pattern into two terms. One term is singular when the soft gluon is collinear with parton 1, the emitting parton, but is not singular when the soft gluon is collinear with parton b, the helper parton. The other term is singular when the soft gluon is collinear with parton b but is not singular when the soft gluon is collinear with parton 1. For our present discussion, let us consider parton 1 to be the emitting parton. On the right hand side of the figure, we have applied the color identity of figure~\ref{fig:splittingidentity} to expand this contribution in our standard color basis. We see that we get two off diagonal contributions.

Parton 3 can also be emitted from parton 1 with parton 2 as the helper parton. This case is illustrated in figure~\ref{fig:splittingidentityketbra3}. When we expand in five-parton color basis states, there are four contributions. One of them is color diagonal.

Let us summarize. Using a few simple examples, we have explored the color content of the splitting operator ${\cal H}_I(t)$ in eq.~(\ref{eq:evolution}). There are two sorts of graph, direct graphs like figure \ref{fig:splittingidentityketbra1} and interference graphs like figures \ref{fig:splittingidentityketbra2} and \ref{fig:splittingidentityketbra3}. The color structure of both kinds of graphs is non-trivial. Even if we start from a color diagonal contribution to the color density operator, each splitting creates off-diagonal contributions. With each new splitting, the color state becomes more complicated.

\subsection{Color structure of the virtual splitting operator}

What about the operator ${\cal V}(t)$ in eq.~(\ref{eq:evolution})? This operator maps a state of the color density operator with $m$ final state partons into another state with $m$ final state partons. Physically, it represents virtual Feynman graphs with hardness scale characteristic of the current shower time $t$. We determine what ${\cal V}(t)$ should be by demanding that shower evolution conserve probability. This means that 
\begin{equation}
\label{eq:unitarity}
\sbra{1} [{\cal H}_I(t) - {\cal V}(t)]\sket{\rho(t)}
= 0
\;\;,
\end{equation}
where multiplying by $\sbra{1}$ is a convenient way of writing the instruction to take the trace of the color statistical operator and integrate over all of the parton variables.

Note that taking the trace of the color statistical operator means using
\begin{equation}
\label{eq:trace}
{\rm Tr}[\rho(\{p,f\}_m,t)]
= \sum_{\{c\}_m ,\{c'\}_m}\rho(\{p,f,c,c'\}_m,t)\,\brax{\{c'\}_m}\ket{\{c\}_m}
\;\;.
\end{equation}
We get the inner product $\brax{\{c'\}_m}\ket{\{c\}_m}$. It seems unfortunate that  $\brax{\{c'\}_m}\ket{\{c\}_m}$ does not vanish when $\{c'\}_m \ne \{c\}_m$. However, this inner product is suppressed by powers of $1/N_\Lc^2$ when $\{c'\}_m\ne \{c\}_m$.

It is possible to satisfy eq.~(\ref{eq:trace}) and, at the same time, match the color structure of virtual Feynman graphs by letting ${\cal V}$ have the form 
\begin{equation}
\label{eq:calVdecomposed0}
{\cal V} = (h + \mi \phi) \otimes 1 + 1 \otimes (h^\dagger - \mi \phi)
\;\;.
\end{equation}
This notation denotes that $(h + \mi \phi)$ is an operator on the ket part of the color density operator and $(h^\dagger - \mi \phi)$ is an operator on the bra part. The operator $\phi$ is hermitian and represents a color phase. We will consider $\phi$ in section~\ref{sec:virtphase}. Until then, we simply set $\phi = 0$. The operator $h$ is hermitian in the full color theory, but the LC+ approximation for it is not. For that reason, we distinguish the roles of $h$ and $h^\dagger$ in our formulas. With $\phi = 0$, we have
\begin{equation}
\label{eq:calVdecomposed}
{\cal V} = h \otimes 1 + 1 \otimes h^\dagger
\;\;.
\end{equation}
We can state this a little more precisely. If $\rho(\{p,f\}_m,t)$ is defined by eq.~(\ref{eq:rhodef}) and $\sket{\rho'} = {\cal V}\sket{\rho}$, then
\begin{equation}
\begin{split}
\label{eq:Vrhodef}
\rho'(\{p,f\}_m,t) 
={}& 
\sum_{\{c\}_m ,\{c'\}_m}\rho(\{p,f,c,c'\}_m,t)
\\& \times
\Big[
h(\{p,f\}_m,t)\ket{\{c\}_m}\bra{\{c'\}_m}
+ \ket{\{c\}_m}\bra{\{c'\}_m}h^\dagger(\{p,f\}_m,t)
\Big]
\;\;.
\end{split}
\end{equation}

\begin{figure}
\centerline{\includegraphics[width = 13.0 cm]{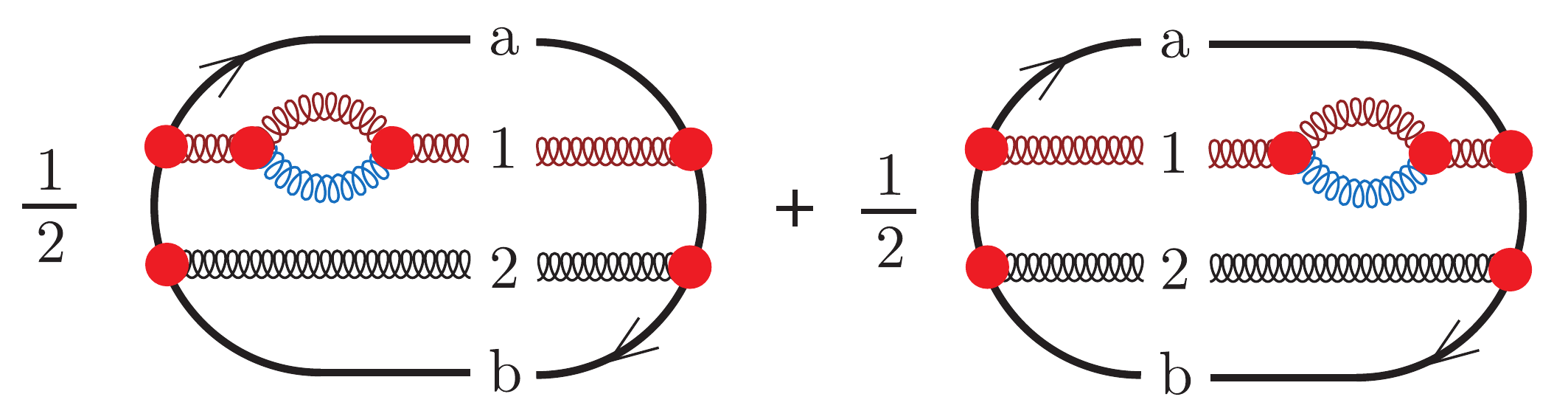}}

\caption{Color structure of the contribution to ${\cal V}(t)$ from the splitting in figure~\ref{fig:splittingidentityketbra1}.
}
\label{fig:Vketbra1}
\end{figure}

\begin{figure}
\centerline{\includegraphics[width = 3.5 cm]{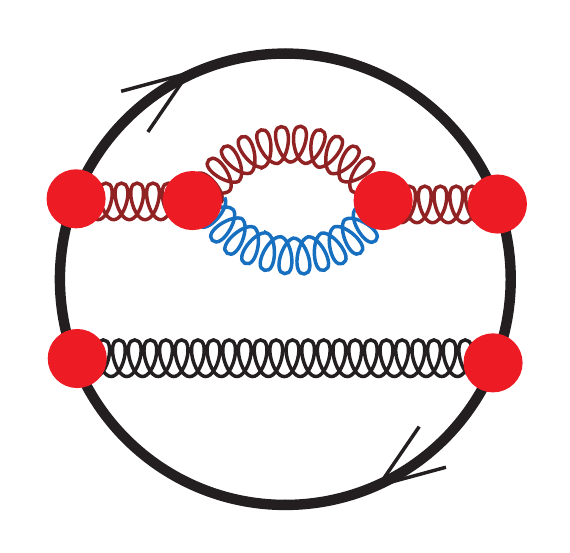}}

\caption{The color trace of figure~\ref{fig:Vketbra1} and of the left hand side of figure~\ref{fig:splittingidentityketbra1}.
}
\label{fig:coloroverlap1}
\end{figure}

Each term in ${\cal H}_I$ determines a contribution to $h$. To see how this works, consider the splitting in figure~\ref{fig:splittingidentityketbra1}. We define $h$ so that the corresponding contribution to $\sket{\rho'}$ has the color factor shown in figure~\ref{fig:Vketbra1}. To verify probability conservation, eq.~(\ref{eq:unitarity}), just take the color trace of figure~\ref{fig:Vketbra1} and compare it to the color trace of the left hand side of figure~\ref{fig:splittingidentityketbra1}. In each case we get the number represented by the color diagram in figure~\ref{fig:coloroverlap1}.

\begin{figure}
\centerline{\includegraphics[width = 4.2 cm]{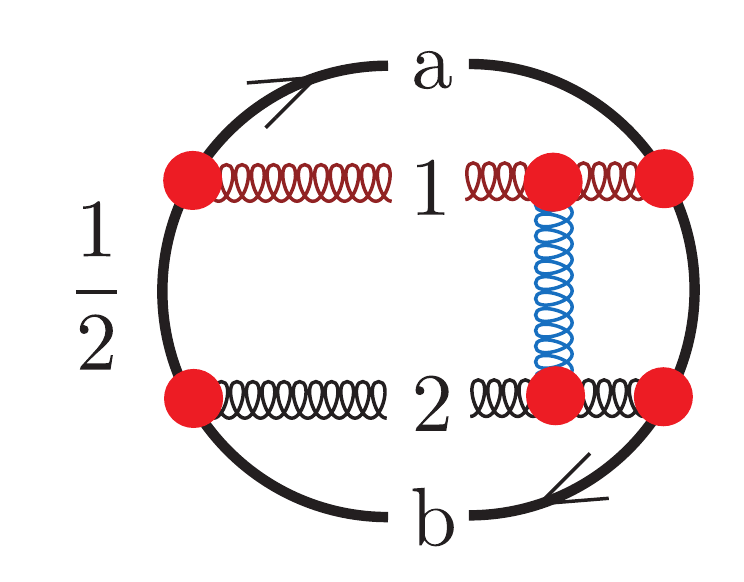}}

\caption{Color structure of the contribution to ${\cal V}(t)$ from the splitting in figure~\ref{fig:splittingidentityketbra3}.
}
\label{fig:Vketbra2}
\end{figure}

The color structure of ${\cal V}$ is more complicated in the case of interference diagrams. Consider the splitting in figure~\ref{fig:splittingidentityketbra3}, in which parton 1 in the ket state splits and we take the interference with the emission of the same gluon from helper parton 2 in the bra state. We can consider this diagram to correspond to a contribution to ${\cal V}$ in which $h^\dagger$ acts on the bra state. We define this contribution to $h^\dagger$ so that $\sket{\rho'}$ has the color factor shown in figure~\ref{fig:Vketbra2}. To verify probability conservation, just take the color trace of figure~\ref{fig:Vketbra2} and compare it to the color trace of the left hand side of figure~\ref{fig:splittingidentityketbra3}. Then the contribution to ${\cal V}$ in which $h$ acts on the ket state corresponds to the complex conjugate of the splitting in figure~\ref{fig:splittingidentityketbra3}.

Note that the contribution to ${\cal V}(t)$ corresponding to a direct splitting term in ${\cal H}_I$ is trivial. Evidently from figure~\ref{fig:Vketbra1}, this contribution to ${\cal V}(t)$ simply multiplies the color state by $C_\LA$. In contrast, the contribution to ${\cal V}(t)$ corresponding to an interference term in ${\cal H}_I$ is not trivial. When one expands the bra state in figure~\ref{fig:Vketbra2} in the color basis states, we get several contributions.\footnote{For the case depicted, there are two contributions. In other cases, there are more.} Thus, in general, ${\cal V}(t)$ is not diagonal in color. This makes an exact accounting for color difficult because ${\cal V}(t)$ enters the generation of parton showers in the form of the Sudakov factor, the time ordered exponential $\mathbb T\exp\left[-\int_{t}^{t'} d\tau\ {\cal V}(\tau)\right]$.

\section{The leading color approximation}
\label{sec:LCintro}

We have now seen something of the structure in color space of a leading order parton shower (of the partitioned dipole variety). Of course, this is not the structure of any existing computer program that models a parton shower with quantum color. There is no such program. The evolution equations make good sense, but implementing more than a couple of splitting steps as a computation is not feasible with current methods. There is, however, a simple approximation that one can apply to get a practical implementation. This is the leading color (LC) approximation.

The general idea of the leading color approximation is to neglect contributions to the probability to get a given final partonic state that are suppressed by powers of $1/N_\Lc^2$. To proceed, one replaces gluons in the $\bm 8$ representation of $SU(3)$ by gluons in the ${\bm 3}\times \bar{\bm 3}$ representation, using
\begin{equation}
\label{eq:fierz}
t^a_{ij} t^a_{i'j'} \to \frac{1}{2}\ \delta_{ij'}\,\delta_{i' j}
\;\;.
\end{equation}
One also notes that all contributions to the color density operator with $\{c\}_m  \ne \{c'\}_m$ can simply be dropped. That is because $\brax{\{c'\}_m}\ket{\{c\}_m}$ is suppressed by a power of $1/N_\Lc^2$ compared to $\brax{\{c\}_m}\ket{\{c\}_m}$ or $\brax{\{c'\}_m}\ket{\{c'\}_m}$. During shower evolution, the ket states and the bra states evolve separately, but at the end of the parton shower, when there are $N$ final state partons, one needs to use $\brax{\{c'\}_N}\ket{\{c\}_N}$ to compute a probability. With a little analysis (see section \ref{sec:colorsuppressionI}), one sees that once one has $\{c\}_m  \ne \{c'\}_m$, one can never get back to a leading $1/N_\Lc^2$ power by further parton splittings. The consequence of this is that as splitting proceeds, one simply drops $\{c\}_m  \ne \{c'\}_m$ contributions. Thus on the right hand side of figure~\ref{fig:splittingidentityketbra1}, one keeps the first two terms and drops the second two terms. On the right hand side of figure~\ref{fig:splittingidentityketbra3}, one keeps the first term and drops the other three. On the right hand side of figure~\ref{fig:splittingidentityketbra2}, one drops both terms. That is, interference between emission of a gluon from parton $l$ and from a helper parton $k$ that is not directly color connected to $l$ is neglected.

With the leading color approximation, the no-splitting operator $\cal V$ is simple. For instance, on the right hand side of figure~\ref{fig:splittingidentityketbra1} we keep the first  two terms. In each of these terms, when we take the color trace we see that adding the emitted gluon  3 creates one more color loop and thus one more factor of $N_\Lc$. Thus we can take the color operator in $\cal V$ to simply multiply the color state we started with by $C_\LA = N_\Lc$ (including two graphs, each with a factor 1/2 from eq.~(\ref{eq:fierz})).

For a splitting $q \to q + g$, the color factor in $\cal V$ would be $N_\Lc /2$ when we calculate this way, but one normally uses $C_\LF$ instead by simply multiplying the splitting probability by $2C_\LF/C_\LA$.

Strictly speaking, one would drop all $g \to q + \bar q$ splittings in the leading color approximation, but one normally keeps the part of this splitting suppressed by $1/N_\Lc$, omitting the parts suppressed by $1/N_\Lc^3$.

The leading color approximation gives a simple shower algorithm. It is also intuitively appealing. The ket states and the bra states in the color density operator always have the same $\{c\}_m$. In $\{c\}_m$, we can think of each quark or antiquark as being connected  by a color string to a neighboring gluon in a color basis state in figure~\ref{fig:strings}. Each gluon is connected to two other partons by a color string. Then emitting a new gluon means connecting it to two of the previous strings. Interference between emitting a new gluon from parton $l$ and helper parton $k$ can only occur if the two partons were color connected; then the new gluon is connected to the string that previously joined partons $l$ and $k$.

One can note two problems with the leading color approximation. First, it neglects terms suppressed by powers of $1/N_\Lc^2$. Second, it cannot start with color density operator contributions $\ket{\{c\}_m}\bra{\{c'\}_m}$ with $\{c\}_m  \ne \{c'\}_m$.

\section{Introduction to the LC+ approximation}
\label{sec:LCplusintro}

We propose in this section an improved ``LC+'' approximation that goes beyond the leading color approximation by including some of the contributions to cross sections that are suppressed by powers of $1/N_\Lc^2$. 

To go beyond the leading color approximation, one has to give up something.\footnote{More precisely, the authors do not know how to go beyond the leading color approximation while giving up nothing.} We choose to give up having an algorithm that can be implemented without having weights for events. In particular, the terms in the splitting probabilities in the LC+ approximation have both plus and minus signs. One cannot generate events with negative probabilities, so in order to include these terms one will have to generate certain splittings with positive probabilities and negative weights. The weights are then carried with the event. Now, having weighted events is not intrinsically a problem for convergence of Monte Carlo integrations to a physical answer. However, there would be a problem that would not be easy to avoid if a physical cross section to be calculated had the form $\sigma_0[1.00 - 9.01/N_\Lc^2 + \cdots]$. Then calculating the result by Monte Carlo integration would not give an accurate answer for $N_\Lc = 3$. We believe that this does not happen. If it does happen for some observable cross section, then the LC+ approximation may not work well. Of course, in this case, the standard LC approximation is giving us the wrong answer and the LC+ result would at least give an indication of problems.

We will state what the LC+ approximation is quite precisely using an operator notation in section~\ref{sec:LCplusfull}. However the main idea can be grasped quite easily using the examples that we have already explored, so we do that in this section.

\subsection{The parton splitting operator}

\begin{figure}
\centerline{\includegraphics[width = 13.6 cm]{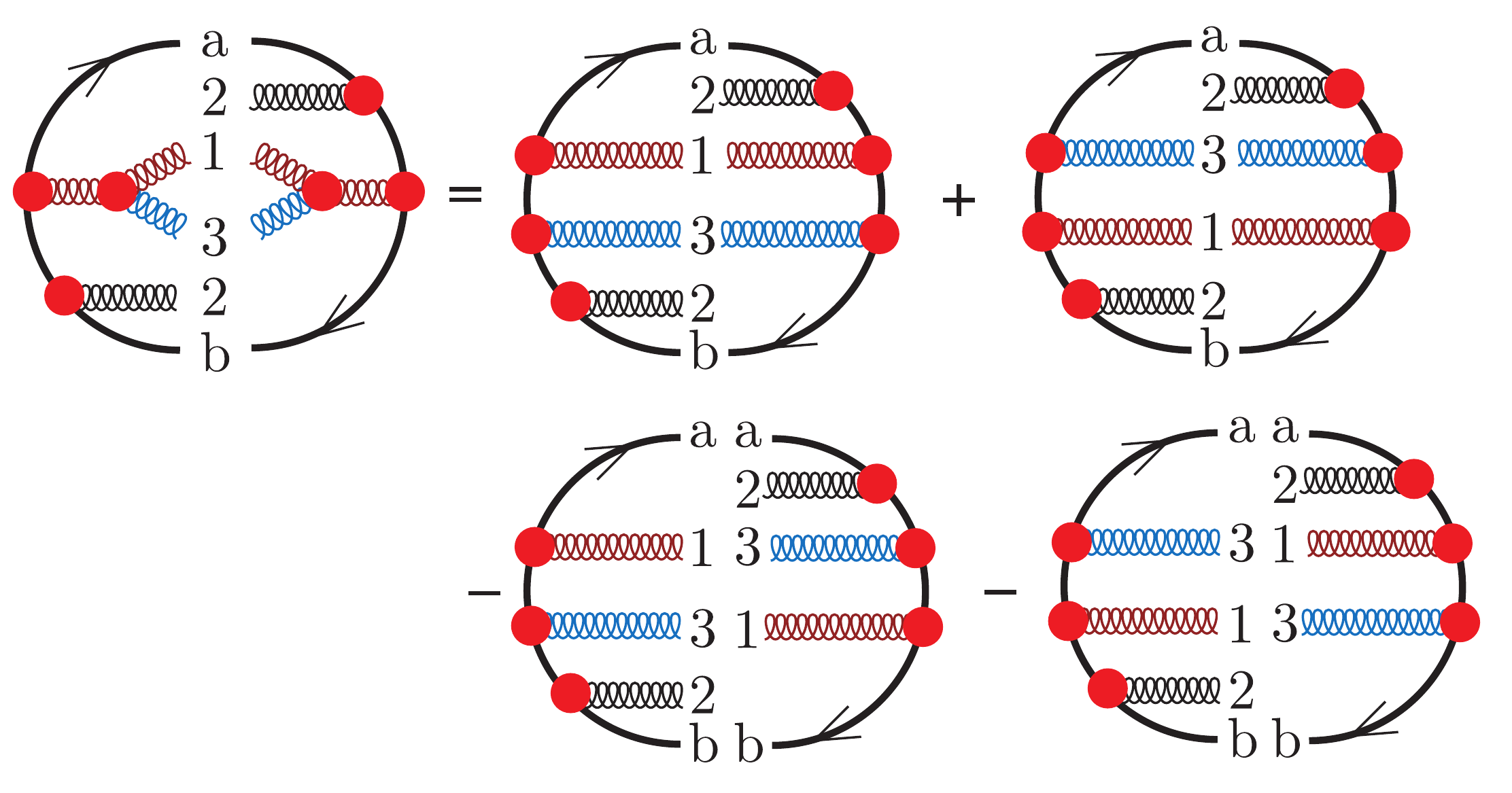}}

\caption{Identity for the color dependence of the splitting of gluon 1 in both the bra state and the ket state in the case $\{c'\}_m \ne \{c\}_m$. In the LC+ approximation, one keeps all four terms.
}
\label{fig:splittingidentityketbra4}
\end{figure}

First, the direct graph for the splitting of parton $l$ by emitting a gluon labeled $m+1$ contains (in a physical gauge) the collinear singularity that occurs when the parton momenta after the splitting obey $\hat p_{m+1} \propto \hat p_l$ and also the double singularity that occurs when $\hat p_{m+1} \to 0$ with $\hat p_{m+1} \propto \hat p_l$. We want the coefficients of the logarithms that arise from these singularities to be exactly right with respect to color. Thus we keep the color structure of these splittings exactly. This means that in the right hand side of figure~\ref{fig:splittingidentityketbra1} we keep all four terms. We can start with $\{c'\}_m \ne \{c\}_m$ as in figure~\ref{fig:splittingidentityketbra4}. Again, there are four terms and we keep them all.

Now consider splittings in which radiation of a gluon $m+1$ from parton $l$ interferes with radiation of gluon $m+1$ from helper parton $k$. Recall that we are using a partitioned dipole shower, in which the splitting functions distinguish the roles of the emitting parton and the helper parton. The emitting parton $l$ can be the parton that radiates gluon $m+1$ in the ket state. Then the helper parton $k$ radiates gluon $m+1$ in the bra state. Alternatively, the emitting parton can be the one that radiates gluon $m+1$ in the bra state, so that the helper parton radiates gluon $m+1$ in the ket state. In our examples in this section, we take the emitting parton $l$ to be the parton that radiates gluon $m+1$ in the ket state, with the helper parton $k$ being the parton that radiates gluon $m+1$ in the bra state. There are, of course, equivalent examples in which the picture is reversed.

\begin{figure}
\centerline{\includegraphics[width = 14.4 cm]{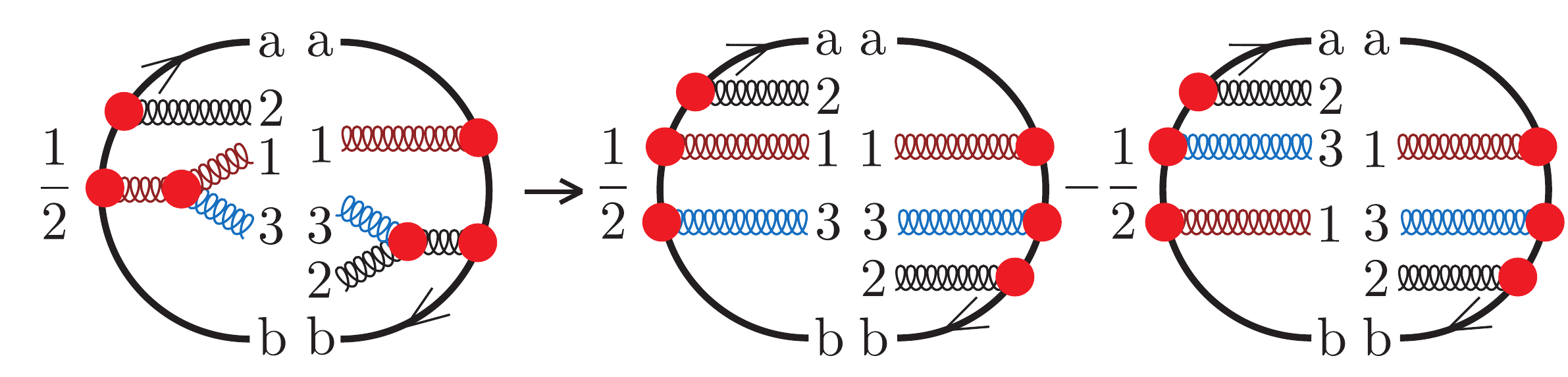}}

\caption{LC+ approximation for the splitting of gluon 1 in the ket state with the participation of helper parton 2 in the bra state in a case with $\{c'\}_m \ne \{c\}_m$.
}
\label{fig:splittingidentityketbra5}
\end{figure}

Our first example was shown in figure~\ref{fig:splittingidentityketbra2}. Here the helper parton ($k = {\rm b}$) is not color connected to the emitting parton ($l = 1$) in the bra state. In this case, the LC+ approximation is to drop the contribution entirely. A second example was shown in figure~\ref{fig:splittingidentityketbra3}. Here the helper parton ($k = 2$) is color connected to the emitting parton ($l = 1$) in the bra state. In this case, the LC+ approximation is to keep the contributions in which the emitted gluon attaches between $l$ and $k$ in the bra state. That is, we keep the first and third terms on the right hand side of figure~\ref{fig:splittingidentityketbra3}. A third example is shown in figure~\ref{fig:splittingidentityketbra5}. In this case, $\{c'\}_m \ne \{c\}_m$. The helper parton ($k = 2$) is color connected to the emitting parton ($l = 1$) in the bra state, so the LC+ approximation is to keep the contributions in which the emitted gluon attaches between $l$ and $k$ in the bra state. The two contributions that are retained in the LC+ approximation are shown on the right hand side of figure~\ref{fig:splittingidentityketbra5}.

\subsection{The virtual splitting operator}

To find the structure of the virtual splitting operator ${\cal V}(t)$ in the LC+ approximation we use the definition (\ref{eq:calVdecomposed}). For a direct splitting diagram, we consider the real emission diagram to correspond to the sum of virtual diagrams in which $h$ acts on the ket state and in which $h^\dagger$ acts on the bra state. For an interference diagram with the helper parton in the bra state, we consider the real emission diagram to correspond to a virtual diagram in which $h^\dagger$ acts on the bra state. An interference diagram with the helper parton in the ket state corresponds to a virtual diagram in which $h$ acts on the ket state.  Then we impose probability conservation, eq.~(\ref{eq:unitarity}), to relate ${\cal V}(t)$ to the definitions of the LC+ approximation for ${\cal H}_I(t)$ outlined above.

\begin{figure}
\centerline{\includegraphics[width = 13.0 cm]{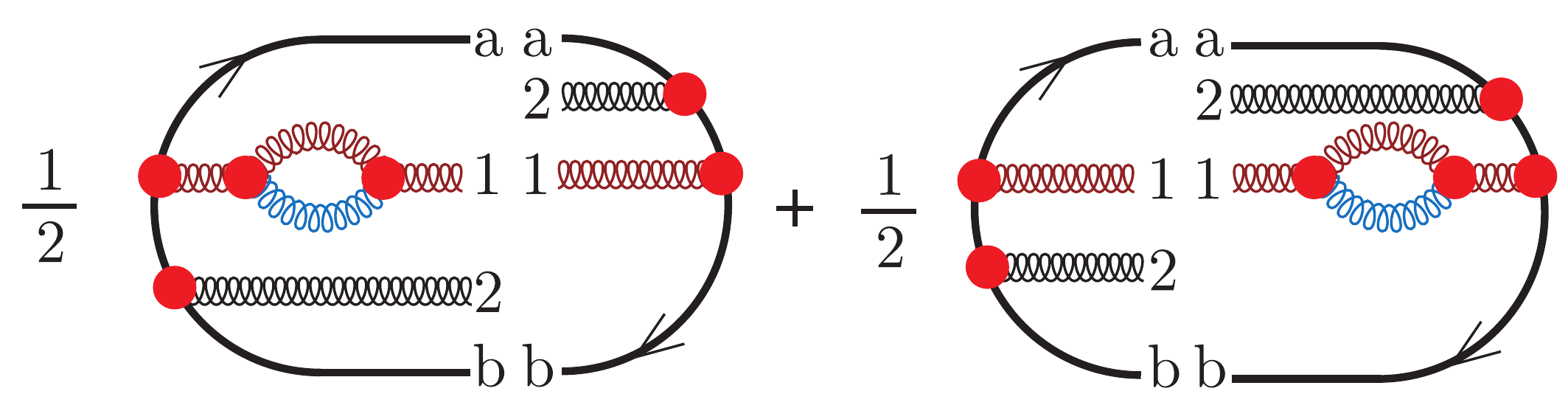}}

\caption{Color structure of the contribution to ${\cal V}(t)$ from the splitting in figure~\ref{fig:splittingidentityketbra4}.
}
\label{fig:Vketbra3}
\end{figure}

For a direct splitting diagram as in figure~\ref{fig:splittingidentityketbra4}, this gives the contribution to ${\cal V}(t)$ illustrated in figure~\ref{fig:Vketbra3}. To see this, we simply note that the color trace of the left hand side of figure~\ref{fig:splittingidentityketbra4} is the same as the color trace of figure~\ref{fig:Vketbra3}.

\begin{figure}
\centerline{\includegraphics[width = 9.8 cm]{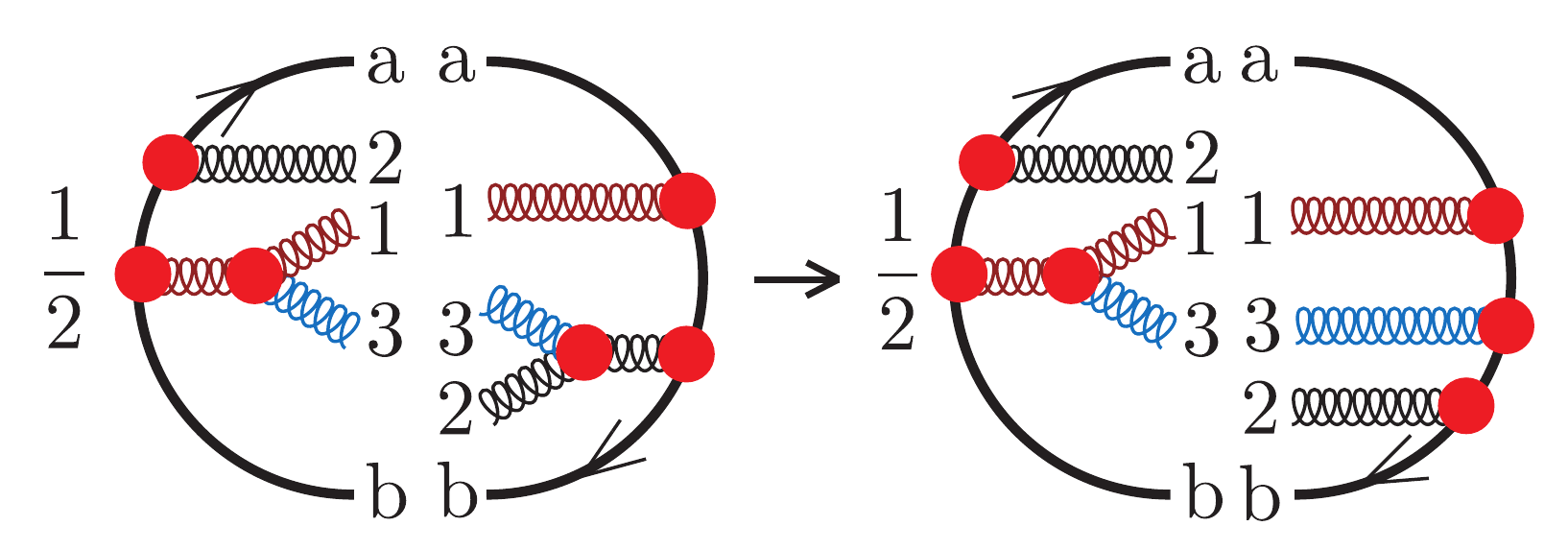}}

\caption{LC+ approximation for the splitting of gluon 1 in the ket state with the participation of helper parton 2 in the bra state in a case with $\{c'\}_m \ne \{c\}_m$. This is the sum of the two terms on the right hand side of figure~\ref{fig:splittingidentityketbra5}.
}
\label{fig:splittingidentityketbra6}
\end{figure}

\begin{figure}
\centerline{\includegraphics[width = 4.4 cm]{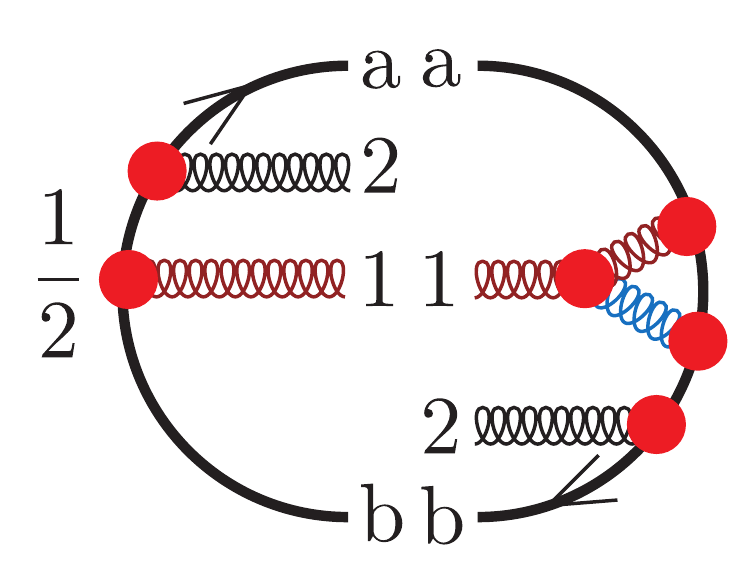}}

\caption{Color structure of the contribution to ${\cal V}(t)$ from the splitting in figure~\ref{fig:splittingidentityketbra6}.
}
\label{fig:Vketbra4}
\end{figure}

For the interference diagram in figure~\ref{fig:splittingidentityketbra5}, we first note that the two terms that are retained in the LC+ approximation sum to the contribution shown in figure~\ref{fig:splittingidentityketbra6}. This gives the contribution to ${\cal V}(t)$ illustrated in figure~\ref{fig:Vketbra4} since the color trace of the right hand side of figure~\ref{fig:splittingidentityketbra6} is the same as the color trace of figure~\ref{fig:Vketbra4}.

\begin{figure}
\centerline{\includegraphics[width = 9.8 cm]{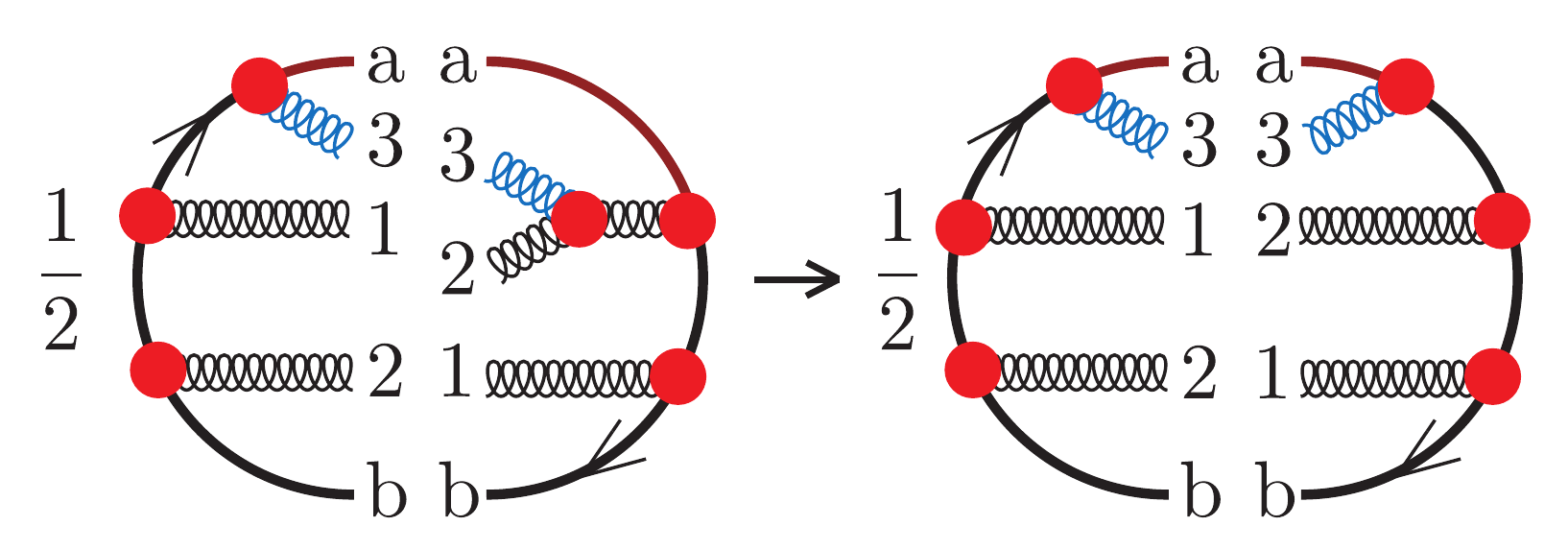}}

\caption{LC+ approximation for the splitting of parton ``a'' in the ket state with the participation of helper parton 2 in the bra state in a case with $\{c'\}_m \ne \{c\}_m$.
}
\label{fig:splittingidentityketbra7}
\end{figure}

We now note something remarkable. The color basis states are eigenstates of the operator $h$ that defines ${\cal V}(t)$. For a direct splitting of a gluon, as in figure~\ref{fig:Vketbra3}, the eigenvalue is $C_\LA/2$. Similarly, for a direct splitting of a quark or antiquark, the eigenvalue is $C_\LF/2$. For an interference diagram in which a gluon splits to two gluons with interference from gluon emission some other parton, as in figure~\ref{fig:Vketbra4}, a simple calculation shows that the eigenvalue is $C_\LA/4$. Similarly, for an interference diagram in which a quark or antiquark splits, emitting a gluon with interference from gluon emission some other parton, the LC+ approximation is shown in figure~\ref{fig:splittingidentityketbra7}. Then the corresponding contribution to ${\cal V}(t)$ has eigenvalue $C_\LF/2$.

\section{Operator based analysis with full color}
\label{sec:operatorbased}

In the preceding sections, we have sketched the main idea of the LC+ approximation. Now we need to make the idea precise. To do that, we first set up a formalism to describe the evolution of generic parton shower that includes quantum color exactly. To keep the presentation simple, we average over spins. More precisely, we sum over the spins of partons after each splitting and average over the spins of the partons that enter each splitting. Then spin is not visible in the evolution equations at all. The formalism is taken from Refs.~\cite{NSshower,NSspinless,NSspin}. However, here we keep the shower quite generic by not specifying exactly the splitting functions, the definition of the shower time $t$ that is used to order successive splittings, or the momentum mapping for each splitting that allows all partons to be on-shell while at the same time exactly conserving momentum.

\subsection{Parton labels}

At each stage of the shower, there are two initial state partons with labels ``a'' and ``b'' together with $m$ final state partons with labels $1,\dots,m$. The parton momenta and flavors are then specified by a list $\{p,f\}_m$ as in eq.~(\ref{eq:pfstate}).\footnote{For the flavors for the initial state partons, it is useful to let $f$ denote the flavor leaving the hard interaction, which is the opposite of the flavor of the physical parton entering the hard interaction.} At each step of the parton shower, any one of the partons can split. This includes an initial state parton, which splits in ``backwards evolution'' to another initial state parton plus a final state parton. In either case, let $l$ be the label of the parton that splits. At a splitting, the parton $l$ remains and one more parton, with label $m+1$ is created. After the splitting, the momenta and flavors are $\{\hat p,\hat f\}_{m+1}$. Whenever a final state gluon is created, we assign the label $m+1$ to the gluon. In the case of a final state $g \to g + g$ splitting, one daughter gluon has label $l$ and the other has label $m+1$. We use the interchange symmetry of the process to rearrange the splitting function so that there is a singularity when gluon $m+1$ becomes soft but no singularity when gluon $l$ becomes soft.

In the case of gluon emission, there are interference graphs. A gluon $m+1$ emitted from parton $l$ in the ket state can be emitted by partner parton $k$ in the bra state. Similarly a gluon $m+1$ emitted from parton $l$ in the bra state can be emitted by partner parton $k$ in the ket state. The interference diagrams are important when gluon $m+1$ is soft.

\subsection{The evolution equation}

The development of a parton shower can be described by giving an evolution equation for the operator $\rho(\{p,f\}_m,t)$ defined by eq.~(\ref{eq:rhodef}) as the density operator on the color space of the evolving partonic system. (It would be an operator on the spin space also except that we average over spins in this paper.) The density operator is a function of the parton momenta and flavors. This function can be regarded as a vector $\sket{\rho(t)}$ in the space of functions of $\{p,f\}_m$ with values in the space of operators on the quantum color space. We take the vector $\sket{\rho(t)}$ to obey a linear evolution equation of the form (\ref{eq:evolution}). 

In eq.~(\ref{eq:evolution}), ${\cal H}_I(t)$ is the splitting operator. Specifying shower evolution means specifying what ${\cal H}_I(t)$ is. To do that, we use basis states $\sket{\{p,f,c',c\}_m}$ that specify a definite momentum, flavor, and color configuration for the partons. The normalizations are such that $\rho(\{p,f,c',c\}_m,t)$ in eq.~(\ref{eq:rhodef}) is
\begin{equation}
\label{eq:rhoketdef}
\rho(\{p,f,c',c\}_m,t) = 
\sbrax{\{p,f,c',c\}_m}\sket{\rho(t)}
\;\;.
\end{equation}
Making use of these basis vectors, we can define ${\cal H}_I(t)$ by specifying the matrix elements of ${\cal H}_I(t)$ between a partonic state after a splitting and the state before the splitting, $\sbra{\{\hat p,\hat f,\hat c',\hat c\}_{m+1}}{\cal H}_I(t) \sket{\{p,f,c',c\}_m}$. We begin this task in the next few subsections by discussing some of the ingredients in this matrix element.

\subsection{Splitting functions}

To describe emission of a soft gluon $m+1$ from parton $l$ with interference from emission from helper parton $k$, we use the dipole splitting function
\begin{equation}
\label{eq:wlkdipole}
\overline w_{lk}^{\,\rm dipole}
= 4\pi\as\
\frac{
-(\hat p_{m+1}\cdot \hat p_l\ \hat p_k
- \hat p_{m+1}\cdot \hat p_k\ \hat p_l)^2}
{(\hat p_{m+1}\cdot \hat p_k\ \hat p_{m+1}\cdot \hat p_l)^2}
\;\;.
\end{equation}
In this expression, partons $l$ and $k$ can have nonzero masses. The expression for  $\overline w_{lk}$ may be more familiar in the massless limit, $\hat p_l^2 \to 0$ and $\hat p_k^2 \to 0$, where it becomes
\begin{equation}
\overline w_{lk}^{\,\rm dipole}
\to 4\pi\as\
\frac{
2\hat p_{k}\cdot \hat p_l}
{\hat p_{m+1}\cdot \hat p_k\ \hat p_{m+1}\cdot \hat p_l}
\;\;.
\end{equation}
Eq.~(\ref{eq:wlkdipole}) includes all four diagrams for emission from either parton $l$ or parton $k$ in both the bra and ket states, calculated in the limit $\hat p_{m+1} = \lambda P_{m+1}$ with $\lambda \to 0$. If we calculate the four contributions separately using the eikonal approximation in a physical gauge, then
\begin{equation}
\label{eq:wdipoleparts}
\overline w_{lk}^{\rm dipole} = 
\overline w_{ll}^{\rm eikonal}
+ \overline w_{kk}^{\rm eikonal}
- 2\overline w_{lk}^{\rm eikonal}
\;\;.
\end{equation}
The first two terms represent the direct graphs while the last term represents the two interference graphs.\footnote{The function $\overline w_{lk}^{\rm eikonal}$, which equals $\overline w_{kl}^{\rm eikonal}$, is denoted by $\overline W_{lk}/(2A_{lk})$ and given in eq.~(4.4) of ref.~\cite{NSspinless}. The function $\overline w_{ll}^{\rm eikonal}$ is denoted by $\overline W_{ll}^{\rm eikonal}$ and given in eq.~(2.10) of ref.~\cite{NSspinless}. These functions include only the leading $\lambda^{-2}$ contribution in the limit $\lambda \to 0$. One could, in principle, add contributions proportional to $\lambda^{-1}$ and higher powers of $\lambda$.}

The splitting function $\overline w_{ll}^{\,\rm eikonal}$ in eq.~(\ref{eq:wdipoleparts}) is singular in the limit in which parton $l$ is massless and parton $m+1$ is collinear with parton $l$: $\hat p_{m+1} \to [(1-z)/z]\, \hat p_l$. In this limit,
\begin{equation}
\overline w_{ll}^{\,\rm eikonal}
\to 
\frac{8\pi\as}{2\hat p_{m+1}\cdot \hat p_l}
\frac{2z}{1-z}
\;\;.
\end{equation}
There is a factor of the virtuality of the splitting, $2\hat p_{m+1}\cdot \hat p_l$ in the denominator and there is a function of the momentum fraction $z$ that is singular in the soft gluon limit $z \to 1$.  The behavior of this function away from the collinear limit depends on the conventions used to define the shower. For instance, it depends on the definition of $z$.

The splitting of parton $l$ into a parton of the same flavor plus a gluon both in the ket state and in the bra state is described by a function $\overline w_{ll}$. Suppose that parton $l$ is a quark. Then in the limit in which parton $l$ is massless and parton $m+1$ is collinear with parton $l$ we have
\begin{equation}
\label{eq:wll}
\overline w_{ll}
\to 
\frac{8\pi\as}{2\hat p_{m+1}\cdot \hat p_l}
\frac{1 + z^2}{1-z}
\;\;.
\end{equation}
Away from this limit, the exact result depends on the conventions used to define the shower.\footnote{For a $g \to g + g$ splitting, the DGLAP splitting kernel appears in $\overline w_{ll}$, as in eq.~(\ref{eq:wll}), once one symmetrizes in $z \leftrightarrow 1-z$. Before symmetrization, the result depends on the conventions used to define the shower.} Note that $\overline w_{ll}^{\,\rm eikonal}$ matches $\overline w_{ll}$ in the limit of a collinear splitting that is also soft, $(1-z) \to 0$.

For other sorts of splittings, there are splitting functions $\overline w_{ll}$ that are proportional to the DGLAP splitting kernels in the limit of massless, collinear splittings.

\subsection{Shower time}

Parton showers are based on evolution of the system as a shower time variable $t$ increases. The idea is to start at the hard interaction and move to a first splitting that is less hard, then move to softer splittings as $t$ increases. For initial state splittings, this means moving backward in physical time.

As a measure of softness, one can use the virtuality of the splitting, the virtuality divided by the energy of the mother parton, or the square of the part of the momentum of one of the daughters that is transverse to the direction of the mother.  The shower program {\sc Herwig} orders splittings according to the angle between the daughters times the energy of the mother, with wide angle splittings first. This ordering has its advantages, but the generic shower scheme outlined here would need some modification to fit with angular ordering. 

\subsection{Momentum mapping}

At each splitting, one starts with $m$ final state partons with momenta $\{p\}_m$ and ends with $m+1$ final state partons with momenta $\{\hat p\}_{m+1}$. One might like to define $\hat p_j = p_j$ for $j \ne l$ and $j\ne m+1$ and to take $\hat p_l + \hat p_{m+1} = p_l$ (or $\hat p_l - \hat p_{m+1} = p_l$ for an initial state splitting). However, this would not allow all three of $\hat p_l$, $\hat p_{m+1}$, and $p_l$ to be on-shell. Accordingly, one needs to take some momentum from the partons other than $l$ in order to supply the needed momentum in the splitting and keep exact momentum conservation. Thus we need a momentum mapping in which the $\{p\}_m$ plus three splitting variables determine the $\{\hat p\}_{m+1}$. The three splitting variables can be, for instance, the virtuality $(\hat p_l \pm \hat p_{m+1})^2 - m^2(f_l)$, a momentum fraction $z$ and an azimuthal angle of the daughters about the direction of the mother. Equivalently, the $\{\hat p\}_{m+1}$ determine three splitting variables and the $\{p\}_m$. The needed momentum mapping can be specified by giving a function $\sbra{\{\hat p,\hat f\}_{m+1}}{\cal P}_{l}\sket{\{p,f\}_m}$ that consists of an integration over the three chosen splitting variables and a product of delta functions that determine the $\{\hat p\}_{m+1}$ from the splitting variables and the $\{p\}_m$. There are many ways to do this, but for our purposes, all we need to know is that defining a shower evolution entails the specification of $\sbra{\{\hat p,\hat f\}_{m+1}}{\cal P}_{l}\sket{\{p,f\}_m}$.

For our generic shower, we take the mapping operator ${\cal P}_{l}$ to depend on the label $l$ of the parton that splits. One can also let the mapping operator depend on the index $k$ of the helping parton involved in interference diagrams. This is a common choice because it allows all of the momentum transfer to come from changing the momentum of parton $k$. That is, one can take $\hat p_l + \hat p_{m+1} - p_l = p_k - \hat p_k$. However, this scheme is a bit awkward for cases in which there is no helper parton $k$, such as $g \to q + \bar q$. Accordingly, we keep the generic shower equations simple by letting ${\cal P}_{l}$ depend only on the index $l$.

\subsection{Splitting operator}

Now we are ready to specify the matrix elements of ${\cal H}_I(t)$:
\begin{equation}
\begin{split}
\label{eq:HIdef}
\big(\{\hat p,\hat f,{}&\hat c',\hat c\}_{m+1}\big|
{\cal H}_I(t)\sket{\{p,f,c',c\}_{m}}
\\={}&
\sum_{l,k}
\delta(t - T(\{\hat p,\hat f\}_{m+1}))\,
(m+1)
\sbra{\{\hat p,\hat f\}_{m+1}}{\cal P}_{l}\sket{\{p,f\}_m}
\\&\times
\frac
{n_\Lc(a) n_\Lc(b)\,\eta_{\La}\eta_{\Lb}}
{n_\Lc(\hat a) n_\Lc(\hat b)\,
 \hat \eta_{\La}\hat \eta_{\Lb}}\,
\frac{
f_{\hat a/A}(\hat \eta_{\La},\mu^{2}_{F})
f_{\hat b/B}(\hat \eta_{\Lb},\mu^{2}_{F})}
{f_{a/A}(\eta_{\La},\mu^{2}_{F})
f_{b/B}(\eta_{\Lb},\mu^{2}_{F})}
\\&\times \frac{1}{2}\bigg[
\theta(k = l)\,
\theta(\hat f_{m+1} \ne g)\,
\overline w_{ll}(\{\hat p,\hat f\}_{m+1})
\\&\qquad +
\theta(k = l)\,
\theta(\hat f_{m+1} = g)\,
[\overline w_{ll}(\{\hat p,\hat f\}_{m+1})
- \overline w_{ll}^{\rm eikonal}(\{\hat p,\hat f\}_{m+1})]
\\&\qquad -
\theta(k\ne l)\,
\theta(\hat f_{m+1} = g)
A'_{lk}(\{\hat p\}_{m+1})\overline w_{lk}^{\rm dipole}(\{\hat p,\hat f\}_{m+1})
\bigg]
\\&\hskip 0.2 cm \times
\bigg[
\sbra{\{\hat c',\hat c\}_{m+1}}
t^\dagger_l(f_l \to \hat f_l + \hat f_{m+1})\otimes 
t_k(f_k \to \hat f_k + \hat f_{m+1})
\sket{\{c',c\}_m}
\\&\quad
+
\sbra{\{\hat c',\hat c\}_{m+1}}
t^\dagger_k(f_k \to \hat f_k + \hat f_{m+1}) \otimes 
t_l(f_l \to \hat f_l + \hat f_{m+1})
\sket{\{c',c\}_m}
\bigg]\;\;.
\end{split}
\end{equation}

In the first line on the right hand side of eq.~(\ref{eq:HIdef}), we have a sum over indices $l$ and $k$ of partons. The parton with label $l$ is the one that splits. There is another label $k$ so that we can include graphs that represent quantum interference between emission of a gluon from parton $l$ and from another parton $k$. We call parton $k$ the helper parton. For the quantum interference terms, we have $k \ne l$. There are also graphs that do not represent quantum interference. For these, $k = l$. 

The first line on the right hand side of eq.~(\ref{eq:HIdef}) also includes a delta function that specifies the definition of the shower time $t$, then the function that defines the momentum mapping.

In the second line, we have a ratio of parton distribution functions. For a final state splitting, this ratio is 1. For an initial state splitting, this ratio replaces the parton distribution functions at the previous momentum fraction $\eta_\La$ or $\eta_\Lb$ by the parton distribution function at the new momentum fraction after the splitting.

In the following lines of eq.~(\ref{eq:HIdef}) there are three terms, corresponding to three types of splittings.

First, there are splittings in which parton $m+1$ is not a gluon. For example, a final state $g \to q + \bar q$ splitting is in this class. For these cases, there is a splitting function $\overline w_{ll}(\{\hat p,\hat f\}_{m+1})$ that is singular for a collinear massless splitting, but does not have a soft parton singularity. 

In the second term in eq.~(\ref{eq:HIdef}), there are splittings in which parton $l$ emits a gluon. The splitting function here, $\overline w_{ll}- \overline w_{ll}^{\rm eikonal}$, is singular for a collinear splitting, but the singularity when the emitted gluon is soft has been removed by the subtraction. 

The most interesting term in eq.~(\ref{eq:HIdef}) is the third, in which a soft gluon $m+1$ is emitted from a dipole consisting of partons $l$ and $k$. The splitting function $\overline w_{lk}^{\rm dipole}$ is calculated using the eikonal approximation according to eq.~(\ref{eq:wlkdipole}). It is singular when the gluon is soft and also contains singularities when the direction of the gluon momentum is collinear to either the direction of $\hat p_l$ (if $\hat p_l^2 = 0$) or the direction of $\hat p_k$ (if $\hat p_k^2 = 0$). It is symmetric under interchange of $\hat p_l$ and $\hat p_k$. 

We have manipulated the third term in order to separate the roles of partons $k$ and $l$. Let $A'_{lk}$ be a function of the momenta $\{\hat p\}_{m+1}$ and $A'_{kl}$ be the same function with the roles of $\hat p_k$ and $\hat p_l$ reversed. Furthermore, let $A'_{lk} + A'_{kl} = 1$. Then, by interchanging the names of the dummy indices $l$ and $k$, we have
\begin{equation}
\begin{split}
\frac{1}{2}
\sum_l\sum_{k\ne l} 
\overline w_{lk}^{\rm dipole}
 ={}&
\frac{1}{2}
\sum_l\sum_{k\ne l} 
[A'_{lk} + A'_{kl}]
\overline w_{lk}^{\rm dipole}
\\={}&
\sum_l\sum_{k\ne l} 
A'_{lk}\overline w_{lk}^{\rm dipole}
\;\;.
\end{split}
\end{equation}
This gives the factor $A'_{lk}$ that multiplies $\overline w_{lk}^{\rm dipole}$ in eq.~(\ref{eq:HIdef}). If we were to take $A'_{lk} = 1/2$, this manipulation would do nothing, but instead we choose $A'_{lk}$ so that $A'_{lk} = 1$ when $\hat p_{m+1}$ is parallel to $\hat p_l$ and $A'_{lk} = 0$ when $\hat p_{m+1}$ is parallel to $\hat p_k$. Then $A'_{lk} \overline w_{lk}^{\rm dipole}$ is singular when the gluon is soft or collinear with parton $l$ but not when it is collinear with parton $k$. Our preferred choice for $A'_{lk}$ is given in eq.~(7.12) of ref.~\cite{NSspin}. One can also use the Catani-Seymour dipole partitioning function \cite{CataniSeymour} for this purpose. 

Finally in eq.~(\ref{eq:HIdef}) there is a color factor, which is the main focus of this paper. There are two terms in the color factor, which are related by interchanging the indices $l$ and $k$. If $k=l$, the two terms are identical. In the first term, there is an operator $t_l^\dagger(f_l \to \hat f_l + \hat f_{m+1})$ that acts on color ket states $\ket{\{c\}_m}$. This operator attaches the proper color matrix for the splitting to the color index for parton $l$. Similarly, we apply the operator $t_k(f_k \to \hat f_k + \hat f_{m+1})$ to the bra color state $\bra{\{c'\}_m}$. The resulting color bras and kets can be expanded in color basis states:
\begin{equation}
\begin{split}
\label{eq:colordef1}
t_l^\dagger(f_l \to \hat f_l + \hat f_{m+1})&\ket{\{c\}_m}\bra{\{c'\}_m}t_k(f_k \to \hat f_k + \hat f_{m+1}) 
\\&= 
\sum_{\{\hat c',\hat c\}_{m+1}}
C(\{\hat c',\hat c\}_{m+1},\{c',c\}_{m})\
\ket{\{\hat c\}_{m+1}}\bra{\{\hat c'\}_{m+1}}
\;\;.
\end{split}
\end{equation}
Then the matrix element that we want is the corresponding expansion coefficient:
\begin{equation}
\begin{split}
\label{eq:colordef2}
\sbra{\{\hat c',\hat c\}_{m+1}}
t^\dagger_l(f_l \to \hat f_l + \hat f_{m+1})\otimes t_k(f_k \to &\hat f_k + \hat f_{m+1})
\sket{\{c',c\}_m}
\\&=
C(\{\hat c',\hat c\}_{m+1},\{c',c\}_{m})
\;\;.
\end{split}
\end{equation}

\subsection{The virtual splitting operator}

The virtual splitting operator ${\cal V}(t)$ in Eq.~(\ref{eq:evolution}) represents the effect of virtual graphs on shower evolution. It reflects the leading infrared and collinear singularities in the virtual graphs. Consequently, ${\cal V}(t)$ leaves the number of partons unchanged and does not change their momenta or flavors. It does, however, multiply by a matrix in color space.\footnote{Thus this paper differs from Ref.~\cite{Platzer:2012qg}, in which Sudakov exponentials are simply numerical factors.} This is because the exchange of a soft virtual gluon changes parton colors. 

The matrix representing ${\cal V}(t)$ has two terms. First, there can be a correction to the ket color state with no change to the bra colors.  Then, there can be a  correction to the bra color state with no change to the ket colors. Thus the color structure of ${\cal V}(t)$ is the color structure of one loop virtual corrections. We write
\begin{equation}
\begin{split}
\label{eq:Vdef1}
{\cal V}(t)\sket{\{p,f,c',c\}_{m}}
={}&
\sum_{\{\hat c\}_m}
H_L(\{p,f\}_{m};\{\hat c\}_m,\{c\}_{m};t)
\sket{\{p,f,c',\hat c\}_{m}}
\\ &
+ \sum_{\{\hat c'\}_m}
H_R(\{p,f\}_{m};\{\hat c'\}_m,\{c'\}_{m};t)
\sket{\{p,f,\hat c', c\}_{m}}
\;\;.
\end{split}
\end{equation}
The color matrices $H_L$ and $H_R$ are the matrices that represent operators $h + \mi\phi$ and $[h + \mi \phi]^\dagger = h^\dagger - \mi \phi$ that act on the ket color vectors and the bra color vectors, respectively:
\begin{equation}
\begin{split}
\label{eq:hdef}
[h(\{p,f\}_{m},t) + \mi \phi(\{p,f\}_{m},t)]\ket{\{c\}_m}
={}& \sum_{\{\hat c\}_m}
H_L(\{p,f\}_{m};\{\hat c\}_m,\{c\}_{m};t)\
\ket{\{\hat c\}_m}
\;\;,
\\
\bra{\{c'\}_m} [h^\dagger(\{p,f\}_{m},t) - \mi \phi(\{p,f\}_{m},t)]
={}& \sum_{\{\hat c'\}_m}
H_R(\{p,f\}_{m};\{\hat c'\}_m,\{c'\}_{m};t)\
\bra{\{\hat c'\}_m}
\;\;.
\end{split}
\end{equation}
Here $\phi = \phi^\dagger$. In the simplest formulation, $\phi = 0$. We consider $\phi \ne 0$ in section~\ref{sec:virtphase}. In the full color treatment, we will have $h^\dagger = h$. However in the LC+ approximation we will define $h^{LC+}$ with $[h^{LC+}]^\dagger \ne h^{LC+}$.

It is useful to solve the shower evolution equation (\ref{eq:evolution}) in the form
\begin{equation}
\sket{\rho(t)} = {\cal U}(t,t_0)\sket{\rho(t_0)}
\;\;,
\end{equation}
where
\begin{equation}
\label{eq:evolutionsolution}
{\cal U}(t,t_0) = {\cal N}(t,t_0)
+ \int_{t_0}^t\!d\tau\ 
{\cal U}(t,\tau)
{\cal H}_I(\tau)
{\cal N}(\tau,t_0) 
\;\;.
\end{equation}
Here ${\cal N}(t_{2},t_1)$ is the no-splitting operator,
\begin{equation}
{\cal N}(t_2,t_1) = \mathbb T \exp\left[
-\int_{t_1}^{t_2} d\tau\ {\cal V}(\tau)\right]
\;\;,
\end{equation}
that provides the Sudakov factor that is usually interpreted as the probability to not have a splitting between time $t_1$ and time $t_{2}$. In ${\cal N}(t_{2},t_1)$, the virtual splitting operators ${\cal V}(\tau)$ are time ordered. Iterating eq.~(\ref{eq:evolutionsolution}) gives  $\sket{\rho(t)}$ as a power series in ${\cal H}_I$, with ${\cal H}_I$ evaluated at times $t_i$ and factors of ${\cal N}(t_{i+1},t_i)$ in between.  We interpret eq.~(\ref{eq:evolutionsolution}) as saying that possibly the system gets from $t_0$ to $t$ without splitting. If not, it goes from $t_0$ to $\tau$ without splitting, then splits according to ${\cal H}_I(\tau)$, then evolves further according to the full evolution operator ${\cal U}(t,\tau)$.

Given the structure (\ref{eq:Vdef1}) and (\ref{eq:hdef}) of ${\cal V}(\tau)$, the operator ${\cal N}(t_{i+1},t_i)$ has the form of a matrix in labels of the color basis elements,
\begin{equation}
\begin{split}
\label{eq:Ndef1}
{\cal N}(t_2,t_1)&\sket{\{p,f,c',c\}_{m}}
\\&=
\sum_{\{\hat c',\hat c\}_m}
N(\{p,f\}_{m};\{\hat c',\hat c\}_m,\{c',c\}_{m};t_2,t_1)
\sket{\{p,f,\hat c',\hat c\}_{m}}
\;\;.
\end{split}
\end{equation}
Here the matrix elements are obtained from the operators $h$ that define ${\cal V}(t)$,
\begin{equation}
\begin{split}
\label{eq:hinSudakov}
\mathbb T \exp\bigg[
-\int_{t_1}^{t_2} d\tau\ &h(\{p,f\}_{m},\tau)\bigg] 
\ket{\{c\}_{m}} 
\bra{\{c'\}_{m}}
\overline{\mathbb T} \exp\bigg[
-\int_{t_1}^{t_2} d\tau\ h^\dagger(\{p,f\}_{m},\tau)\bigg]
\\
={}& 
\sum_{\{\hat c',\hat c\}_m}
N(\{p,f\}_{m};\{\hat c',\hat c\}_m,\{c',c\}_{m};t_2,t_1)
\ket{\{\hat c\}_{m}} 
\bra{\{\hat c'\}_{m}}
\;\;.
\end{split}
\end{equation}
Thus the Sudakov factor breaks into two factors, one for the ket state and one for the bra state. That is, there is Sudakov factor for each color ordered quantum amplitude. That is remarkably simple. However, the structure of the Sudakov factor for each quantum amplitude is remarkably complicated since it is an operator that mixes color basis states.

\subsection{Probability conservation}

We relate ${\cal V}(t)$ to ${\cal H}_I(t)$ using the requirement that showering not change the probability of the hard scattering event that initiates the shower. Probability conservation requires eq.~(\ref{eq:unitarity}),
\begin{equation}
\label{eq:unitarity2}
\sbra{1}{\cal H}_I(t)\sket{\{p,f,c',c\}_{m}}
=
\sbra{1}{\cal V}(t)\sket{\{p,f,c',c\}_{m}}
\;\;.
\end{equation}
Here $\sbra{1}$ stands for the measurement of inclusive probability,
\begin{equation}
\label{eq:1def}
\sbrax{1}\sket{\{p,f,c',c\}_{m}}
= \brax{\{c'\}_{m}}\ket{\{c\}_{m}}
\;\;.
\end{equation}
The inner product $\brax{\{c'\}_{m}}\ket{\{c\}_{m}}$ corresponds to taking the trace of the color density operator $\ket{\{c\}_{m}}\bra{\{c'\}_{m}}$. 

For a general statistical state $\sket{\rho}$, we measure the inclusive probability that the partons are in any configuration by using the completeness sum for the statistical states $\sket{\{p,f,c',c\}_{m}}$ (as defined in ref.~\cite{NSshower})
\begin{equation}
\label{eq:completenesssum}
1 = \sum_m \frac{1}{m!}\int\!d\{p,f\}_m \sum_{\{c,c'\}_m}
\sket{\{p,f,c',c\}_{m}}\sbra{\{p,f,c',c\}_{m}}
\;\;,
\end{equation}
where, to be precise,
\begin{equation}
\begin{split}
\int\!d\{p,f\}_m ={}& \sum_{f_\La}\int_0^1\!d\eta_\La 
\sum_{f_\Lb}\int_0^1\!d\eta_\Lb
\prod_{i=1}^m
\left\{
\sum_{f_i}
\int\!\frac{d^4 p_i}{(2\pi)^4}\,2\pi\delta(p_i^2 - m^2(f_i))\right\}
\\&\times
(2\pi)^4\delta\left(p_\La + p_\Lb - \sum_{i=1}^m p_i\right)
\;\;.
\end{split}
\end{equation}
Thus the inclusive probability that the partons described by $\sket{\rho}$ are in any configuration is
\begin{equation}
\label{eq:inclusiveprobability}
\sbrax{1}\sket{\rho}
=
\sum_m \frac{1}{m!}\int\!d\{p,f\}_m \sum_{\{c,c'\}_m}
\brax{\{c'\}_{m}}\ket{\{c\}_{m}}\sbrax{\{p,f,c',c\}_{m}}\sket{\rho}
\;\;.
\end{equation}

\subsection{The inclusive splitting probability}

Using eq.~(\ref{eq:inclusiveprobability}), the inclusive probability for a splitting starting from $\sket{\{p,f,c',c\}_{m}}$ is
\begin{equation}
\begin{split}
\label{eq:inclusiveH}
\sbra{1}{\cal H}_I(t)\sket{\{p,f,c',c\}_{m}}
={}&
\frac{1}{(m+1)!}\int\!d\{\hat p,\hat f\}_{m+1} \sum_{\{\hat c,\hat c'\}_{m+1}}
\brax{\{\hat c'\}_{m+1}}\ket{\{\hat c\}_{m+1}}
\\&\times
\sbra{\{\hat p,\hat f,\hat c',\hat c\}_{m+1}}
{\cal H}_I(t)\sket{\{p,f,c',c\}_{m}}
\;\;.
\end{split}
\end{equation}
We are particularly interested in the color structure of this. We note by examining eq.~(\ref{eq:HIdef}) that on the right hand side of this equation the following generic color factors will occur:
\begin{equation*}
\sum_{\{\hat c,\hat c'\}_{m+1}}
\brax{\{\hat c'\}_{m+1}}\ket{\{\hat c\}_{m+1}}
\sbra{\{\hat c',\hat c\}_{m+1}}
t^\dagger_A \otimes t^{\vphantom{\dagger}}_B
\sket{\{c',c\}_m}
\;\;.
\end{equation*}
If we take the color trace of eq.~(\ref{eq:colordef1}) and compare it to the trace of eq.~(\ref{eq:colordef2}) we have
\begin{equation}
\begin{split}
\label{eq:coloridentity1}
\sum_{\{\hat c,\hat c'\}_{m+1}}
\brax{\{\hat c'\}_{m+1}}\ket{\{\hat c\}_{m+1}}
\sbra{\{\hat c',\hat c\}_{m+1}}
t^\dagger_A \otimes t^{\vphantom{\dagger}}_B
\sket{\{c',c\}_m}
=
\bra{\{c'\}_m}t^{\vphantom{\dagger}}_B
t_A^\dagger\ket{\{c\}_m} 
\;\;.
\end{split}
\end{equation}
If we use eq.~(\ref{eq:HIdef}) in eq.~(\ref{eq:inclusiveH}) and use the identity (\ref{eq:coloridentity1}), we find
\begin{equation}
\begin{split}
\label{eq:HI1}
\sbra{1}
{\cal H}_I(t)&\sket{\{p,f,c',c\}_{m}}
\\={}&
\sum_{l,k}
\frac{1}{m!}\int\!d\{\hat p,\hat f\}_{m+1}
\delta(t - T(\{\hat p,\hat f\}_{m+1}))\,
\sbra{\{\hat p,\hat f\}_{m+1}}{\cal P}_{l}\sket{\{p,f\}_m}
\\&\times
\frac
{n_\Lc(a) n_\Lc(b)\,\eta_{\La}\eta_{\Lb}}
{n_\Lc(\hat a) n_\Lc(\hat b)\,
 \hat \eta_{\La}\hat \eta_{\Lb}}\,
\frac{
f_{\hat a/A}(\hat \eta_{\La},\mu^{2}_{F})
f_{\hat b/B}(\hat \eta_{\Lb},\mu^{2}_{F})}
{f_{a/A}(\eta_{\La},\mu^{2}_{F})
f_{b/B}(\eta_{\Lb},\mu^{2}_{F})}
\\&\times \frac{1}{2}\bigg[
\theta(k = l)\,
\theta(\hat f_{m+1} \ne g)\,
\overline w_{ll}(\{\hat p,\hat f\}_{m+1})
\\&\qquad +
\theta(k = l)\,
\theta(\hat f_{m+1} = g)\,
[\overline w_{ll}(\{\hat p,\hat f\}_{m+1})
- \overline w_{ll}^{\rm eikonal}(\{\hat p,\hat f\}_{m+1})]
\\&\qquad -
\theta(k\ne l)\,
\theta(\hat f_{m+1} = g)
A'_{lk}(\{\hat p\}_{m+1})\overline w_{lk}^{\rm dipole}(\{\hat p,\hat f\}_{m+1})
\bigg]
\\&\hskip 0.2 cm \times
\bra{\{c'\}_m} 
t_k(f_k \to f_k + g)\,t^\dagger_l(f_l \to f_l + g)
\\&\hskip 1.6 cm
+ t_l(f_l \to f_l + g)\,t^\dagger_k(f_k \to f_k + g)
\ket{\{c\}_m}
\;\;.
\end{split}
\end{equation}

\subsection{Structure of the virtual splitting operator}
\label{sec:virtualstructure}

Given the structure of ${\cal V}(t)$ in eq.~(\ref{eq:Vdef1}) and the definition (\ref{eq:1def}) of $\sbra{1}$, the right hand side of eq.~(\ref{eq:unitarity2}) is 
\begin{equation}
\begin{split}
\sbra{1}{\cal V}(t)\sket{\{p,f,c',c\}_{m}}
={}&
\sum_{\{\hat c\}_m}
H_L(\{p,f\}_{m};\{\hat c\}_m,\{c\}_{m};t)
\brax{\{c'\}_{m}}\ket{\{\hat c\}_{m}}
\\ &
+ \sum_{\{\hat c'\}_m}
H_R(\{p,f\}_{m};\{\hat c'\}_m,\{c'\}_{m};t)
\brax{\{\hat c'\}_{m}}\ket{\{c\}_{m}}
\;\;.
\end{split}
\end{equation}
Using the definition (\ref{eq:hdef}) of the color operator $h$, this can be written
\begin{equation}
\begin{split}
\label{eq:V1}
\sbra{1}{\cal V}(t)\sket{\{p,f,c',c\}_{m}}
={}&
\bra{\{c'\}_{m}}h(\{p,f\}_{m},t) + h^\dagger\{p,f\}_{m},t)\ket{\{c\}_{m}}
\;\;.
\end{split}
\end{equation}
Here the operator on the right hand side is $(h + \mi\phi) + (h^\dagger - \mi \phi) = h + h^\dagger$. Thus this relation determines $h + h^\dagger$ but not $\phi$. We have assumed that $\phi = 0$. In section~\ref{sec:virtphase}, we will explore the color phase $\phi$.

We demand probability conservation, eq.~(\ref{eq:unitarity2}). Comparing eq.~(\ref{eq:V1}) to eq.~(\ref{eq:HI1}), we see that probability conservation holds if 
\begin{equation}
\begin{split}
\label{eq:Vresult}
h(\{p,f\}_{m},t)
={}&
\sum_{l,k}
\frac{1}{m!}\int\!d\{\hat p,\hat f\}_{m+1}
\delta(t - T(\{\hat p,\hat f\}_{m+1}))\,
\sbra{\{\hat p,\hat f\}_{m+1}}{\cal P}_{l}\sket{\{p,f\}_m}
\\&\times
\frac
{n_\Lc(a) n_\Lc(b)\,\eta_{\La}\eta_{\Lb}}
{n_\Lc(\hat a) n_\Lc(\hat b)\,
 \hat \eta_{\La}\hat \eta_{\Lb}}\,
\frac{
f_{\hat a/A}(\hat \eta_{\La},\mu^{2}_{F})
f_{\hat b/B}(\hat \eta_{\Lb},\mu^{2}_{F})}
{f_{a/A}(\eta_{\La},\mu^{2}_{F})
f_{b/B}(\eta_{\Lb},\mu^{2}_{F})}
\\&\times \frac{1}{2}\bigg[
\theta(k = l)\,
\theta(\hat f_{m+1} \ne g)\,
\overline w_{ll}(\{\hat p,\hat f\}_{m+1})
\\&\qquad +
\theta(k = l)\,
\theta(\hat f_{m+1} = g)\,
[\overline w_{ll}(\{\hat p,\hat f\}_{m+1})
- \overline w_{ll}^{\rm eikonal}(\{\hat p,\hat f\}_{m+1})]
\\&\qquad -
\theta(k\ne l)\,
\theta(\hat f_{m+1} = g)
A'_{lk}(\{\hat p\}_{m+1})\overline w_{lk}^{\rm dipole}(\{\hat p,\hat f\}_{m+1})
\bigg]
\\&\hskip 0.2 cm \times
t_l(f_l \to \hat f_l + \hat f_{m+1})
t^\dagger_k(f_k \to \hat f_k + \hat f_{m+1})
\;\;.
\end{split}
\end{equation}
Thus we define ${\cal V}(t)$ according to eqs.~(\ref{eq:Vdef1}) and (\ref{eq:hdef}) with $h(\{p,f\}_{m},t)$ given by eq.~(\ref{eq:Vresult}). 

In the color factor, we have a factor $t^{\vphantom{\dagger}}_l t^\dagger_k$. Then $h^\dagger$ is given by the same expression with a factor $t^{\vphantom{\dagger}}_k t^\dagger_l$. In fact, $h = h^\dagger$ because $t^{\vphantom{\dagger}}_k t^\dagger_l = t^{\vphantom{\dagger}}_l t^\dagger_k$. For $k=l$, this is obvious. For $k \ne l$, we are dealing with a gluon with color index $a$ exchanged between the lines $k$ and $l$. We note that $t^{\vphantom{\dagger}}_k t^\dagger_l$ inserts $SU(3)$ generator matrices $T^a$ in the appropriate representations on lines $l$ and $k$, then sums over $a$. The generator matrices are self adjoint and they commute with each other because they act on different parton lines, so that $t^{\vphantom{\dagger}}_k t^\dagger_l = t^{\vphantom{\dagger}}_l t^\dagger_k$.

This result might perhaps be regarded as elegant, but it poses difficulties for practical implementation in a parton shower Monte Carlo program. The difficulty comes from the fact that the operator $t_l(f_l \to f_l + g)\, t^\dagger_k(f_k \to f_k + g)$ is represented by a non-diagonal matrix in the standard color basis. In the full $h(\{p,f\}_{m},t)$ we have a sum of such operators with momentum dependent coefficients. For the Sudakov factor (\ref{eq:hinSudakov}), we need the exponential of an integral over $t$ of this matrix. Thus the Sudakov factors are complicated and it is not easy to see what to do with them.

\subsection{Color suppression index}
\label{sec:colorsuppressionI}

To help with our analysis, we define a quantity $P$ that we can call the color suppression power and a related but more useful quantity $I$ that we can call the color suppression index. Both $P$ and $I$ are related to the number of powers of $1/N_\Lc$ associated with the current color state at any stage in shower evolution.

To define $P$ and $I$, we must first deal with certain exceptional sources of factors $1/N_\Lc$. We call $p_\LE$ the number of powers of $1/N_\Lc$ coming from these exceptional sources of color suppression. For all of this paper except section \ref{sec:virtphase}, the sole source of non-zero $p_\LE$ is $g \to q + \bar q$ splittings. In a $g \to q + \bar q$ splitting, there are two ways to connect the new $q$ and $\bar q$ to the previous color state. The original gluon color was defined by inserting $t^a_{ij}$ along a color $\bm 3$ line. Now with the $g \to q + \bar q$ splitting we have a color matrix $t^a_{ij}t^a_{i'j'}$ in the color amplitude. We can use the Fierz identity,
\begin{equation}
t^a_{ij}t^a_{i'j'}
= \frac{1}{2}\, \delta_{ij'}\delta_{i'j} 
- \frac{1}{2N_\Lc}\,\delta_{ij}\delta_{i'j'}
\;\;,
\end{equation}
to write the result expanded in color basis states. At each $g \to q + \bar q$ splitting, shower evolution picks either the first, leading color, term or else the second, color suppressed, term. If the second term is chosen, further evolution uses the second color state, $\delta_{ij}\delta_{i'j'}$, and incorporates the factor $-1/(2N_\Lc)$ into the weight factor for the event. We let $p_\LE$ represent the number of times during the shower evolution that we pick up a $1/N_\Lc$ factor by using the second term in the Fierz identity. 

Now we can define the color suppression power. Suppose that after enough splitting steps to obtain $m$ final state partons, we reach a color density operator $\ket{\{c\}_m}\bra{\{c'\}_m}$. Then we can define $P(m)$ as the number of powers of $1/N_\Lc$ in the color overlap function $\brax{\{c'\}_m}\ket{\{c\}_m}$ plus the sum of the explicit powers, $p_\LE$, of $1/N_\Lc$ that multiply the color amplitudes $\ket{\{c\}_m}$ and $\bra{\{c'\}_m}$:
\begin{equation}
\label{eq:colorpower}
\left(\frac{1}{N_\Lc}\right)^{\!p_\LE}
\brax{\{c'\}_m}\ket{\{c\}_m}
= \frac{c_P(m)}{N_\Lc^{P(m)}}
\left\{
1 + {\cal O}\left(\frac{1}{N_\Lc}\right)
\right\}
\;\;,
\end{equation}
where $c_P(m)$ is non-zero and independent of $N_\Lc$. The color suppression power is of obvious interest, but is less useful than we would like because cancellation among terms in the expansion of $\brax{\{c'\}_m}\ket{\{c\}_m}$ can lead to $P$ being larger than it is for individual terms in the expansion.

We can improve on the definition by defining a color suppression index $I$ according to
\begin{equation}
\label{eq:colorindex}
\left(\frac{1}{N_\Lc}\right)^{\!p_\LE}
\brax{\{c'\}_m}\ket{\{c\}_m}_{U(N_\Lc)}
= \frac{c_I(m)}{N_\Lc^{I(m)}}
\left\{
1 + {\cal O}\left(\frac{1}{N_\Lc}\right)
\right\}
\;\;,
\end{equation}
where $c_I(m)$ is non-zero and independent of $N_\Lc$. The change here is that we calculate the color overlap function by using the color group $U(N_\Lc)$ instead of $SU(N_\Lc)$. To calculate $\brax{\{c'\}_m}\ket{\{c\}_m}_{U(N_\Lc)}$, one uses the Fierz identity once for each gluon line, omitting the $1/N_\Lc$ term.

A simple example (with $p_\LE = 0$) may be helpful. Suppose the hard process at the start of the shower is $q + \bar q \to  q + \bar q + g$ by means of a $Z$ boson exchange.  One possible $q\bar q q\bar q g$ configuration is $\ket{\{c\}}$ represented by $t^a_{ij} \delta_{kl}/\sqrt{N_\Lc^2 C_\LF}$. Another possible configuration is $\bra{\{c'\}}$ represented by $\delta_{ji} t^a_{lk} /\sqrt{N_\Lc^2 C_\LF}$. The overlap of these is
\begin{equation}
\brax{\{c'\}_m}\ket{\{c\}_m} =
\frac{1}{N_\Lc^2 C_\LF}\ t^a_{ij}\delta_{ji}\ t^a_{lk}\delta_{kl}
= 0
\;\;.
\end{equation}
Thus $P(m) = \infty$. Using the $U(N_\Lc)$ approximation, with $t^a_{ij}t^a_{lk} \to \delta_{ik}\delta_{lj}$, the overlap is 
\begin{equation}
\brax{\{c'\}_m}\ket{\{c\}_m}_{U(N_\Lc)}
=\frac{1}{N_\Lc^2 - 1}
\;\;.
\end{equation}
Thus $I(m) = 2$.\footnote{The trace of $\ket{\{c\}_m}\bra{\{c'\}_m}$ in this example vanishes, but this does not mean that this color state should be dropped in evolution with full color. With one more gluon emission, one can get a color state with nonzero $\brax{\{c'\}_m}\ket{\{c\}_m}$.}

The color suppression index has two properties that make it quite useful. First, as the number $m$ of final state partons in the shower increases, we always have
\begin{equation}
\label{eq:PgtI}
P(m) \ge I(m)
\;\;.
\end{equation}
That is, if we use $I(m)$ to estimate the amount of color suppression, we can never overestimate. Second, at each stage of the shower, the color suppression index either stays the same or else it increases:
\begin{equation}
\label{eq:Igrows}
I(m+1) \ge I(m)
\;\;.
\end{equation}
Thus we can think of $I$ as measuring color disorder, like entropy: it can never decrease as the shower evolves. Both of eqs.~(\ref{eq:PgtI}) and (\ref{eq:Igrows}) can be proved in a fairly straightforward way. We omit the proofs.

There is a useful concept that helps us track the way that $I(m)$ changes as we add gluons. This concept is outlined in appendix \ref{sec:growingI}, but we can give some flavor of it here in a few sentences. We assign a two valued parameter to each gluon according to how its color $\bm 3$ and $\bar{\bm 3}$ lines are connected in the $SU(N_\Lc)$ approximation: each gluon can be in a ``healthy'' or ``frail'' configuration. If a new gluon is added in a healthy configuration, then $I(m+1) = I(m)$, while if the new gluon is in a frail configuration, then $I(m+1) = I(m) + 2$. That is, the $SU(N_\Lc)$ color connections of the new gluon determine whether or not $I$ increases. Furthermore, when $I(m+1) = I(m)$, previous gluons that were healthy remain healthy and previous gluons that were frail may remain frail or may become healthy. When $I(m+1) = I(m) + 2$, previous gluons that were frail remain frail and previous gluons that were healthy may remain healthy or may become frail. Keeping track of the health status of gluons is simple and enables one to easily track changes in the color suppression index, as we will see in appendices \ref{sec:probabilities} and \ref{sec:nottoomuch}.

\section{The LC+ approximation}
\label{sec:LCplusfull}

With the generic structure of a (spin averaged) leading order shower set up, it is pretty simple to define the LC+ approximation. In the definition (\ref{eq:HIdef}) of ${\cal H}_I(t)$, in the terms that represent quantum interference between emitting a gluon from parton $l$ and emitting the same gluon from helper parton $k$ we replace
\begin{equation}
\begin{split}
t^\dagger_k(f_k \to f_k + g)\ket{\{c\}_{m}} \to{}& 
C(l,m+1)t^\dagger_k(f_k \to f_k + g)\ket{\{c\}_{m}}
\;\;,
\\
\bra{\{c'\}_{m}}t_k(f_k \to f_k + g) \to{}& 
\bra{\{c'\}_{m}}t_k(f_k \to f_k + g) C^\dagger(l,m+1)
\;\;.
\end{split}
\end{equation}
Here $C(l,m+1)$ acting on a state $\ket{\{\hat c\}_{m+1}}$ gives 1 if partons $l$ and $m+1$ are color connected in $\{\hat c\}_{m+1}$ and 0 otherwise. Thus in the LC+ approximation we keep only the color states in which partons $l$ and $m+1$ after the splitting are color connected. We generalize this to include the possibility that $k = l$ by replacing
\begin{equation}
\begin{split}
t^\dagger_k(f_k \to \hat f_k + \hat f_{m+1})\ket{\{c\}_{m}} \to{}& 
C(l,m+1)t^\dagger_k(f_k \to \hat f_k + \hat f_{m+1})\ket{\{c\}_{m}}
\;\;,
\\
\bra{\{c'\}_{m}}t_k(f_k \to \hat f_k + \hat f_{m+1}) \to{}& 
\bra{\{c'\}_{m}}t_k(f_k \to \hat f_k + \hat f_{m+1}) C^\dagger(l,m+1)
\;\;,
\end{split}
\end{equation}
with a generalized definition of $C(l,m+1)$. For $k=l$, whenever parton $m+1$ is a gluon, partons $m+1$ and $l$ after the splitting are always color connected, so we want $C(l,m+1)$ to return the color state unchanged. Also, for $k=l$ whenever $\hat f_l = g$, partons $m+1$ and $l$ after the splitting are also color connected, so we want $C(l,m+1)$ to return the color state unchanged.\footnote{This case can occur for an initial state splitting with $f_l = \hat f_{m+1} = q$ with $q$ being a quark or antiquark flavor.} In the case $k=l$ with $\hat f_l = q$ and $\hat f_{m+1} = \bar q$, or vice versa, the two daughter partons may not be color connected. We simply define $C(l,m+1)$ to give one in this case. Thus we define
\begin{equation}
C(i,j)\ket{\{\hat c\}_{m+1}} =
\begin{cases}
\ket{\{\hat c\}_{m+1}} 
& \text{partons $i$ and $j$ carry color $\{\bm 3,\bar{\bm 3}\}$ or $\{\bar{\bm 3}, \bm 3\}$} \\
\ket{\{\hat c\}_{m+1}} 
& i \text{ and } j \text{ color connected in } \{\hat c\}_{m+1} \\
0 
& \text{otherwise} \\
\end{cases}
\;\;.
\end{equation}
The operator $C(i,j)$ is a projection operator but it is not an orthogonal projection operator because the basis states are not orthogonal. Thus $C(i,j) \ne C(i,j)^\dagger$.

Thus the LC+ approximation for ${\cal H}_I$ is only slightly modified from the full ${\cal H}_I$ in eq.~(\ref{eq:HIdef}):
\begin{equation}
\begin{split}
\label{eq:HIdefLCplus}
\big(\{\hat p,\hat f,{}&\hat c',\hat c\}_{m+1}\big|
{\cal H}_I^{{\rm LC+}}(t)\sket{\{p,f,c',c\}_{m}}
\\={}&
\sum_{l,k}
\delta(t - T(\{\hat p,\hat f\}_{m+1}))\,
(m+1)
\sbra{\{\hat p,\hat f\}_{m+1}}{\cal P}_{l}\sket{\{p,f\}_m}
\\&\times
\frac
{n_\Lc(a) n_\Lc(b)\,\eta_{\La}\eta_{\Lb}}
{n_\Lc(\hat a) n_\Lc(\hat b)\,
 \hat \eta_{\La}\hat \eta_{\Lb}}\,
\frac{
f_{\hat a/A}(\hat \eta_{\La},\mu^{2}_{F})
f_{\hat b/B}(\hat \eta_{\Lb},\mu^{2}_{F})}
{f_{a/A}(\eta_{\La},\mu^{2}_{F})
f_{b/B}(\eta_{\Lb},\mu^{2}_{F})}
\\&\times \frac{1}{2}\bigg[
\theta(k = l)\,
\theta(\hat f_{m+1} \ne g)\,
\overline w_{ll}(\{\hat p,\hat f\}_{m+1})
\\&\qquad +
\theta(k = l)\,
\theta(\hat f_{m+1} = g)\,
[\overline w_{ll}(\{\hat p,\hat f\}_{m+1})
- \overline w_{ll}^{\rm eikonal}(\{\hat p,\hat f\}_{m+1})]
\\&\qquad +
\theta(k\ne l)\,
\theta(\hat f_{m+1} = g)
A'_{lk}(\{\hat p\}_{m+1})\overline w_{lk}^{\rm dipole}(\{\hat p,\hat f\}_{m+1})
\bigg]
\\&\hskip 0.2 cm \times
M_\Lc(\{\hat c',\hat c\}_{m+1},\{c',c\}_m,l,k,\{\hat f\}_{m+1})
\;\;.
\end{split}
\end{equation}
Here $M_\Lc$ is the color matrix defined by
\begin{equation}
\begin{split}
\label{eq:Mcdef}
&M_\Lc(\{\hat c',\hat c\}_{m+1},\{c',c\}_m,l,k,\{\hat f\}_{m+1})
\\&=
[\theta(k=l) - \theta(k\ne l)]
\\&\quad \times
\bigg\{
\sbra{\{\hat c',\hat c\}_{m+1}}
t^\dagger_l(f_l \to \hat f_l + \hat f_{m+1})\otimes 
t_k(f_k \to \hat f_k + \hat f_{m+1})
C^\dagger(l,m+1)
\sket{\{c',c\}_m}
\\&\quad\
+
\sbra{\{\hat c',\hat c\}_{m+1}}
C(l,m+1)t^\dagger_k(f_k \to \hat f_k + \hat f_{m+1}) \otimes 
t_l(f_l \to \hat f_l + \hat f_{m+1})
\sket{\{c',c\}_m}
\bigg\}
.
\end{split}
\end{equation}
The effect of the projection operator $C^\dagger(l,m+1)$ was illustrated in figure~\ref{fig:splittingidentityketbra3}. With full color, the interference graph shown gives four contributions $\ket{\{\hat c\}_{m+1}}\bra{\{\hat c'\}_{m+1}}$, but the projection operator $C^\dagger(l,m+1)$ removes the second and fourth contributions, in which gluon $m+1 = 3$ is not color connected to parton $l = 1$ in the bra state. Only the first and third contributions remain. Similarly in figure~\ref{fig:splittingidentityketbra2} there are two contributions with full color but both are eliminated in the LC+ approximation.

There are two terms in eq.~(\ref{eq:HIdefLCplus}) with $k=l$, one with $\hat f_{m+1} \ne g$ and one with $\hat f_{m+1} = g$. The corresponding splitting functions have collinear singularities but not soft singularities. In both these terms, the projection operator $C(l,m+1)$ acts as the unit operator, so no approximation is made in these terms. 

In the $k \ne l$ term, $\hat f_{m+1} = g$ and the effective splitting function $A'_{lk}\overline w_{lk}^{\rm dipole}$ is singular when the gluon is soft, collinear with parton $l$, or both soft and collinear.\footnote{In a reference frame in which $\hat p_l$ and $\hat p_k$ are fixed, ``soft'' means $|\vec {\hat p}_{m+1}| \to 0$ with $\theta_{m+1,l}$ fixed, while ``collinear'' means $\theta_{m+1,l} \to 0$ with $|\vec {\hat p}_{m+1}|$ fixed and ``both soft and collinear'' means $|\vec {\hat p}_{m+1}| \to 0$ and $\theta_{m+1,l} \to 0$ independently. In terms of dot products, ``soft'' means $\hat p_{m+1}\cdot \hat p_l \to 0$  and $\hat p_{m+1}\cdot \hat p_k \to 0$ with with $\hat p_{m+1}\cdot \hat p_l/\hat p_{m+1}\cdot \hat p_k$ fixed, while ``collinear'' means $\hat p_{m+1}\cdot \hat p_l \to 0$ with $\hat p_{m+1}\cdot \hat p_k$ fixed and ``both soft and collinear'' means $\hat p_{m+1}\cdot \hat p_l \to 0$ and $\hat p_{m+1}\cdot \hat p_k \to 0$ with also $\hat p_{m+1}\cdot \hat p_l/\hat p_{m+1}\cdot \hat p_k \to 0$.} In the limit that the gluon is collinear with parton $l$ or both collinear and soft, the LC+ approximation becomes exact, even though these terms contain the operator $C(l,m+1)$. To see this, note that $A'_{lk}\overline w_{lk}^{\rm dipole}$ is independent of $\hat p_k$ in the limit $\hat p_{m+1} \to \lambda \hat p_l$. Thus all of the terms with different indices $k$ have the same coefficient $A'_{lk}\overline w_{lk}^{\rm dipole}$. Because of that, we can use the color identity $\sum_l t_l = 0$ to write
\begin{equation}
\begin{split}
\sum_{k\ne l}&
\Big\{
t^\dagger_l(f_l \to f_l + g)\otimes t_k(f_k \to f_k + g)\,C^\dagger(l,m+1)
\\ & \qquad
+ C(l,m+1)\,t^\dagger_k(f_k \to f_k + g) \otimes t_l(f_l \to f_l + g)
\Big\}
\\&=
-\Big\{
t^\dagger_l(f_l \to f_l + g)\otimes t_l(f_l \to f_k + g)\,C^\dagger(l,m+1)
\\ & \qquad
+ C(l,m+1)\,t^\dagger_l(f_l \to f_l + g) \otimes t_l(f_l \to f_l + g)
\Big\}
\;\;.
\end{split}
\end{equation}
However, after a gluon emission from line $l$, parton $m+1$ is always color connected to parton $m+1$ in the new color state. Thus
\begin{equation}
\begin{split}
\sum_{k\ne l}&
\Big\{
t^\dagger_l(f_l \to f_l + g)\otimes t_k(f_k \to f_k + g)\,C^\dagger(l,m+1)
\\ & \qquad
+ C(l,m+1)\,t^\dagger_k(f_k \to f_k + g) \otimes t_l(f_l \to f_l + g)
\Big\}
\\&=
-2\,
t^\dagger_l(f_l \to f_l + g)\otimes t_l(f_l \to f_l + g)
\;\;.
\end{split}
\end{equation}
That is, the operator $C$ has no effect in the collinear or soft$\times$collinear limit. We conclude that the LC+ approximation becomes exact in the collinear or soft$\times$collinear limit. It is only an approximation when an emitted gluon is soft but not collinear to the emitting parton.

We can state this in a different way. Suppose that the integration over the momentum of gluon $m+1$ is limited not by the Sudakov factor but by imposing cuts $\theta_{l,m+1} > \theta_{\rm min}$ and $|\vec{\hat p}_{m+1}| > z_{\rm min}\,|\vec{\hat p}_{l}|/(1-z_{\rm min})$. Then the integration over $\hat p_{m+1}$ will produce logarithms $\alpha \log\theta_{\rm min}\,\log z_{\rm min} + \beta \log\theta_{\rm min} + \gamma\log z_{\rm min}$. The coefficients $\alpha$, $\beta$, $\gamma$ depend on the color states $\{c\}_{m+1}$, $\{c'\}_{m+1}$ after the splitting. For fixed $(\{c\}_{m+1},\{c'\}_{m+1})$, the double log coefficient $\alpha$ will match between the full theory and the LC+ approximation. The single log coefficient $\beta$ will also match between the full theory and the LC+ approximation. The single log coefficient $\gamma$ will miss contributions corresponding to subleading color states when calculated in the LC+ approximation.

For the virtual splitting operator, we use the the definition specified in eqs.~(\ref{eq:Vdef1}) and (\ref{eq:hdef}) to write ${\cal V}^{\rm LC+}(t)$ in terms of an operator $h^{\rm LC+}(\{p,f\}_{m},t)$ on the space of color vectors, as in eq.~(\ref{eq:V1}). With full color, $h(\{p,f\}_{m},t)$ was given by eq.~(\ref{eq:Vresult}). Now with the color operator $C(l,m+1)$, eq.~(\ref{eq:HIdef}) for ${\cal H}_I$ is replaced by eqs.~(\ref{eq:HIdefLCplus}) and (\ref{eq:Mcdef}). Thus we get the LC+ approximation for $h(\{p,f\}_{m},t)$,
\begin{equation}
\begin{split}
\label{eq:hLCplus}
h^{\rm LC+}(\{p,f\}_{m},t)
={}&
\sum_{l,k}
\frac{1}{m!}\int\!d\{\hat p,\hat f\}_{m+1}
\\&\times
\delta(t - T(\{\hat p,\hat f\}_{m+1}))\,
\sbra{\{\hat p,\hat f\}_{m+1}}{\cal P}_{l}\sket{\{p,f\}_m}
\\&\times
\frac
{n_\Lc(a) n_\Lc(b)\,\eta_{\La}\eta_{\Lb}}
{n_\Lc(\hat a) n_\Lc(\hat b)\,
 \hat \eta_{\La}\hat \eta_{\Lb}}\,
\frac{
f_{\hat a/A}(\hat \eta_{\La},\mu^{2}_{F})
f_{\hat b/B}(\hat \eta_{\Lb},\mu^{2}_{F})}
{f_{a/A}(\eta_{\La},\mu^{2}_{F})
f_{b/B}(\eta_{\Lb},\mu^{2}_{F})}
\\&\times \frac{1}{2}\bigg[
\theta(k = l)\,
\theta(\hat f_{m+1} \ne g)\,
\overline w_{ll}(\{\hat p,\hat f\}_{m+1})
\\&\qquad +
\theta(k = l)\,
\theta(\hat f_{m+1} = g)\,
[\overline w_{ll}(\{\hat p,\hat f\}_{m+1})
- \overline w_{ll}^{\rm eikonal}(\{\hat p,\hat f\}_{m+1})]
\\&\qquad -
\theta(k\ne l)\,
\theta(\hat f_{m+1} = g)
A'_{lk}(\{\hat p\}_{m+1})\overline w_{lk}^{\rm dipole}(\{\hat p,\hat f\}_{m+1})
\bigg]
\\&\hskip 0.2 cm \times
t_l(f_l \to \hat f_l + \hat f_{m+1})\,
C(l,m+1)\,
t^\dagger_k(f_k \to \hat f_k + \hat f_{m+1})
\;\;.
\end{split}
\end{equation}

Now the operator $h^{\rm LC+}$ contains operators that act on the color space. However, the color basis vectors are eigenfunctions of these operators. In the case $k=l$, for which $C(l,m+1)$ acts as the unit operator, the operator $t_l(f_l \to \hat f_l + \hat f_{m+1})\,t^\dagger_l(f_l \to \hat f_l + \hat f_{m+1})$ has eigenvalue $T_\LR$, $C_\LF$, or $C_\LA$ depending on the flavors of $f_l$ and $\hat f_{m+1}$ and $\hat f_l$. (The $C_\LA$ case is illustrated in figure~\ref{fig:Vketbra3}.) For the case $k \ne l$, the operator $t_l(f_l \to f_l + g)\,C(l,m+1)\, t^\dagger_k(f_k \to f_k + g)$ is a little more complicated. Because of the projection operator $C(l,m+1)$, this operator gives zero unless the helper parton $k$ is color connected to the emitting parton $l$ in the state $\ket{\{c\}_m}$ on which $h^{\rm LC+}$ acts. When $l$ and $k$ are color connected, one gets an eigenvalue $C_\LA/2$ or $C_\LF$ depending on the flavor of parton $l$. (The $C_\LA/2$ case is illustrated for $h^\dagger$ in figure~\ref{fig:Vketbra4}.) Thus 
\begin{equation}
h^{\rm LC+}(\{p,f\}_{m},t)
\ket{\{c\}_{m}}
= \lambda^{\rm LC+}(\{p,f,c\}_{m},t)\ket{\{c\}_{m}}
\end{equation}
where the eigenvalue $\lambda^{\rm LC+}$ is 
\begin{equation}
\begin{split}
\label{eq:hLCplusfinal}
\lambda^{\rm LC+}(&\{p,f,c\}_{m},t)
\\={}&
\sum_{l,k}
\frac{1}{m!}\int\!d\{\hat p,\hat f\}_{m+1}
\\&\times
\delta(t - T(\{\hat p,\hat f\}_{m+1}))\,
\sbra{\{\hat p,\hat f\}_{m+1}}{\cal P}_{l}\sket{\{p,f\}_m}
\\&\times
\frac
{n_\Lc(a) n_\Lc(b)\,\eta_{\La}\eta_{\Lb}}
{n_\Lc(\hat a) n_\Lc(\hat b)\,
 \hat \eta_{\La}\hat \eta_{\Lb}}\,
\frac{
f_{\hat a/A}(\hat \eta_{\La},\mu^{2}_{F})
f_{\hat b/B}(\hat \eta_{\Lb},\mu^{2}_{F})}
{f_{a/A}(\eta_{\La},\mu^{2}_{F})
f_{b/B}(\eta_{\Lb},\mu^{2}_{F})}
\\&\times \frac{1}{2}\bigg[
\theta(k = l)\,
\theta(\hat f_{m+1} \ne g)\,
\overline w_{ll}(\{\hat p,\hat f\}_{m+1})
\\&\qquad +
\theta(k = l)\,
\theta(\hat f_{m+1} = g)\,
[\overline w_{ll}(\{\hat p,\hat f\}_{m+1})
- \overline w_{ll}^{\rm eikonal}(\{\hat p,\hat f\}_{m+1})]
\\&\qquad +
\theta(k\ne l)\,
\theta(\hat f_{m+1} = g)
A'_{lk}(\{\hat p\}_{m+1})\overline w_{lk}^{\rm dipole}(\{\hat p,\hat f\}_{m+1})
\bigg]
\\&\hskip 0.2 cm \times
\chi(k,l,\{c\}_{m})\,N(k,l,\{\hat f\}_{m+1})
\;\;.
\end{split}
\end{equation}
The color eigenvalue specified in the last line is zero unless partons $k$ and $l$ are color connected in $\{c\}_{m}$ or $k=l$:
\begin{equation}
\label{eq:chikl}
\chi(k,l,\{c\}_{m}) =
\begin{cases}
1 & k=l \\
1 
& k \text{ and } l \text{ color connected in } \{c\}_{m} \\
0 
& \text{otherwise} \\
\end{cases}
\;\;.
\end{equation}
When $l$ and $k$ are color connected, the eigenvalue depends on whether $k = l$ and on the flavors in the splitting:
\begin{equation}
\begin{split}
\label{eq:Nlkdef}
N(k,l,\{\hat f\}_{m+1})
={}&
\begin{cases}
T_\LR & k=l,      \hat f_l \ne g, \hat f_{m+1} \ne g \\
C_\LF & k=l,      \hat f_l = g,   \hat f_{m+1} \ne g \\
C_\LF & k=l,      \hat f_l \ne g, \hat f_{m+1} = g   \\
C_\LA & k=l,      \hat f_l = g,   \hat f_{m+1} = g   \\
C_\LF & k\ne l,   \hat f_l \ne g, \hat f_{m+1} = g   \\
C_\LA/2 & k\ne l, \hat f_l = g,   \hat f_{m+1} = g  
\end{cases}
\;\;.
\end{split}
\end{equation}

Since the color basis vectors are eigenvectors of $h^{\rm LC+}$, the Sudakov factors for the color ordered amplitudes, eq.~(\ref{eq:hinSudakov}), are simply numerical factors:  
\begin{equation}
\begin{split}
\mathbb T \exp\bigg[-\int_{t_1}^{t_2} d\tau\ &h^{\rm LC+}(\{p,f\}_{m},\tau)\bigg] 
\ket{\{c\}_{m}} 
\\&=
\exp\left[-\int_{t_1}^{t_2} d\tau\ \lambda^{\rm LC+}(\{p,f,c\}_{m},\tau)\right] 
\ket{\{c\}_{m}}
\;\;.
\end{split}
\end{equation}
This is just the same as in the leading color approximation. In fact, with a suitable adjustment of what numerical factors one uses in the leading color approximation, the LC+ Sudakov factors are exactly the square root of the Sudakov factor for the leading color approximation. 
 
\section{Weights}
\label{sec:weights}

Let us now look at a splitting step in the evolution equation (\ref{eq:evolutionsolution}) in some detail, using the LC+ approximation. With a starting state $\sket{\{p,f,c',c\}_{m}}$ at time $t_0$, we have (using eq.~(\ref{eq:completenesssum}) in the LC+ version of eq.~(\ref{eq:evolutionsolution}))
\begin{equation}
\begin{split}
\label{eq:evolution2}
{\cal U}^{{\rm LC+}}(t',t_0)&\sket{\{p,f,c',c\}_{m}}
\\&=
{\cal N}^{{\rm LC+}}(t',t_0)\sket{\{p,f,c',c\}_{m}}
\\ & + 
\int_{t_0}^{t'}\!dt\
\frac{1}{(m+1)!}
\int\!d\{\hat p,\hat f\}_{m+1} 
\sum_{l,k}
\sum_{\{\hat c,\hat c'\}_{m+1}}
{\cal U}^{{\rm LC+}}(t',t)\sket{\{\hat p,\hat f,\hat c',\hat c\}_{m+1}}
\\&\times
\sbra{\{\hat p,\hat f,\hat c',\hat c\}_{m+1}}
{\cal H}^{{\rm LC+}}_{l,k}(t)\,
{\cal N}^{{\rm LC+}}(t,t_0)\sket{\{p,f,c',c\}_{m}}
\;\;.
\end{split}
\end{equation}
Here we have explicitly displayed the sum over parton indices in ${\cal H}_I$:
\begin{equation}
{\cal H}_I^{{\rm LC+}}(t)
= \sum_{l,k}
{\cal H}_{l,k}^{{\rm LC+}}(t)
\;\;.
\end{equation}
The $l$ and $k$ dependent splitting operator is, from eq.~(\ref{eq:HIdefLCplus}),
\begin{equation}
\begin{split}
\label{eq:HIdefLCpluslk}
\big(\{\hat p,\hat f,{}&\hat c',\hat c\}_{m+1}\big|
{\cal H}_{l,k}^{{\rm LC+}}(t)\sket{\{p,f,c',c\}_{m}}
\\={}&
\delta(t - T(\{\hat p,\hat f\}_{m+1}))\,
(m+1)
\sbra{\{\hat p,\hat f\}_{m+1}}{\cal P}_{l}\sket{\{p,f\}_m}
\\&\times
\frac
{n_\Lc(a) n_\Lc(b)\,\eta_{\La}\eta_{\Lb}}
{n_\Lc(\hat a) n_\Lc(\hat b)\,
 \hat \eta_{\La}\hat \eta_{\Lb}}\,
\frac{
f_{\hat a/A}(\hat \eta_{\La},\mu^{2}_{F})
f_{\hat b/B}(\hat \eta_{\Lb},\mu^{2}_{F})}
{f_{a/A}(\eta_{\La},\mu^{2}_{F})
f_{b/B}(\eta_{\Lb},\mu^{2}_{F})}
\\&\times \frac{1}{2}\bigg[
\theta(k = l)\,
\theta(\hat f_{m+1} \ne g)\,
\overline w_{ll}(\{\hat p,\hat f\}_{m+1})
\\&\qquad +
\theta(k = l)\,
\theta(\hat f_{m+1} = g)\,
[\overline w_{ll}(\{\hat p,\hat f\}_{m+1})
- \overline w_{ll}^{\rm eikonal}(\{\hat p,\hat f\}_{m+1})]
\\&\qquad +
\theta(k\ne l)\,
\theta(\hat f_{m+1} = g)
A'_{lk}(\{\hat p\}_{m+1})\overline w_{lk}^{\rm dipole}(\{\hat p,\hat f\}_{m+1})
\bigg]
\\&\hskip 0.2 cm \times
M_c(\{\hat c',\hat c\}_{m+1},\{c',c\}_m,l,k,\{\hat f\}_{m+1},\{f\}_{m})\;\;.
\end{split}
\end{equation}
Here $M_\Lc$ is the color matrix defined in eq.~(\ref{eq:Mcdef}).

Now consider the no-splitting operator ${\cal N}^{{\rm LC+}}(t,t_0)$. The color basis states are eigenstates of this operator, 
\begin{equation}
\begin{split}
{\cal N}^{{\rm LC+}}&(t,t_0)\sket{\{p,f,c',c\}_{m}} 
\\&= 
\exp\bigg[- \int_{t_0}^t \! d\tau\
\big(
\lambda^{\rm LC+}(\{p,f,c\}_{m},\tau)
+ \lambda^{\rm LC+}(\{p,f,c'\}_{m},\tau)
\big)
\bigg]
\\&\qquad\times
\sket{\{p,f,c',c\}_{m}}
\;\;.
\end{split}
\end{equation}
Using eq.~(\ref{eq:hLCplusfinal}), the integrand in the exponent can be written as an integral over momenta, a sum over flavors, and a sum over parton labels $l$ and $k$:
\begin{equation}
\begin{split}
\label{eq:lambda2}
\lambda^{\rm LC+}&(\{p,f,c\}_{m},\tau)
+ \lambda^{\rm LC+}(\{p,f,c'\}_{m},\tau)
\\&=
\frac{1}{m!}\int\!d\{\hat p,\hat f\}_{m+1} 
\sum_{l,k}
\\&\times
\delta(\tau - T(\{\hat p,\hat f\}_{m+1}))\,
\sbra{\{\hat p,\hat f\}_{m+1}}{\cal P}_{l}\sket{\{p,f\}_m}
\\&\times
\lambda(\{p,f,c\}_{m},l,k,\{\hat p,\hat f\}_{m+1})
\;\;.
\end{split}
\end{equation}
where
\begin{equation}
\begin{split}
\label{eq:partiallambda}
\lambda(\{p,f,c\}_{m},& l,k,\{\hat p,\hat f\}_{m+1})
\\&=
\frac
{n_\Lc(a) n_\Lc(b)\,\eta_{\La}\eta_{\Lb}}
{n_\Lc(\hat a) n_\Lc(\hat b)\,
 \hat \eta_{\La}\hat \eta_{\Lb}}\,
\frac{
f_{\hat a/A}(\hat \eta_{\La},\mu^{2}_{F})
f_{\hat b/B}(\hat \eta_{\Lb},\mu^{2}_{F})}
{f_{a/A}(\eta_{\La},\mu^{2}_{F})
f_{b/B}(\eta_{\Lb},\mu^{2}_{F})}
\\&\times\frac{1}{2}\Big[
\theta(k=l)
\theta(\hat f_{m+1} \ne g)\,
\overline w_{ll}(\{\hat p,\hat f\}_{m+1})
\\&\quad +
\theta(k=l)
\theta(\hat f_{m+1} = g)\,
[\overline w_{ll}(\{\hat p,\hat f\}_{m+1})
- \overline w_{ll}^{\rm eikonal}(\{\hat p,\hat f\}_{m+1})]
\\&\quad +
\theta(\hat f_{m+1} = g)
\theta(k\ne l)
A'_{lk}(\{\hat p\}_{m+1})\overline w_{lk}^{\rm dipole}(\{\hat p,\hat f\}_{m+1})
\Big]
\\&\times
[\chi(k,l,\{c\}_{m}) 
+ \chi(k,l,\{c'\}_{m})]
N(k,l,\{\hat f\}_{m+1})
\;\;.
\end{split}
\end{equation}
Here, $\chi(k,l,\{c\}_{m})$ was defined in eq.~(\ref{eq:chikl}) and $N(k,l,\{\hat f\}_{m+1})$ was defined in eq.~(\ref{eq:Nlkdef}).

We see that the function $\lambda(\{p,f,c\}_{m},l,k,\{\hat p,\hat f\}_{m+1})$ that appears in the Sudakov exponent is almost the same as the integrand in the matrix element of ${\cal H}_{l,k}^{{\rm LC+}}(t)$. In fact
\begin{equation}
\begin{split}
\label{eq:HIdefLCpluslkbis}
\big(\{\hat p,\hat f,{}&\hat c',\hat c\}_{m+1}\big|
{\cal H}_{l,k}^{{\rm LC+}}(t)\sket{\{p,f,c',c\}_{m}}
\\={}&
\delta(t - T(\{\hat p,\hat f\}_{m+1}))\,
(m+1)
\sbra{\{\hat p,\hat f\}_{m+1}}{\cal P}_{l}\sket{\{p,f\}_m}
\\&\times
\lambda(\{p,f,c\}_{m},l,k,\{\hat p,\hat f\}_{m+1})\,
C(\{\hat c',\hat c\}_{m+1},\{c',c\}_m,l,k,\{\hat f\}_{m+1})
\;\;,
\end{split}
\end{equation}
where
\begin{equation}
\begin{split}
\label{eq:Cdef}
C(\{\hat c',\hat c\}_{m+1},&\{c',c\}_m,l,k,\{\hat f\}_{m+1})
\\&
=
\frac{
M_\Lc(\{\hat c',\hat c\}_{m+1},\{c',c\}_m,l,k,\{\hat f\}_{m+1})}
{[\chi(k,l,\{c\}_{m}) + \chi(k,l,\{c'\}_{m})]
N(k,l,\{\hat f\}_{m+1})}
\;\;.
\end{split}
\end{equation}
Note that $M_\Lc(\{\hat c',\hat c\}_{m+1},\{c',c\}_m,l,k,\{\hat f\}_{m+1})$, eq.~(\ref{eq:Mcdef}), in the numerator of $C$ is nonzero only if one or both of $\chi(k,l,\{c\}_{m})$ and $\chi(k,l,\{c'\}_{m})$ are nonzero, so that we are never dividing by zero in a nonzero contribution to the sum over $l,k$. 

When we insert eq.~(\ref{eq:HIdefLCpluslkbis}) into eq.~(\ref{eq:evolution2}), we see that one can generate a splitting in standard Monte Carlo style by choosing the new momenta and flavors together with $l$ and $k$ with a probability proportional to $\lambda(\{p,f,c\}_{m},l,k,\{\hat p,\hat f\}_{m+1})$. This leaves a sum over the choices of colors,
\begin{equation*}
\sum_{\{\hat c,\hat c'\}_{m+1}}
\sket{\{\hat p,\hat f,\hat c',\hat c\}_{m+1}}
C(\{\hat c',\hat c\}_{m+1},\{c',c\}_m,l,k,\{\hat f\}_{m+1})
\;\;.
\end{equation*}
For each choice of $\{\hat c',\hat c\}_{m+1}$, there is a color factor $C$ and a statistical state vector $\sket{\{\hat p,\hat f,\hat c',\hat c\}_{m+1}}$ that is the input to the next splitting. In a computer program, one could imagine implementing the sum by summing the results returned by a splitting function that is called recursively. However, this is not really practical. Instead, one can perform the color sum Monte Carlo Style. One chooses $\{\hat c',\hat c\}_{m+1}$ with a probability $\rho_\Lc$, normalized to
\begin{equation}
\sum_{\{\hat c',\hat c\}_{m+1}}
\rho_\Lc(\{\hat c',\hat c\}_{m+1},\{c',c\}_m,l,k,\{\hat f\}_{m+1})
= 1
\;\;.%
\end{equation}
Then one assigns a weight factor to the splitting equal to
\begin{equation}
\begin{split}
\label{eq:colorwc}
w_\Lc(\{\hat c',\hat c\}_{m+1}&,\{c',c\}_m,l,k,\{\hat f\}_{m+1}) 
=
\frac{C(\{\hat c',\hat c\}_{m+1},\{c',c\}_m,l,k,\{\hat f\}_{m+1})}
{\rho_c(\{\hat c',\hat c\}_{m+1},\{c',c\}_m,l,k,\{\hat f\}_{m+1})}
\;\;.
\end{split}
\end{equation}
Averaged over many trials, this reproduces the desired sums.

The shower starts with a color weight factor of $C_0(\{c',c\}_{N_0})$ from the calculated color density matrix for producing an initial color configuration of $N_0$ final state partons at the start of the shower. At each splitting, we multiply by the weight factor $w_\Lc$ from eq.~(\ref{eq:colorwc}). The shower evolves with successive splittings until it reaches some cutoff hardness value. At that point, let us say that we measure the expectation value of some color operator ${\cal O}_\Lc$. Then this expectation value is
\begin{equation}
\label{eq:measurement}
{\rm Tr} \Big({\cal O}_\Lc \ket{\{c\}_{N}}\bra{\{c'\}_{N}}\Big)
= \bra{\{c'\}_{N}}{\cal O}_\Lc \ket{\{c\}_{N}}
\;\;. 
\end{equation}
We simply multiply the weight by this factor. In particular, making no measurement of color corresponds to multiplying by just the color overlap function $\brax{\{c'\}_{N}} \ket{\{c\}_{N}}$. Thus if the final color is not measured, the complete generated event comes with a weight
\begin{equation}
\label{eq:wtot}
w_{\rm tot} = \brax{\{c'\}_{N}}\ket{\{c\}_{N}}
\left(\prod_{m=I}^{N-1} 
\frac{C(\{\hat c',\hat c\}_{m+1},\{c',c\}_m,l_m,k_m,\{\hat f\}_{m+1})}
{\rho_c(\{\hat c',\hat c\}_{m+1},\{c',c\}_m,l_m,k_m,\{\hat f\}_{m+1})}
\right)
C_0(\{c',c\}_{N_0})
\;\;.
\end{equation}

In eq.~(\ref{eq:wtot}), the color overlap function $\brax{\{c'\}_{N}}\ket{\{c\}_{N}}$ is of crucial importance. If $\{c'\}_{N} = \{c\}_{N}$, this factor is 1 or very close to 1. If $\{c'\}_{N} \ne \{c\}_{N}$, this factor is proportional to $1/N_\Lc^{[P(N)-p_\LE]}$, where $p_\LE$ is the explicit power of $1/N_\Lc$ in the weight as defined in section \ref{sec:colorsuppressionI} and $P(N)$ is at least as big as the color suppression index $I(N)$ from eq.~(\ref{eq:colorindex}). Recall that the color suppression index either stays the same or else increases at each parton splitting.

One can use the behavior of the color overlap function to supply a hint about how to choose  the probabilities $\rho_c$. For numerical efficiency, one wants the dispersion of the weights not to be large. Consequently, it is sensible to make $\rho_\Lc$ small for any choice of color configuration $\{\hat c',\hat c\}_{m+1}$ that increases the color suppression index. For instance, for the splitting shown in figure~\ref{fig:splittingidentityketbra3}, the LC+ approximation retains the first and third terms on the right hand side; one would pick the color configuration shown in the first term with probability close to 1/2 and one would pick the color configurations shown in the third term with probability proportional to $1/N_\Lc^2$. We explore this idea further in appendix \ref{sec:probabilities} using information from appendix \ref{sec:growingI} about how the color suppression index grows.

It is also possible to choose $\rho_\Lc = 0$ for any color choice $\{\hat c',\hat c\}_{m+1}$ that makes $I(m+1)$ greater than some predetermined limit $I_{\rm max}$ on the color suppression index. This is an approximation beyond the LC+ approximation. We simply throw away configurations that have too many powers of $1/N_\Lc^2$. We explore this idea further in appendix \ref{sec:nottoomuch}

\section{The end of the shower}
\label{sec:end}

At some point, the perturbative shower should end because the hardness scale encoded in the shower time $t$ is not hard enough for a perturbative treatment to be reliable. At this point, one has a definite color density operator $\ket{\{c\}_{N}}\bra{\{c'\}_{N}}$ and one should multiply the color weight factor by the expectation value $\bra{\{c'\}_{N}}{\cal O}_\Lc \ket{\{c\}_{N}}$ of whatever operator corresponds to the measurement to be made on the final state, according to eq.~(\ref{eq:measurement}).

In the simplest case, the operator in question is simply the unit operator and one multiplies by $\brax{\{c'\}_{N}}\ket{\{c\}_{N}}$. However, one may want to apply a hadronization model to the final state. Commonly, the hadronization model starts from the assumption that color strings join the outgoing partons. A string configuration is labeled by a color label $\{c_\Lf\}_{N}$ of the same form as the labels for our color basis states. Then the string state should be chosen with a probability proportional to the expectation value $\bra{\{c'\}_{N}}{\cal O}(\{c_\Lf\}_{N}) \ket{\{c\}_{N}}$ of the operator ${\cal O}(\{c_\Lf\}_{N})$ that measures whether the quantum system is in color string state $\{c_\Lf\}_{N}$.

What is this operator? The leading color guess is that it is  $\ket{\{c_\Lf\}_{N}}\bra{\{c_\Lf\}_{N}}$. However, this guess can be correct only in the leading color approximation because our color basis states are not exactly orthogonal to one another (and also the closed string states are not exactly normalized). Thus we need another set of basis states $\ket{\{c\}_{N},\perp}$ that equal our color basis states to leading order in $1/N_\Lc$ but are exactly orthonormal,
\begin{equation}
\label{eq:orthonormal}
\brax{\{c'\}_{N},\perp}\ket{\{c\}_{N},\perp}
= 
\begin{cases}
1 & \{c'\}_{N} = \{c\}_{N} \\
0 & \{c'\}_{N} \ne \{c\}_{N} \\
\end{cases}
\;\;.
\end{equation}
The new states should be related to our basis states by a matrix equation
\begin{equation}
\ket{\{c\}_{N}} = \sum_{\{\hat c\}_{N}}
A(\{\hat c\}_{N},\{c\}_{N}) \ket{\{\hat c\}_{N},\perp}
\;\;,
\end{equation}
where $A$ is a unit matrix up to $1/N_\Lc^2$ corrections. Eq.~(\ref{eq:orthonormal}) implies that
\begin{equation}
\sum_{\{\hat c\}_{N}}
A^\dagger(\{c'\}_{N},\{\hat c\}_{N})\
A(\{\hat c\}_{N},\{c\}_{N}) 
= G(\{\hat c'\}_{N},\{\hat c\}_{N})
\;\;,
\end{equation}
where
\begin{equation}
G(\{\hat c'\}_{N},\{\hat c\}_{N}) = 
\brax{\{\hat c'\}_{N}}\ket{\{\hat c\}_{N}}
\;\;.
\end{equation}
There is a minimal solution to the equation $A^\dagger A =  G$ above. We let $A$ be a real symmetric matrix equal to
\begin{equation}
A =  \sqrt{G}
\;\;.
\end{equation}
To find $A$ exactly, one can diagonalize $G$ and replace each eigenvalue $\lambda$ by $\sqrt \lambda$. Alternatively, the expansion
\begin{equation}
A = 1 + \frac{1}{2}\,(G-1) - \frac{1}{8}\,(G-1)^2 + \cdots
\end{equation}
allows one to write $A$ as a power series in $1/N_\Lc^2$.

We conclude that it is sensible to define
\begin{equation}
{\cal O}(\{c_\Lf\}_{N}) =
\ket{\{c_\Lf\}_{N},\perp}
\bra{\{c_\Lf\}_{N},\perp}
\;\;.
\end{equation}
Then we choose the color string configuration $\{c_\Lf\}_{N}$ for hadronization with the color weight factor $\brax{\{c'\}_{N}}\ket{\{c_\Lf\}_{N},\perp}
\brax{\{c_\Lf\}_{N},\perp}\ket{\{c\}_{N}}$. One needs to sum over the choices for color string configuration $\{c_\Lf\}_{N}$. As in previous steps in the shower development, one can perform the sum Monte Carlo style, choosing configuration $\{c_\Lf\}_{N}$ with a probability $\rho(\{c_\Lf\}_{N})$ and multiplying by a weight factor
\begin{equation}
w_\Lc(\{c_\Lf\}_{N},\{c',c\}_{N})
=
\frac{\brax{\{c'\}_{N}}\ket{\{c_\Lf\}_{N},\perp}
\brax{\{c_\Lf\}_{N},\perp}\ket{\{c\}_{N}}}
{\rho(\{c_\Lf\}_{N})}
\;\;.
\end{equation}

Note that the numerator of $w_\Lc$ involves matrix elements of $A$. It is presumably sufficient to calculate $w_\Lc$ to leading order in $1/N_\Lc^2$.  One possible string configuration is always $\{c_\Lf\}_{N} = \{c\}_{N}$, for which $\brax{\{c_\Lf\}_{N},\perp}\ket{\{c\}_{N}} \approx 1$. Another is $\{c_\Lf\}_{N} = \{c'\}_{N}$. These are the leading possibilities if $\brax{\{c'\}_{N}}\ket{\{c\}_{N}} \propto 1/N_\Lc^{p}$ with $p = 0$ or 2. For $p = 4$, there can be other important configurations such that both $\brax{\{c_\Lf\}_{N},\perp}\ket{\{c\}_{N}} \propto 1/N_\Lc^2$ and $\brax{\{c'\}_{N}}\ket{\{c_\Lf\}_{N},\perp} \propto 1/N_\Lc^2$.

One does not necessarily need to follow the LC+ shower all the way to a $(1\ \text{GeV})^2$ virtuality scale. One could use the LC+ shower for a few splitting steps, down to, say, a $(100\ \text{GeV})^2$ scale. Then one could assign a classical color string configuration $\{c_\Lf\}_{N}$ to the partonic state as outlined above. After that, one could use a leading color shower to get from the $(100\ \text{GeV})^2$ virtuality scale to the $(1\ \text{GeV})^2$ scale. This would be appropriate if the measurement to be made on the final state is only minimally sensitive to what happens at the softer scales.

\section{Full color perturbatively}
\label{sec:perturbativecolor}

The shower evolution equation with full color has the form 
\begin{equation}
\sket{\rho(t)} = {\cal U}(t,t_0)\sket{\rho(t_0)}
\;\;,
\end{equation}
where ${\cal U}(t,t_0)$ obeys the evolution equation eq.~(\ref{eq:evolution}),
\begin{equation}
\label{eq:evolutionU}
\frac{d}{dt}\,{\cal U}(t,t_0)
= [{\cal H}_I(t) - {\cal V}(t)]\,{\cal U}(t,t_0)
\;\;.
\end{equation}
We have approximated ${\cal U}(t,t_0)$ by ${\cal U}^{{\rm LC+}}(t,t_0)$ where
\begin{equation}
\label{eq:evolutionLCplusbis}
\frac{d}{dt}\,{\cal U}^{{\rm LC+}}(t,t_0)
= [{\cal H}^{{\rm LC+}}_I(t) - {\cal V}^{{\rm LC+}}(t)]\,
{\cal U}^{{\rm LC+}}(t,t_0)
\;\;.
\end{equation}
This differential equation can be solved iteratively in the form
eq.~(\ref{eq:evolutionsolution})
\begin{equation}
\label{eq:evolutionsolutionLCplus}
{\cal U}^{{\rm LC+}}(t,t_0) = 
{\cal N}^{{\rm LC+}}(t,t_0)
+ \int_{t_0}^t\!d\tau\ 
{\cal U}^{{\rm LC+}}(t,\tau)\,
{\cal H}_I^{{\rm LC+}}(\tau)\,
{\cal N}^{{\rm LC+}}(\tau,t_0) 
\;\;.
\end{equation}
Here ${\cal N}^{{\rm LC+}}(t_{2},t_1)$ is the no-splitting operator,
\begin{equation}
{\cal N}^{{\rm LC+}}(t_2,t_1) = \exp\left[
-\int_{t_1}^{t_2} d\tau\ {\cal V}^{{\rm LC+}}(\tau)\right]
\;\;.
\end{equation}
It is well to recall here the essential point: the operator ${\cal V}^{{\rm LC+}}(\tau)$ is diagonal in the standard color basis that we use, so that it is practical to calculate its exponential.

Now, what if we want shower evolution with full color? Then we need
\begin{equation}
\label{eq:evolutionU2}
\frac{d}{dt}\,{\cal U}(t,t_0)
= [{\cal H}_I^{{\rm LC+}}(t) - {\cal V}^{{\rm LC+}}(t) + \Delta {\cal H}_I(t) - \Delta{\cal V}(t)]\,
{\cal U}(t,t_0)
\;\;,
\end{equation}
where
\begin{equation}
\begin{split}
\Delta {\cal H}_I(t) ={}& {\cal H}_I(t) - {\cal H}_I^{{\rm LC+}}(t)
\;\;,
\\
\Delta {\cal V}(t) ={}& {\cal V}(t) - {\cal V}^{{\rm LC+}}(t)
\;\;.
\end{split}
\end{equation}
This evolution equation is equivalent to
\begin{equation}
\label{eq:evolutionsolutionfull}
{\cal U}(t,t_0) = 
{\cal U}^{{\rm LC+}}(t,t_0)
+ \int_{t_0}^t\!d\tau\ 
{\cal U}(t,\tau)\,
\left[
\Delta {\cal H}_I(\tau)
-\Delta {\cal V}(\tau)
\right]
{\cal U}^{{\rm LC+}}(\tau,t_0) 
\;\;,
\end{equation}
which can be solved iteratively:
\begin{equation}
\begin{split}
\label{eq:softexpansion}
{\cal U}(t_\Lf,t_0) ={}& {\cal U}^{{\rm LC+}}(t_\Lf,t_0)
\\&+
 \int_{t_0}^{t_\Lf}\! d\tau\ 
{\cal U}^{{\rm LC+}}(t,\tau_1)\,
\big[\Delta {\cal H}_I(\tau_1)
- \Delta {\cal V}(\tau_1)\big]\,
{\cal U}^{{\rm LC+}}(\tau_1,t_0)
\\&+
 \int_{t_0}^{t_\Lf}\! d\tau_2
 \int_{t_0}^{\tau_2}\! d\tau_1\
{\cal U}^{{\rm LC+}}(t_\Lf,\tau_2)\,
\big[\Delta {\cal H}_I(\tau_2)
- \Delta {\cal V}(\tau_2)\big]\,
{\cal U}^{{\rm LC+}}(\tau_2,\tau_1)
\\&\qquad\times
\big[\Delta {\cal H}_I(\tau_1)
- \Delta {\cal V}(\tau_1)\big]\,
{\cal U}^{{\rm LC+}}(\tau_1,t_0)
\\ &+\cdots
\;\;.
\end{split}
\end{equation}
One can have any (small) maximum number of insertions of $\big[\Delta {\cal H}_I(\tau)- \Delta {\cal V}(\tau)\big]$. For instance, to start with one might test whether one such insertion makes a significant difference. To make one insertion of $\big[\Delta {\cal H}_I(\tau_1)- \Delta {\cal V}(\tau_1)\big]$, one would generate $\tau_1$ at random. Then one would run an LC+ shower from the starting scale $t_0$ up to scale $\tau_1$. After that, application of $\Delta {\cal H}_I(\tau_1)$ produces a weight and a new shower state, which is the starting point for an LC+ shower from $\tau_1$ to the final shower time $t_\Lf$. Similarly, application of $\Delta {\cal V}(\tau_1)$ produces a weight and a new shower state, which is the starting point for a second LC+ shower from $\tau_1$ to $t_\Lf$. The resulting values for the measurement function from the two second stage showers would then be summed.

It is interesting to note the counting of logarithms in eq.~(\ref{eq:softexpansion}). Suppose that the parton shower is used to calculate an observable ${\cal O}$ in which there is a large logarithm $L$, so that $\langle {\cal O} \rangle$ has the form $\sum_n c(n,2n)\as^n L^{2n} + \sum_n c(n,2n-1)\as^n L^{2n-1} + \cdots $. Suppose further that a shower with full color, generated by ${\cal U}(t_\Lf,t_0)$, correctly calculates all coefficients $c(n,2n)$ and $c(n,2n-1)$. Then the first term in eq.~(\ref{eq:softexpansion}) correctly generates the coefficients $c(n,2n)$ since the LC+ approximation is exact with respect to color for the soft$\times$collinear singularities. In the second term in eq.~(\ref{eq:softexpansion}), the one insertion of $\big[\Delta {\cal H}_I(\tau_2) - \Delta {\cal V}(\tau_2)\big]$ generates a factor $\as L$ by correcting the color content of a wide angle soft splitting. This factor multiplies factors $\as^{n-1} L^{2n-2}$. Thus the second term  makes contributions to the coefficients $c(n,2n-1)$. The third and higher terms in eq.~(\ref{eq:softexpansion}) contribute to coefficients $c(n,j)$ with $j \le 2n-2$ only. Thus to get the coefficients $c(n,2n-1)$, one needs only the first two terms in eq.~(\ref{eq:softexpansion}).

In general, by using one or two terms beyond the LC+ approximation in eq.~(\ref{eq:softexpansion}), one can see whether the splitting operators beyond the LC+ approximation have an important influence on whatever observable is being investigated using the parton shower. One may expect that for many observables, the operators $\Delta {\cal H}_I(t)$ or $\Delta {\cal V}(t)$ are not important.  Including some factors of $[\Delta {\cal H}_I(\tau) -\Delta {\cal V}(\tau)]$ can test this hypothesis.

\section{Soft gluon exchange phase}
\label{sec:virtphase}

Up until now, we have presented the evolution equations for a parton shower in which the virtual splitting operator ${\cal V}$ is determined by the real splitting operator ${\cal H}_I$ together with an assumption about the structure of ${\cal V}$.

However, the structure assigned to ${\cal V}(t)$ leaves out an important physical effect: the wave function of two partons emerging from a hard interaction can accumulate an SU(3) phase factor $U \approx 1 + \mi\phi(t) dt$ by exchanging a soft gluon.\footnote{The phase is sometimes called the Coulomb gluon phase by analogy with the Coulomb phase in non-relativistic quantum scattering. However, in a relativistic calculation in Coulomb gauge, only part of the phase comes from the Coulomb force between colored partons; the rest comes from the exchange of physically polarized gluons.} Here $\phi$ is an operator on the quantum color state. In an inclusive cross section, this phase cancels between the virtual graphs for the bra state and for the ket state. That is, the phase cancels in a completely inclusive measurement because
\begin{equation}
{\rm Tr}\!\left[U\ket{\{c\}_m}\bra{\{c'\}_m} U^\dagger\right] 
= \bra{\{c'\}_m} U^\dagger U\ket{\{c\}_m} 
= \bra{\{c'\}_m} 1\ket{\{c\}_m}
\;\;.
\end{equation}
However, the color matrix $\phi(t)$ changes the color state, which influences the further evolution of the shower, so that the effects of the color phase do not cancel from all observables measured at the end of the shower. These effects can be important \cite{Forshaw:2005sx, SuperleadingLogs, Dokshitzer:2005ig, Kidonakis:1997gm, Catani:2011st}.

Recall the structure of eqs.~(\ref{eq:Vdef1}) and (\ref{eq:hdef}) for ${\cal V}(t)$, which we can summarize in a shorthand notation as
\begin{equation}
{\cal V}(t) \to 
[h(\{p,f\}_{m},t) + \mi \phi (\{p,f\}_{m},t)]\otimes 1
+ 1 \otimes [h^\dagger(\{p,f\}_{m},t) - \mi \phi (\{p,f\}_{m},t)]
\;\;.
\end{equation}
The probability conservation equation determines $h$ through\footnote{Recall that $h = h^\dagger$ with full color but that in the LC+ approximation we defined $h^{LC+}$ with $[h^{LC+}]^\dagger \ne h^{LC+}$.}
\begin{equation}
\begin{split}
\label{eq:V1better}
\sbra{1}{\cal H}_I(t)\sket{\{p,f,c',c\}_{m}}
={}&
\bra{\{c'\}_{m}} h(\{p,f\}_{m},t) + h^\dagger(\{p,f\}_{m},t)\ket{\{c\}_{m}}
\;\;.
\end{split}
\end{equation}
This gave us $h$ from ${\cal H}_I(t)$, as in section \ref{sec:virtualstructure}. That is, we could find $h$ without looking at any details of virtual graphs by simply knowing that the singularities of real emission graphs have to cancel with corresponding singularities of virtual exchange graphs.

Previously, we made the assumption that $\phi = 0$. However, when a soft gluon is exchanged between partons $l$ and $k$, the graph has a part proportional to $\mi$ times a hermitian matrix $\phi$ in the color space. This color phase is not zero. A simple calculation using the eikonal approximation, similar to other calculations in the literature (for example, ref.~\cite{Kidonakis:1997gm}), gives
\begin{equation}
\begin{split}
\label{eq:phiresult}
\phi(\{p,f\}_{m},t)
={}&
- 2\pi
\sum_{l,k}
\theta(k\ne l)\,\frac{\theta_{kl}}{v_{kl}}\frac{\as(\mu_{\rm R}^{kl}(t))}{4\pi}\ 
\bm{T}_k\cdot \bm{T}_l
\;\;.
\end{split}
\end{equation}
In order to make the connection to parton showers clear, we present our calculation of this result in appendix \ref{sec:phase}. In eq.~(\ref{eq:phiresult}), we sum over partons $l$ and over partons $k$ with $k\ne l$. Thus each pair of partons appears twice in the sum. Not all combinations contribute:
\begin{equation}
\label{eq:thetakl}
\theta_{kl} =
\begin{cases}
1 & \text{$k$ and $l$ are final state partons}
\\
1 & \text{$k$ and $l$ are initial state partons}
\\
0 & \text{otherwise}
\end{cases}
\;\;.
\end{equation}
The phase depends on the relative velocities of partons $k$ and $l$:
\begin{equation}
v_{kl} = \sqrt{1 - \frac{m(f_k)^2\,m(f_l)^2}{(p_k\cdot p_l)^2}}
\;\;.
\end{equation}
Note that if either parton is massless, $v_{kl} = 1$. There is a factor $\alpha_s$ evaluated at a scale $\mu_\LR$ that depends on the shower time and potentially depends on the momenta of partons $k$ and $l$, depending the physical meaning of the shower time used in the parton shower (see appendix \ref{sec:phase}). Finally, there is a color operator $\bm{T}_k\cdot \bm{T}_l$. The operator $T_k$ inserts a color matrix $T^b$ on line $k$: if $\Psi$ is the vector representing the color state $\{c\}_m$ before the virtual exchange as in eq.~(\ref{eq:openstring}) or eq.~(\ref{eq:closedstring}), then the effect of the $T_k$ is to map
\begin{equation}
\Psi^{a_1 \cdots a_k \cdots a_n} \to 
 T^b_{a^{}_k,a'_k} \Psi^{a_1 \cdots a'_k \cdots a_n}
\;\;.
\end{equation}
Here $b$ is the color index of the exchanged gluon. Similarly, $T_l$ inserts a color matrix $T^b$ on line $l$. Finally, we sum over the color index $b$, as indicated by the dot in $\bm{T}_k\cdot \bm{T}_l$. The operator $\bm{T}_k\cdot \bm{T}_l$ maps $\ket{\{c\}_m}$ into a linear combination of color basis states $\ket{\{\hat c\}_m}$.

Let us denote the part of ${\cal V}(t)$ that comes from $\phi(t)$ by ${\cal V}_{\!\phi}(t)$. Imagine calculating the expectation value of some observable ${\cal O}$ by using a parton shower in which we calculate perturbatively in powers of ${\cal V}_{\!\phi}(t)$. We note immediately that contributions proportional to an odd number of factors of ${\cal V}_{\!\phi}(t)$ do not contribute to $\langle {\cal O}\rangle$ because $\langle {\cal O}\rangle$ is real and these contributions have an odd number of powers of $\mi$. However with an even number of powers of ${\cal V}_{\!\phi}(t)$ we have a factor $\pm (\mi\pi)^{2n} = \pm (-1)^n \pi^{2n}$ and have a generally nonvanishing contribution. How big these contributions are depends on how sensitive the observable is to the color flow of the event. This is a question that deserves further study that is beyond the scope of the present paper. However, one can note immediately that the development of a parton shower depends crucially on the color structure of the shower state because this color structure determines the preferred emission direction for relatively soft gluons. Thus a sudden change in color structure caused by inserting two occurrences of ${\cal V}_{\!\phi}(t)$ at some early shower time $t$ can have a substantial influence on the flow of momentum at the end of the shower. For instance, a gap in the rapidity of emitted partons could be created or could be filled in.

Can we say more about the color structure of the phase operator $\phi$? We find that the color operator $\bm{T}_k\cdot \bm{T}_l$ applied to $\ket{\{c\}_m}$ gives
\begin{equation}
\begin{split}
\label{eq:Tklexpansion}
\bm{T}_k\cdot \bm{T}_l \ket{\{c\}_m}
={}& \frac{C_\LA}{2}
\bigg\{\lambda(l,k,\{c\}_{m})\,\chi(k,l,\{c\}_{m})
\ket{\{c\}_m}
\\&
+\frac{1}{N_\Lc^2}\,\zeta(l,k,\{c\}_{m})\ket{\{c\}_m}
+ \frac{1}{N_\Lc}\,\bm{T}_{kl}^{\rm R}\ket{\{c\}_m}
\bigg\}
\;\;.
\end{split}
\end{equation}
The first term does not change the color state and does not change the color suppression index. In the first term, $\chi(k,l,\{c\}_{m})$ is one if $l$ and $k$ are color connected, zero otherwise, as in eq.~(\ref{eq:chikl}). The eigenvalue $\lambda$ is
\begin{equation}
\lambda(l,k,\{c\}_{m}) = 
\begin{cases}
- 2 & \parbox{7 cm}{$k,l$ carry colors $\{\bm 8,{\bm 8}\}$  connected to a two parton closed string}\\
\displaystyle{-1} & \text{otherwise}
\end{cases}
\;\;.
\end{equation}
The second term is a color suppressed contribution, with a factor $1/N_\Lc^2$ and a factor
\begin{equation}
\zeta(l,k,\{c\}_{m}) = 
\begin{cases}
1 & \text{$k,l$ carry colors $\{\bm 3,\bar{\bm 3}\}$ or $\{\bar{\bm 3}, \bm 3\}$} \\
-1 & \text{$k,l$ carry colors $\{\bm 3,{\bm 3}\}$ or $\{\bar{\bm 3}, \bar{\bm 3}\}$}\\
0 & \text{otherwise}
\end{cases}
\;\;.
\end{equation}
The corresponding contribution to the gluon exchange phase cancels exactly between the phase $\phi(\{p,f\}_{m},t)\ket{\{c\}_m}$ of the amplitude and the phase $-\bra{\{c'\}_m}\phi(\{p,f\}_{m},t)$ of the conjugate amplitude. The remaining term, $\bm{T}_{kl}^{\rm R}$, changes the state $\ket{\{c\}_m}$ to a linear combination of other color basis states.\footnote{We count the $1/N_\Lc$ factor that multiplies $\bm{T}_{kl}^{\rm R}$ as increasing the color suppression power $p_\LE$ by 1 in eqs.~(\ref{eq:colorpower}) and (\ref{eq:colorindex}). With this definition, the term $({1}/{N_\Lc})\,\bm{T}_{kl}^{\rm R}$ either increases the color suppression index, eq.~(\ref{eq:colorindex}), of the color state $\{c,c'\}_m$ or leaves it unchanged.}

The contribution to ${\cal V}_\phi$ from the first term in $\bm{T}_k\cdot \bm{T}_l$ in eq.~(\ref{eq:Tklexpansion}) can be included in the LC+ approximation ${\cal V}^{\rm LC+}$ for ${\cal V}$. Then the remaining part of ${\cal V}_\phi$ becomes part of $\Delta{\cal V}$ and can be treated perturbatively as in eq.~(\ref{eq:evolutionsolutionfull}).

The color phase operator $\phi$ depends on the parton masses. There are two places where the mass dependence might be important for the analysis of TeV scale processes at the LHC. First, in the later stages of showering, one can produce gluons with transverse momenta not too far above the $b$-quark mass. Sometimes such a gluon can split to $b + \bar b$. Then the $b$ and $\bar b$ are non-relativistic, so the mass dependence matters. Second, the hard process under investigation can produce top quarks or perhaps squarks, gluinos, and other very massive particles. In these cases, the particle masses matter.

Even though particle masses can matter, it is of interest to understand what happens when all parton masses can be neglected, so that $v_{lk} \to 1$. There can still be dependence on the parton labels $k,l$ if the relation between the shower time and the renormalization scale in $\as$ depends on the parton kinematics. Let us suppose that we neglect any such dependence. Then
\begin{equation}
\begin{split}
\label{eq:simplephiresult}
\phi(\{p,f\}_{m},t)
={}&
- 2\pi\ \frac{\as}{4\pi}\,
\sum_{l,k}
\theta(k\ne l)\,\theta_{kl}\,
\bm{T}_k\cdot \bm{T}_l
\;\;.
\end{split}
\end{equation}
Since the whole shower is invariant under color rotations, we have used color basis states that are overall color singlets. Applied to color singlet states, we have 
\begin{equation}
\sum_{k}\bm{T}_k\cdot \bm{T}_l = 0
\end{equation}
for any $l$. From this, we derive
\begin{equation}
\begin{split}
\label{eq:phydecomposition}
\phi(\{p,f\}_{m},t)
={}&
\phi_0(\{p,f\}_{m},t)
+ \phi_{\La \Lb}(\{p,f\}_{m},t)
\;\;.
\end{split}
\end{equation}
where
\begin{equation}
\begin{split}
\label{eq:simplephiresultA}
\phi_0(\{p,f\}_{m},t)
={}&
2\pi\ \frac{\as}{4\pi}\,
\left[\sum_{i = 1}^m \bm{T}_i\cdot \bm{T}_i 
- \bm{T}_\La\cdot \bm{T}_\La
- \bm{T}_\Lb\cdot \bm{T}_\Lb
\right]
\;\;,
\\
\phi_{\La\Lb}(\{p,f\}_{m},t)
={}&
-2\pi\ \frac{\as}{4\pi}\,
4\bm{T}_\La\cdot \bm{T}_\Lb
\;\;.
\end{split}
\end{equation}
In the first term in $\phi_0$, we sum over final state partons while in the next two terms we sum over the initial state partons. In either case, we have a color operator $\bm{T}_i\cdot \bm{T}_i$, which is just the Casimir operator, with eigenvalue $C_\LA$ or $C_\LF$ depending on whether parton $i$ is a quark or a gluon. Thus
\begin{equation}
\begin{split}
\phi_0(\{p,f\}_{m},t) \ket{\{c\}_m}
={}& 2\pi\ \frac{\as}{4\pi}\,
\big[
C_\LA\, (N^{\rm F}_g - N^{\rm I}_g) 
+ C_\LF\, (N^{\rm F}_q - N^{\rm I}_q)
\big]
\ket{\{c\}_m}
\;\;,
\end{split}
\end{equation}
where $N^{\rm F}_g$ is the number of gluons in the final state of $\{p,f\}_{m}$ while $N^{\rm I}_g$ is the number of gluons in the initial state and $N^{\rm F}_q$ is the number of quarks and antiquarks in the final state while $N^{\rm I}_q$ is the number of quarks and antiquarks in the initial state. When we apply $-\phi_0$ to the bra state $\bra{\{c'\}_m}$, we get exactly the opposite phase. Thus the term $\phi_0$ in the color phase contributes nothing to the development of the shower and can be simply dropped. This leaves the single contribution $\phi_{\La\Lb}$, representing an effective double strength color exchange between the initial state partons. Notice that with the approximation used here there is no color phase for $e^+ e^-$ annihilation or for deeply inelastic scattering.

Using eq.~(\ref{eq:Tklexpansion}) for the color operator $\bm{T}_\La\cdot \bm{T}_\Lb$, we obtain a leading color term that leaves the color state unchanged plus a term that changes the color state. The leading color term provides a phase that occurs if partons a and b are color connected in the state $\{c\}_m$. There is an exactly opposite phase that occurs if partons a and b are color connected in the state $\{c'\}_m$. Thus there is no net phase if partons a and b are color connected both in state $\{c\}_m$ and in state $\{c'\}_m$ and there is no net phase if partons a and b are not color connected in state $\{c\}_m$ or in state $\{c'\}_m$. If partons a and b are color connected in one of $\{c\}_m$ or $\{c'\}_m$ but not the other, then there is a net phase that appears in every evolution interval until the color connection situation changes. These phase factors will tend to reduce the contribution of this sort of state to whatever observable is to be measured.

There are also  contributions to $\bm{T}_\La\cdot \bm{T}_\Lb$ that change the color configuration. These terms have the potential to change the energy flow in the final state by changing the color flow.

\section{Conclusions}
\label{sec:conclusions}

In general, a parton shower Monte Carlo event generator should generate contributions to the density operator in color space in which the color in the ket vector $\ket{\{c\}_m}$ and the color in the bra vector $\bra{\{c'\}_m}$ are different. Standard parton shower generators based on evolution from hard splittings to soft splittings work in the leading color (LC) approximation, which is the leading order in an expansion in powers of $1/N_\Lc^2$. In the leading color approximation, only states with $\{c'\}_m = \{c\}_m$ occur.

In this paper, we have introduced the LC+ approximation, a generalization of the leading color approximation. Going beyond the leading color approximation inevitably involves sums over color choices. These sums can performed with Monte Carlo summation: selecting a choice at random according to a prescribed probability function. At the end of the shower each event comes with a weight. 

The LC+ approximation has several nice features:
\begin{itemize}

\item For each splitting, the leading soft$\times$collinear singularity and the leading collinear singularity are treated exactly with respect to color. There is an approximation with respect to color, but it occurs only in wide angle soft splittings.

\item Evolution can start with any state $\sket{\{c',c\}_m}$. Thus if one starts with a hard scattering process, one can fully use the color subamplitudes that multiply color states $\ket{\{c\}_m}$ for the hard scattering, including in the calculation interference between ket states $\ket{\{c\}_m}$ and bra states $\bra{\{c'\}_m}$ with different color configurations.

\item The Sudakov factors are numbers. That is to say, the standard color basis states $\sket{\{c',c\}_m}$ are eigenstates of the Sudakov operator. One does not have to exponentiate non-diagonal matrices in the color space.

\item In fact, not only are the Sudakov factors numbers, but also there is a separate Sudakov factor for the ket state $\ket{\{c\}_m}$ and for the bra state $\bra{\{c'\}_m}$. This feature may prove useful for matrix element matching when the color structure of the amplitudes is treated exactly.

\item With a simple extension of the formalism, one can include in the LC+ approximation the phase induced by exchange of soft gluons.

\item The LC+ approximation is still approximate: remainder terms in the generators of shower evolution are left over. However, the remainder terms can be included in a perturbative calculation up to some order.

\item The LC+ approximation can provide an efficient tool to sum large logarithms with full color at leading and next-to-leading log level for a certain class of observables if one uses the first perturbative correction as described in section~\ref{sec:perturbativecolor}.

\end{itemize}

The inclusion of weights in the shower generation has the potential to produce numerical problems. If the dispersion in the weights is large, then the number of events needed to calculate the expectation value of some observable with reasonable accuracy will be large. The total weight for an event is the product an initial weight, a final weight, and weights for individual splittings. By multiplying a large number of individual weights, one has the potential to produce large weights. For instance, $(1+ 1/N_\Lc^2)^N$ can be large if $N$ is large. For this reason, it seems likely that there is a practical maximum to the number of splittings that can be generated using the LC+ approximation.

Fortunately, it is possible to turn the LC+ approximation off before it goes too far. In fact, we have found two ways to do that. First, one can simply run the LC+ algorithm for some number $N_{\rm max}$ of splittings and then return to the LC approximation. For that, one must replace mixed states $\sket{\{c',c\}_m}$ by diagonal states $\sket{\{\tilde c, \tilde c\}_m}$. We have presented a plausible model for doing that. Second, one can continue using the LC+ approximation, but not allow the generation of states with more than a certain amount of color suppression as measured by what we have called the color suppression index. 

We are thus encouraged that the LC+ approximation can prove useful. The authors do not have at immediate hand a dipole based Monte Carlo event generator suitable for implementing the approximation.\footnote{We are, however, working on such a program.} However, the LC+ approximation is quite general and could, we think, be implemented in an existing generator.

Finally, one can ask if one really needs to go beyond the LC approximation. The answer is that we do not know if one needs to go beyond the LC approximation. The LC approximation is the first term in an expansion in powers of $1/N_\Lc^2$. We do not know if the contributions of order $1/N_\Lc^2$ or beyond are important. One could imagine that these contributions are not important for some observables but are important for others. If that is the case, we do not know which observables are sensitive to $1/N_\Lc^2$ effects.\footnote{We might suspect, for instance, that effects assocciated with requiring that no jets above a small $P_T$ cut appear in a given rapidity region are sensitive to non-leading color contributions, but we do not know that.} An implementation of the LC+ approximation would give us the means to investigate these questions.

\acknowledgments{ 
This work was supported in part by the United States Department of Energy and by the Helmoltz Alliance ``Physics at the Terascale." We thank Mrinal Dasgupta, Jeff Forshaw, and Mike Seymour for enlightening conversations about color in parton showers and about the role of the phase from soft gluon exchange.
}
\appendix

\section{How the color suppression index grows}
\label{sec:growingI}

In this appendix, we investigate how the color suppression index $I(m)$ grows.

Imagine calculating $\brax{\{c'\}_m}\ket{\{c\}_m}_{U(N_\Lc)}$. Look at any one of the gluons in the state, say gluon $l$. Its color factor was $\sum _a t^a_{i'i} t^a_{j'j}$. With the $U(N_\Lc)$ approximation, this becomes $(1/2)\, \delta_{i'j}\, \delta_{j'i}$. Counting the rest of the graph, the indices can be connected in two possible ways
\begin{equation}
\label{eq:goodgluon}
\sum \delta_{i'j}\, \delta_{j'i} \times \delta_{i'j}\,\delta_{j'i}
\end{equation}
or
\begin{equation}
\label{eq:badgluon}
\sum \delta_{i'j}\, \delta_{j'i} \times \delta_{i'i}\,\delta_{j'j}
\;\;.
\end{equation}
In the first case, let us call gluon $l$ a healthy gluon. In the second case, let us call gluon $l$ a frail gluon.\footnote{We call the gluon frail because if we keep the full gluon color matrices we get $\sum t^a_{i'i} t^a_{j'j} \times \delta_{i'i}\,\delta_{j'j} = 0$.} With a healthy gluon, we get a factor $N_\Lc^2$ for the sum in eq.~(\ref{eq:goodgluon}). With a frail gluon, we get a factor $N_\Lc^1$ for the sum in eq.~(\ref{eq:badgluon}).

Let us look at this in more detail. 

Consider eq.~(\ref{eq:goodgluon}) for the healthy gluon. If the healthy gluon were not there, we would have a factor $\sum \delta_{i'i}\,\delta_{i i'} = N_\Lc$. Now insert the healthy gluon and divide by $C_\LF \propto N_\Lc$ to normalize the states with one more gluon according to eqs.~(\ref{eq:openstring}) and (\ref{eq:closedstring}). The new overlap then has a factor $N_\Lc^2/N_\Lc = N_\Lc$. That is, the color suppression index has not changed.

Now consider eq.~(\ref{eq:badgluon}) for the frail gluon. If the frail gluon were not there, we would have a factor $\sum \delta_{ii}\,\delta_{jj} = N_\Lc^2$. Now insert the frail gluon and divide by $C_\LF \propto N_\Lc$ to normalize the states with one more gluon. The new overlap then has a factor $N_\Lc/N_\Lc = 1$. That is, the color suppression index has increased by 2.

We can now look at what happens when a frail gluon splits and what happens when a healthy gluon splits.

Suppose first that gluon $l$ is a frail gluon and splits, creating another gluon $m+1$. Within the LC+ approximation, there are two configurations for the two gluons. The first is the parallel configuration, $\sum_{a,b} [t^a t^b]_{i'i}\, [t^b t^a]_{j'j}$. With this configuration, an easy calculation shows that both gluons in the new state are healthy gluons. Furthermore, $I(m+1) = I(m)$. The two gluons can also be in the crossed configuration, $\sum_{a,b} [t^a t^b]_{i'i}\, [t^a t^b]_{j'j}$.  With this configuration, an easy calculation shows that both gluons in the new state are also healthy gluons and again $I(m+1) = I(m)$. Thus splitting turns frail gluons into healthy gluons and leaves the color suppression index unchanged.

Suppose next that gluon $l$ is a healthy gluon and splits. Within the LC+ approximation, there are again two configurations for the two gluons. The first is the parallel configuration, $\sum_{a,b} [t^a t^b]_{i'i}\, [t^b t^a]_{j'j}$. With this configuration, an easy calculation shows that both gluons in the new state are healthy gluons and the color suppression index is unchanged: $I(m+1) = I(m)$. The two gluons can also be in the crossed configuration, $\sum_{a,b} [t^a t^b]_{i'i}\, [t^a t^b]_{j'j}$.  With this configuration, we find that both gluons in the new state are now frail gluons and the color suppression index of the state increases: $I(m+1) = I(m) + 2$.

When gluon $l$ splits, creating another gluon $m+1$, there will normally be other gluons already present and the splitting of gluon $l$ can change the health status of some of these other gluons. Let us examine this in a general way, not just for insertions of the new gluon $m+1$ in the ways allowed by the LC+ approximation. Using the $U(N_\Lc)$ treatment of all gluons as carrying the ${\bm 3}\times \bar{\bm 3}$ representation of $U(N_\Lc)$, a color graph consists simply of a set of quark loops. As detailed in eqs.~(\ref{eq:goodgluon}) and (\ref{eq:badgluon}), when a new gluon $m+1$ is added, the color suppression index grows by 2 when the new gluon joins two of the previous quark loops, so that the new gluon is a frail gluon. The color suppression index remains the same when both ends of the new gluon line connect to the same quark loop, so that the new gluon is a healthy gluon.

Consider, then, another gluon with label $s$. Suppose that gluon $s$ is healthy, so that its ${\bm 3}$ and $\bar{\bm 3}$ lines form part of two separate loops. When a new gluon $m+1$ joins these two loops, $I(m+1) = I(m)+2$ and gluon $s$ becomes frail. In all other instances, gluon $s$ remains healthy. Next, suppose that gluon $s$ is frail, so that its ${\bm 3}$ and $\bar{\bm 3}$ lines form part of just one loop. When a new gluon $m+1$ connects this loop to itself, two loops are created from one, so $I(m+1) = I(m)$. It is possible that the ${\bm 3}$ line of gluon $s$ is part of one of the new loops and the $\bar{\bm 3}$ line is part of the other. Then gluon $s$ is now healthy. In other ways of connecting gluon $m+1$, gluon $s$ remains frail. We conclude that splittings that increase the color suppression index may change healthy gluons to frail gluons but never change frail gluons to healthy gluons, while splittings that leave the color suppression index unchanged may change frail gluons to healthy gluons, but never healthy gluons to frail gluons.

We will use these observations in appendices \ref{sec:probabilities} and \ref{sec:nottoomuch}.

\section{Choice of probabilities}
\label{sec:probabilities}

As outlined in section~\ref{sec:weights}, a practical shower evolution computer algorithm implementing the LC+ approximation can use the same method as with a leading color shower to chose the index $l$ of the parton that splits and the index $k$ of the helper parton as well as the momenta and flavors of the daughter partons in the splitting. It remains to chose the color configuration $\{\hat c',\hat c\}_{m+1}$ of the daughter partons. Here, standard methods do not suffice because the functions that play the role of quantum probabilities in the leading color case are now not always positive. One can, however, still use the Monte Carlo method for implementing the sums over colors, choosing the new color configurations at random. The program should chose color configurations $\{\hat c',\hat c\}_{m+1}$ at each splitting according to a probability function $\rho_c(\{\hat c',\hat c\}_{m+1},\{c',c\}_m,l,k,\{\hat f\}_{m+1})$. This function is a probability, so it must be non-negative and normalized to
\begin{equation}
\label{eq:rhocnorm}
\sum_{\{\hat c',\hat c\}_{m+1}}
\rho_c(\{\hat c',\hat c\}_{m+1},\{c',c\}_m,l,k,\{\hat f\}_{m+1})
=
1
\;\;.
\end{equation}
Then, following eq.~(\ref{eq:colorwc}), at each splitting we multiply the accumulated weight for the shower history by a weight factor $w_\Lc$ for the splitting given by $w_\Lc = C/\rho_\Lc$, where $C$ is the color factor given in eq.~(\ref{eq:Cdef}). 

There is an initial weight factor $C_0(\{c',c\}_{N_0})$ from the calculated color density operator for producing an initial color configuration for $N_0$ final state partons at the start of the shower.

At the end of the shower with $N$ final state partons, there is an final weight factor corresponding to the observable to be measured. If the observable is not sensitive to color, then the part of this final weight factor that is associated with color is the color overlap function $\brax{\{c'\}_{N}}\ket{\{c\}_{N}}$.

Thus the complete generated event comes with a weight as given in eq.~(\ref{eq:wtot}),
\begin{equation*}
w_{\rm tot} = \brax{\{c'\}_{N}}\ket{\{c\}_{N}}
\left(\prod_{m=I}^{N-1} 
\frac{C(\{\hat c',\hat c\}_{m+1},\{c',c\}_m,l_m,k_m,\{\hat f\}_{m+1})}
{\rho_c(\{\hat c',\hat c\}_{m+1},\{c',c\}_m,l_m,k_m,\{\hat f\}_{m+1})}
\right)
C_0(\{c',c\}_{N_0})
\;\;,
\end{equation*}
where the color factor $C$ is given in eq.~(\ref{eq:Cdef}). The expectation value of the observable calculated after many shower events is independent of the probability function $\rho_\Lc$ used in the calculation. However, the dispersion of results from the many shower events does depend on the choice of $\rho_c$. In order to have a low dispersion of values and thus an efficient calculation, one wants the net weights $w_{\rm tot}$ not to vary too much. In particular, the weights $w_{\rm tot}$ should never be too large.

In order to keep $w_{\rm tot}$ from being large, it seems that the factors $\rho_\Lc$ in the denominator of eq.~(\ref{eq:wtot}) should be large, but there is a constraint from the normalization condition eq.~(\ref{eq:rhocnorm}): the probability budget should not be spent on configurations where it is not really needed. For that reason, one would set $\rho_\Lc = 0$ for color configurations $\{\hat c',\hat c\}_{m+1}$ for which $C = 0$.

One can also make $\rho_\Lc$ small for color configurations that will, in the end, make $\brax{\{c'\}_{N}}\ket{\{c\}_{N}}$ small. Now, $\brax{\{c'\}_{N}}\ket{\{c\}_{N}} \propto (1/N_\Lc)^{P(m) - p_\LE}$, where $p_\LE$ is the power in the weight defined in section \ref{sec:colorsuppressionI} and $P(m)$ is the color suppression power defined by this relation. Recall that $P(m) \ge I(m)$, where $I(m)$ is the color suppression index defined in eq.~(\ref{eq:colorindex}). The color suppression index either stays the same or grows at each splitting step, so it is sensible to make $\rho_\Lc$ small for any choice of color configuration $\{\hat c',\hat c\}_{m+1}$ that makes $I$ grow.

We consider a $g \to g + g$ splitting for the case that $k \ne l$ and that parton $k$ is also a gluon. Look at $M_\Lc$, the numerator of $C(\{\hat c',\hat c\}_{m+1},\{c',c\}_m,l_m,k_m,\{\hat f\}_{m+1})$, as defined in eq.~(\ref{eq:Mcdef}):
\begin{equation}
\begin{split}
\label{eq:Mcdefgluons}
&M_\Lc(\{\hat c',\hat c\}_{m+1},\{c',c\}_m,l,k,\{\hat f\}_{m+1})
\\&= 
-\sbra{\{\hat c',\hat c\}_{m+1}}
t^\dagger_l(g \to g + g)\otimes 
t_k(g \to g + g)
C^\dagger(l,m+1)
\sket{\{c',c\}_m}
\\&\quad
-
\sbra{\{\hat c',\hat c\}_{m+1}}
C(l,m+1)t^\dagger_k(g \to g + g) \otimes 
t_l(g \to g + g)
\sket{\{c',c\}_m}
\;\;.
\end{split}
\end{equation}
Recall that $C(l,m+1)$ selects color states in which partons $l$ and $m+1$ are color connected. The operators $t^\dagger_l$ and $t^\dagger_k$ here insert the new gluon into the color states according to
\begin{equation}
t^\dagger_l(g \to g + g) = 
\sqrt{C_\LF}\,(a^\dagger_+(l) - a^\dagger_-(l))
\;\;.
\end{equation}
Here $a^\dagger_+(l)$ inserts gluon $m+1$ just to the right of gluon $l$ on its color string in $\ket{\{c\}_m}$ and $a^\dagger_-(l)$ inserts gluon $m+1$ just to the left.\footnote{Here ``left'' and ``right'' refer to the listing of gluon attachments in eqs.~(\ref{eq:openstring}) and (\ref{eq:closedstring}). The adjoint operators $a_\pm(l)$ act analogously on the states $\bra{\{c'\}_m}$. In drawing a picture to represent the bra color state $\bra{\{c'\}_m}$, one normally reverses the direction of the arrow on the color {\bf 3} line. However, this does not change the definition of ``left'' and ``right.''} This notation is explained in more detail in sections 7.2 and 7.3 of ref.~\cite{NSshower}. The factor $\sqrt{C_\LF}$ comes from the normalization of the color states, eqs.~(\ref{eq:openstring}) and (\ref{eq:closedstring}). Thus
\begin{equation}
\begin{split}
M_\Lc&(\{\hat c',\hat c\}_{m+1},\{c',c\}_m,l,k,\{\hat f\}_{m+1})
\\&
= -C_\LF 
\sbra{\{\hat c',\hat c\}_{m+1}}
(a^\dagger_+(l) - a^\dagger_-(l)) \otimes 
(a_+(k) - a_-(k))\,C^\dagger(l,m+1)
\sket{\{c',c\}_m}
\\&\qquad 
-C_\LF 
\sbra{\{\hat c',\hat c\}_{m+1}}
C(l,m+1)\,
(a^\dagger_+(k) - a^\dagger_-(k)) \otimes 
(a_+(l) - a_-(l))
\sket{\{c',c\}_m}
\;\;.
\end{split}
\end{equation}
It will be useful to define  
\begin{equation}
\begin{split}
\label{chiplusminusdef}
\chi_+(k,l,\{c\}_{m}) ={}&
\begin{cases}
 1 
&\text{if $k$ lies just to the right of $l$ on a string in $\{c\}_{m}$}
\\
0  
&\text{otherwise}
\end{cases}
\;\;,
\\
\chi_-(k,l,\{c\}_{m}) ={}& 
\begin{cases}
 1 
&\text{if $k$ lies just to the left of $l$ on a string in $\{c\}_{m}$}
\\
0  
&\text{otherwise}
\end{cases}
\;\;.
\end{split}
\end{equation}
Then
\begin{equation}
\begin{split}
\bra{\{c'\}_m}
a_\pm(k)\, C^\dagger(l,m+1) ={}& 
\bra{\{c'\}_m}a_\mp(l)\, \chi_\mp(k,l,\{c'\}_{m})
\;\;,
\\
C(l,m+1)\,a^\dagger_\pm(k)\ket{\{c\}_m} ={}& \chi_\mp(k,l,\{c\}_{m})\,
a^\dagger_\mp(l)\ket{\{c\}_m}
\;\;.
\end{split}
\end{equation}
Using this notation, we have
\begin{equation}
\begin{split}
\label{eq:Mcstructure}
M_\Lc&(\{\hat c',\hat c\}_{m+1},\{c',c\}_m,l,k,\{\hat f\}_{m+1})
\\&
=\sbra{\{\hat c',\hat c\}_{m+1}}
a^\dagger_+(l) \otimes a_+(l)\sket{\{c',c\}_m}
C_\LF
\big[
\chi_+(k,l,\{c\}_{m}) 
+\chi_+(k,l,\{c'\}_{m})
\big]
\\&\quad  +
\sbra{\{\hat c',\hat c\}_{m+1}}
a^\dagger_-(l) \otimes a_-(l)\sket{\{c',c\}_m}
C_\LF
\big[
\chi_-(k,l,\{c\}_{m}) 
+\chi_-(k,l,\{c'\}_{m})
\big]
\\&\quad  -
\sbra{\{\hat c',\hat c\}_{m+1}}
a^\dagger_+(l) \otimes a_-(l)\sket{\{c',c\}_m}
C_\LF
\big[
\chi_+(k,l,\{c\}_{m}) 
+\chi_-(k,l,\{c'\}_{m})
\big]
\\&\quad  -
\sbra{\{\hat c',\hat c\}_{m+1}}
a^\dagger_-(l) \otimes a_+(l)\sket{\{c',c\}_m}
C_\LF
\big[
\chi_-(k,l,\{c\}_{m}) 
+\chi_+(k,l,\{c'\}_{m})
\big]
\;\;.
\end{split}
\end{equation}

What does eq.~(\ref{eq:Mcstructure}) tell us? There are functions $\chi_\pm(k,l,\{c\}_{m})$ and $\chi_\pm(k,l,\{c\}_{m})$ that describe the color structure of the initial state. One of the functions $\chi_\pm(k,l,\{c\}_{m})$ equals 1 if parton $k$ is connected to parton $l$ in the state $\ket{\{c\}_m}$. One of the functions $\chi_\pm(k,l,\{c'\}_{m})$ equals 1 if parton $k$ is connected to parton $l$ in the state $\bra{\{c'\}_m}$. Potentially, parton $k$ is color connected to parton $l$ in both the bra and the ket state. If parton $k$ is not color connected to parton $l$ in either the bra and the ket state, then $M_\Lc = 0$. Then there is a color factor $C_\LF$. Finally, there is a factor that describes the color structure of the state after the splitting. For instance, $\sbra{\{\hat c',\hat c\}_{m+1}} a^\dagger_+(l) \otimes a_+(l)\sket{\{c',c\}_m}$ is 1 if gluon $m+1$ is inserted to the right of gluon $l$ in the ket state and in the bra state and this factor is zero otherwise. There are four possible ways to insert the new gluon, corresponding to the four terms in eq.~(\ref{eq:Mcstructure}). The first two terms correspond to parallel configurations while the last two terms correspond to crossed configurations. The fact that the crossed configurations come with minus signs tells us that we cannot use the coefficients in eq.~(\ref{eq:Mcstructure}) as Monte Carlo probabilities for the respective insertions.

We recommend defining the probabilities $\rho_\Lc$ according to the status of gluon $l$ in the color configuration $\{c',c\}_m$. If gluon $l$ is a healthy gluon, then the color suppression index remains unchanged if we chose one of the parallel configurations for the insertion of the new gluon, but if we chose one of the crossed configurations, then the color suppression index increases by 2. Thus we can chose parallel configurations with a probability close to 1 times an overall normalization factor and we can choose crossed configurations with a probability proportional to $1/N_\Lc^2$ times the overall normalization factor. We define
\begin{equation}
\begin{split}
\label{eq:healthyrho}
\rho_\Lc&(\{\hat c',\hat c\}_{m+1},\{c',c\}_m,l,k,\{\hat f\}_{m+1})
\\&
=\sbra{\{\hat c',\hat c\}_{m+1}}
a^\dagger_+(l) \otimes a_+(l)\sket{\{c',c\}_m}
\frac{2 C_\LF}{C_\LA\lambda_\rho}
\big[
\chi_+(k,l,\{c\}_{m}) 
+\chi_+(k,l,\{c'\}_{m})
\big]
\\&\quad  +
\sbra{\{\hat c',\hat c\}_{m+1}}
a^\dagger_-(l) \otimes a_-(l)\sket{\{c',c\}_m}
\frac{2 C_\LF}{C_\LA\lambda_\rho}
\big[
\chi_-(k,l,\{c\}_{m}) 
+\chi_-(k,l,\{c'\}_{m})
\big]
\\&\quad  +
\sbra{\{\hat c',\hat c\}_{m+1}}
a^\dagger_+(l) \otimes a_-(l)\sket{\{c',c\}_m}
\frac{1}{\lambda_\rho\,N_\Lc^2}
\big[
\chi_+(k,l,\{c\}_{m}) 
+\chi_-(k,l,\{c'\}_{m})
\big]
\\&\quad  +
\sbra{\{\hat c',\hat c\}_{m+1}}
a^\dagger_-(l) \otimes a_+(l)\sket{\{c',c\}_m}
\frac{1}{\lambda_\rho\,N_\Lc^2}
\big[
\chi_-(k,l,\{c\}_{m}) 
+\chi_+(k,l,\{c'\}_{m})
\big]
,
\end{split}
\end{equation}
where the normalizing factor $\lambda_\rho$ is
\begin{equation}
\label{eq:healthynorm}
\lambda_\rho
= 
\big[\chi(k,l,\{c\}_{m}) + \chi(k,l,\{c'\}_{m})\big]
\;\;.
\end{equation}
Here $\chi(k,l,\{c\}_{m})$, defined in eq.~(\ref{eq:chikl}), is
\begin{equation}
\chi(k,l,\{c\}_{m})
=
\chi_+(k,l,\{c\}_{m})
+
\chi_-(k,l,\{c\}_{m})
\;\;.
\end{equation}
This is the counting factor that appears in the denominator of $C$ in eq.~(\ref{eq:Cdef}).

If gluon $l$ is a frail gluon, then the color suppression index remains unchanged for both parallel and crossed insertions of the new gluons. Therefore, we allow all of the color choices with equal probabilities:
\begin{equation}
\begin{split}
\label{eq:frailrho}
\rho_\Lc&(\{\hat c',\hat c\}_{m+1},\{c',c\}_m,l,k,\{\hat f\}_{m+1})
\\&
=\sbra{\{\hat c',\hat c\}_{m+1}}
a^\dagger_+(l) \otimes a_+(l)\sket{\{c',c\}_m}
\frac{1}{\lambda'_\rho}
\big[
\chi_+(k,l,\{c\}_{m}) 
+\chi_+(k,l,\{c'\}_{m})
\big]
\\&\quad  +
\sbra{\{\hat c',\hat c\}_{m+1}}
a^\dagger_-(l) \otimes a_-(l)\sket{\{c',c\}_m}
\frac{1}{\lambda'_\rho}
\big[
\chi_-(k,l,\{c\}_{m}) 
+\chi_-(k,l,\{c'\}_{m})
\big]
\\&\quad  +
\sbra{\{\hat c',\hat c\}_{m+1}}
a^\dagger_+(l) \otimes a_-(l)\sket{\{c',c\}_m}
\frac{1}{\lambda'_\rho}
\big[
\chi_+(k,l,\{c\}_{m}) 
+\chi_-(k,l,\{c'\}_{m})
\big]
\\&\quad  +
\sbra{\{\hat c',\hat c\}_{m+1}}
a^\dagger_-(l) \otimes a_+(l)\sket{\{c',c\}_m}
\frac{1}{\lambda'_\rho}
\big[
\chi_-(k,l,\{c\}_{m}) 
+\chi_+(k,l,\{c'\}_{m})
\big]
\;\;.
\end{split}
\end{equation}
Here the normalizing factor $\lambda'_\rho$ is
\begin{equation}
\label{eq:frailnorm}
\lambda'_\rho
= 2
\big[
\chi(k,l,\{c\}_{m})
+\chi(k,l,\{c'\}_{m})
\big]
\;\;.
\end{equation}

\section{Avoiding too high color suppression}
\label{sec:nottoomuch}

The LC+ approximation has the good feature that it can generate a shower starting from the color density operator terms $\ket{\{c\}_{m}}\bra{\{c'\}_{m}}$ with $\{c'\}_{m} \ne \{c\}_{m}$. Such a state inevitably leads to $\{c'\}_{N} \ne \{c\}_{N}$ at the end of the shower and thus to a weight factor $\brax{\{c'\}_{N}}\ket{\{c\}_{N}}$ proportional to $1/N_\Lc^2$ to some power. However, the LC+ approximation can generate color states with overlaps proportional to $1/N_\Lc^2$ to a very large power. Suppose that we are content to limit the accuracy of shower evolution (within the LC+ approximation) to $1/N_\Lc$ to a certain power $I_{\rm max}$. For instance, one might set $I_{\rm max}=4$. Then we can expand weight factors $\brax{\{c'\}_{m}}\ket{\{c\}_{m}}$ as a power series in $1/N_\Lc$ and not calculate contributions with too many powers of $1/N_\Lc$. Furthermore, one can track the color suppression index $I(m)$ of generated states. One can arrange not to generate any splitting that makes $I(m+1) > I_{\rm max}$. 

Let us see how to avoid generating states with $I(m+1) > I_{\rm max}$. Suppose that we have a color state with $I(m) = I_{\rm max}$. There may well be some frail gluons in the state. Within the LC+ approximation, these gluons can split. As we have seen in section \ref{sec:growingI}, the splitting of the frail gluons leaves $I(m) = I_{\rm max}$ and turns the frail gluons into healthy gluons.

There will likely also be some healthy gluons in the state. As we have seen, when a healthy gluon splits in the parallel configuration, the color suppression index is left at $I(m+1) = I_{\rm max}$. However, when a healthy gluon splits in the crossed configuration, the color suppression index increases to $I(m+1) = I_{\rm max}+ 2$. This is beyond the limit that we want to maintain. Thus, when $I(m) = I_{\rm max}$, the splitting of a healthy gluon in the crossed configuration should not be allowed.

This analysis suggests that one should modify the probability function $\rho_\Lc$ for a healthy gluon splitting when $I(m) = I_{\rm max}$ by dropping the last two terms in eq.~(\ref{eq:healthyrho}), so that one now uses
\begin{equation}
\begin{split}
\label{eq:healthyrhomod}
\tilde\rho_\Lc&(\{\hat c',\hat c\}_{m+1},\{c',c\}_m,l,k,\{\hat f\}_{m+1})
\\&
=\sbra{\{\hat c',\hat c\}_{m+1}}
a^\dagger_+(l) \otimes a_+(l)\sket{\{c',c\}_m}
\frac{1}{\tilde\lambda_\rho}
\big[
\chi_+(k,l,\{c\}_{m}) 
+\chi_+(k,l,\{c'\}_{m})
\big]
\\&\quad  +
\sbra{\{\hat c',\hat c\}_{m+1}}
a^\dagger_-(l) \otimes a_-(l)\sket{\{c',c\}_m}
\frac{1}{\tilde\lambda_\rho}
\big[
\chi_-(k,l,\{c\}_{m}) 
+\chi_-(k,l,\{c'\}_{m})
\big]
\;\;.
\end{split}
\end{equation}
The corresponding normalizing factor $\tilde\lambda_\rho$ is
\begin{equation}
\label{eq:healthynormmod}
\tilde\lambda_\rho
= 
\chi(k,l,\{c\}_{m})
+\chi(k,l,\{c'\}_{m})
\;\;.
\end{equation}
This leads to a weight factor $w_\Lc = 2 C_\LF/C_\LA = 1 - 1/N_\Lc^2$. Since this procedure in any case drops color suppressed contributions once $I(m)$ has reached $I_{\rm max}$, it makes sense to simply change the weight factor to $w_\Lc = 1$.

If one were to set the probability $\rho_\Lc$ to generate a splitting of a healthy gluon in the crossed configuration to $\rho_\Lc = \epsilon \ll 1$, then one would very rarely generate an event with $I(N) > I_{\rm max}$ with a very large weight. We set $\epsilon = 0$ so that no such events are actually generated. One should note, however, that then there is no control over contributions to the calculated observable at order greater than $1/N_\Lc^{I_{\rm max}}$. Thus, for instance, one can calculate the color overlap $\brax{\{c'\}_{N}}\ket{\{c\}_{N}}$ at the end of the shower as a power series in $1/N_\Lc$ and drop terms of order greater than $1/N_\Lc^{I_{\rm max} - p_\LE}$. (See eq.~(\ref{eq:colorindex}) for $p_\LE$.)

\section{Calculation of gluon exchange phase}
\label{sec:phase}

To define the phase from soft gluon exchange, we consider a gluon with momentum $q = (E,\vec q)$ exchanged between possibly massive final state partons $l$ and $k$. Using the eikonal approximation, and supplying an ultraviolet cutoff $m_1$ and an infrared cutoff $m_2$, the amplitude for this process is
\begin{equation}
G = \mi\, \frac{4\pi\as}{(2\pi)^{4}}\,\bm{T}_l\cdot \bm{T}_k 
\int\! d\vec q \int\! dE\
\frac{\theta(m_2 <|\vec q| < m_1)\ \hat p_l \cdot \hat p_k}
{(\hat p_l\cdot q + \mi\epsilon)(\hat p_k\cdot q - \mi \epsilon)(q^2 + \mi\epsilon)}
\;\;.
\end{equation}
Here $\hat p_l$ and $\hat p_k$ are the momenta of partons $l$ and $k$ as they enter the final state. We let
$\hat p_l =  E_l v_l$
and
$\hat p_k =  E_k v_k$ where
\begin{equation}
\begin{split}
v_l ={}& (1,\vec v_l)
\;\;,
\\
v_k ={}& (1,\vec v_k)
\;\;.
\end{split}
\end{equation}
Thus $\vec v_l$ and $\vec v_k$ are the 3-velocities of the particles in the reference frame that we are using. If either of particles has a non-zero mass, then the absolute value of the corresponding velocity is less than 1. Thus our integral is 
\begin{equation}
G = 
\frac{\mi\, \as\,v_l\cdot v_k}{(2\pi)^{3}}
\, 2 \bm{T}_l\cdot \bm{T}_k \int\! d\vec q \int\! dE\
\frac{\theta(m_2 <|\vec q| < m_1)\ }
{(E - \vec v_l\cdot \vec q + \mi\epsilon)
(E - \vec v_k\cdot \vec q - \mi\epsilon)
(E^2 - \vec q^{\,2} + \mi\epsilon)}
\;\;.
\end{equation}

Now we can perform the $E$ integration to give
\begin{equation}
\begin{split}
\label{eq:G1}
G ={}& \frac{\as\,v_l \cdot v_k}{(2\pi)^{2}} \, 
2 \bm{T}_l\cdot \bm{T}_k
\int\! \frac{d\vec q}{2|\vec q|}\ \theta(m_2 <|\vec q| < m_1)
\\&\times
\frac{1}
{(\vec v_k - \vec v_l)\cdot \vec q + \mi\epsilon}
\left(
\frac{1}{|\vec q| + \vec v_l\cdot \vec q}
+ \frac{1}{|\vec q| - \vec v_k\cdot \vec q}
\right)
\;\;.
\end{split}
\end{equation}
We are interested in the imaginary part of $G$, which comes from replacing $1/[(\vec v_k - \vec v_l)\cdot \vec q + \mi\epsilon]$ by $-\mi\pi\, \delta((\vec v_k - \vec v_l)\cdot \vec q)$:
\begin{equation}
\begin{split}
\label{eq:ImG1}
\mi\, \mathrm{Im}\, G ={}& -\mi\,\frac{\as\,v_l \cdot v_k}{4\pi} \, 
2 \bm{T}_l\cdot \bm{T}_k
\int\! \frac{d\vec q}{2|\vec q|}\ \theta(m_2 <|\vec q| < m_1)
\\&\times
\delta((\vec v_k - \vec v_l)\cdot \vec q)
\left(
\frac{1}{|\vec q| + \vec v_l\cdot \vec q}
+ \frac{1}{|\vec q| - \vec v_k\cdot \vec q}
\right)
\;\;.
\end{split}
\end{equation}

Now choose coordinates with both $\vec v_k$ and $\vec v_l$ in the $x$-$z$ plane, with $\hat z$ in the direction of $\vec v_k - \vec v_l$ and $v_{l,x} > 0$, $v_{k,x} > 0$. Then with $\vec q_\perp = (q_x,q_y)$ and $q_x = -|\vec q_\perp| \cos \theta$, we have
\begin{equation}
\begin{split}
\label{eq:ImG2}
\mi\, \mathrm{Im}\, G
={}& -\mi\,\frac{\as\,v_l \cdot v_k}{4\pi |\vec v_k - \vec v_l|}\,
2 \bm{T}_l\cdot \bm{T}_k
\int_{m_2}^{m_1}\! \frac{d|\vec q_\perp|}{|\vec q_\perp|}
\int_0^{2\pi}\!d\theta\
\frac{1}{1 - v_{l,x}\,\cos\theta}
\;\;.
\end{split}
\end{equation}
In eq.~(\ref{eq:ImG1}), there are two terms that are related by $\vec q \to -\vec q$ and $\vec v_l \leftrightarrow \vec v_k$, but since $v_{l,x} = v_{k,x}$, these two terms are equal. We have just written one of the terms here and multiplied by 2. 

The integrals are trivial:
\begin{equation}
\begin{split}
\label{eq:ImG3}
\mi\, \mathrm{Im}\, G
={}&  -2\pi\mi\,
\frac{\as\,v_l \cdot v_k}{4\pi|\vec v_k - \vec v_l|
[1 - v_{l,x}^2]^{1/2}}\,
2 \bm{T}_l\cdot \bm{T}_k\,
\log(m_1/m_2)
\;\;.
\end{split}
\end{equation}
The velocity dependent prefactor can be simplified using
\begin{equation}
|\vec v_k - \vec v_l|^2[1 - v_{l,x}^2]
= (v_k\cdot v_l)^2 v_{kl}^2
\;\;,
\end{equation}
where
\begin{equation}
v_{kl} = \sqrt{1 - \frac{v_k^2\,v_l^2}{(v_k\cdot v_l)^2}}
=  \sqrt{1 - \frac{m_k^2\,m_l^2}{(\hat p_k\cdot \hat p_l)^2}}
\;\;.
\end{equation}
Then
\begin{equation}
\begin{split}
\label{eq:ImG4}
\mi\, \mathrm{Im}\, G
={}& -2\pi\mi\,
\frac{\as}
{4\pi\,v_{kl}} \,2 \bm{T}_l\cdot \bm{T}_k\,
\log(m_1/m_2)
\;\;.
\end{split}
\end{equation}

If the soft gluon is exchanged between initial state partons $l$ and $k$, there are some sign differences in the derivation but the the result is the same. However, if the soft gluon is exchanged between an initial state parton and a final state parton then we find $\mathrm{Im}\, G = 0$. Thus the general result is 
\begin{equation}
\begin{split}
\label{eq:ImG5}
\mi\, \mathrm{Im}\, G
={}& -2\pi\mi\,\frac{\theta_{kl}}{v_{kl}}\,
\frac{\as}
{4\pi} \,2 \bm{T}_l\cdot \bm{T}_k\,
\log(m_1/m_2)
\;\;,
\end{split}
\end{equation}
where $\theta_{kl}$ is 1 if $k$ and $l$ are both final state partons or both initial state partons and zero otherwise, as in eq.~(\ref{eq:thetakl}).

The logarithm in eq.~(\ref{eq:ImG5}) results from an integration $d|\vec q|/|\vec q|$ between $m_2$ and $m_1$. If we say that $|\vec q|$ is related to shower time $t$ by 
\begin{equation}
\label{eq:tdef}
|\vec q| = C\, Q_0\, e^{-t}
\;\;,
\end{equation}
where $Q_0$ represents the hard momentum scale corresponding to shower time zero and $C$ is any dimensionless constant, then $d|\vec q|/|\vec q| = -dt$, so
\begin{equation}
\begin{split}
\label{eq:ImG6}
\mi\, \mathrm{Im}\, G
={}& -2\pi\mi\,\frac{\theta_{kl}}{v_{kl}}\,
\frac{\as}
{4\pi} \,2 \bm{T}_l\cdot \bm{T}_k\,
\int_{t_1}^{t_2}\! dt
\;\;,
\end{split}
\end{equation}
where $m_1 = C\, Q_0 e^{-t_1}$ and $m_2 = C\, Q_0 e^{-t_2}$. The coefficient of $dt$ here is the differential gluon exchange phase,
\begin{equation}
\begin{split}
\label{eq:phiresult0}
\mi\phi_{lk}(\{p,f\}_{m},t)
={}&
- 2\pi\mi\,
\frac{\theta_{kl}}{v_{kl}}\frac{\as}{4\pi}\ 
2\bm{T}_k\cdot \bm{T}_l
\;\;,
\end{split}
\end{equation}
as reported in eq.~(\ref{eq:phiresult}).\footnote{Note that in eq.~(\ref{eq:phiresult}) the term corresponding to gluon exchange between partons $l$ and $k$ occurs twice in the sum over parton indices. That is why an explicit factor 2 disappears between eq.~(\ref{eq:phiresult0}) and eq.~(\ref{eq:phiresult}).}

We have related the shower time $t$ to $|\vec q|$ in eq.~(\ref{eq:tdef}) with the aid of an unspecified parameter $C$.  Different parton shower schemes use different choices for shower time. For our present purposes, we do not need to make a definite choice. Nevertheless, it is useful to present at least one choice in order to illustrate the physics issues involved.

Note that there is an energy denominator for the parton $l$ plus gluon state before parton $l$ absorbs the gluon, previously emitted from parton $k$: to first order in $|\vec  q|/E_l$, we have
\begin{equation}
\begin{split}
\Delta E ={}& 
\sqrt{\vec p_l^{\,2} + m_l^2}
- |\vec q|
- \sqrt{(\vec p_l + \vec q)^2 + m_l^2}
\sim - \big(|\vec q| + \vec v_l \cdot \vec q \big)
\;\;.
\end{split}
\end{equation}
This is the denominator in the first term in eq.~(\ref{eq:ImG1}). One sensible way to define the shower time for the first term in eq.~(\ref{eq:ImG1}) would be to set 
\begin{equation}
|\vec q| + \vec v_l \cdot \vec q \approx  Q_0\, e^{- t}
\;\;.
\end{equation}
That is, integrating over a range of shower time up to $t_2$ would mean integrating over $\vec q$ subject to $|\vec q| + \vec v_l \cdot \vec q > Q_0 \exp(- t_2)$. Using the change of variables in eq.~(\ref{eq:ImG2}), the relation between shower time and the integration variables is
\begin{equation}
|\vec q\,|(1 - v_{l,x}\cos\theta) \approx Q_0\, e^{- t}
\;\;.
\end{equation}
It is convenient to define $\tilde\theta$ by
\begin{equation}
v_{l,x} = \cos\tilde\theta
\;\;.
\end{equation}
(For massless partons, $\tilde\theta$ is half the angle between $\vec v_l$ and $\vec v_k$.) Then the relation is
\begin{equation}
2|\vec q|
\left[\sin^2(\tilde \theta/2) + \cos\tilde\theta\, \sin^2(\theta/2)\right]
\approx Q_0\, e^{- t}
\;\;.
\end{equation}
We integrate over $\theta$. If $\tilde \theta$ is of order 1, then the integration over $\theta$ is dominated by $\theta$ values of order 1. If $\tilde \theta \ll 1$, then the integration over $\theta$ is dominated by $\theta$ values of order $\tilde \theta$. Thus, after the $\theta$ integration, the effective relation between $|\vec q|$ and the shower time is
\begin{equation}
2|\vec q|\sin^2(\tilde \theta/2) \sim Q_0\, e^{- t}
\;\;.
\end{equation}
Thus one possible relation between $t$ and $|\vec q|$ is eq.~(\ref{eq:tdef}), with $C = 1/[2\sin^2(\tilde \theta/2)]$.

What about the renormalization scale, $\mu_\LR^2(t)$? One might base this on the denominators of covariant Feynman diagrams instead of the energy denominators. Define $D_l(q) = (\hat p_l + q)^2 - m_l^2$, where $\hat p_l$ is the final state momentum of parton $l$, $\hat p_l = (E_l,\vec p_s)$ with $E_l = [\vec p_l^{\,2} + m_l^2]^{-1/2}$. The exchanged gluon is virtual, but let us consider the on-shell contribution with $E = -|\vec q|$. Then 
\begin{equation}
D_l(q) = 2 \hat p_l\cdot q = - 2 E_l \big(|\vec q| + \vec v_l\cdot \vec q\, \big)
\;\;.
\end{equation}
If we use $-D_l$ for $\mu_\LR^2$ we would take $\mu_\LR^2 = 2 E_l Q_0 \exp(-t)$. Similarly, we might take $\mu_\LR^2 = 2 E_k Q_0 \exp(-t)$. For the graph as a whole,
\begin{equation}
\mu_\LR^2 = (E_l + E_k) Q_0\,e^{-t}
\end{equation}
might seem sensible as long as $E_l$ and $E_k$ are not too different.


\end{document}